\documentclass[aps,prd,showpacs,showkeys,notitlepage,preprintnumbers]{revtex4-1}
\usepackage{amsmath,amssymb,amsthm,bm,bbm,color,hyperref}
\usepackage{graphics,subfigure,rotating,etoolbox}
\usepackage[toc,page]{appendix}
\newcommand{\nocontentsline}[3]{}
\newcommand{\tocless}[2]{\bgroup\let\addcontentsline=\nocontentsline#1{#2}\egroup}

\begin{document} 

\title{Approaching $SU(2)$ gauge dynamics with smeared $Z(2)$ vortices}

\author{Roman H\"ollwieser\footnote{Funded by an Erwin Schr\"odinger Fellowship of the Austrian Science Fund under Contract No. J3425-N27.}}
\email{hroman@kph.tuwien.ac.at}
\affiliation{Institute of Atomic and Subatomic Physics, Vienna University of
Technology,\\ Operngasse 9, 1040 Vienna, Austria}
\affiliation{Department of Physics, New Mexico State University, PO Box 30001, Las Cruces, NM 88003-8001, USA}
\author{Michael Engelhardt}
\email{engel@nmsu.edu}
\affiliation{Department of Physics, New Mexico State University, PO Box 30001, Las Cruces, NM 88003-8001, USA}

\date{\today}

\begin{abstract}
We present a method to smear (center projected) $Z(2)$ vortices in lattice gauge
configurations such as to embed vortex physics into a full $SU(2)$ gauge
configuration framework. In particular, we address the problem that using $Z(2)$
configurations in conjunction with overlap  (or chirally improved) fermions is
problematic due to their lack of smoothness. Our method allows us to regain this
smoothness and simultaneously maintain the center vortex structure. We test
our method with various gluonic and fermionic observables and investigate to
what extent we are able to approach SU(2) gauge dynamics without destroying the
original vortex structure.
\end{abstract}



\maketitle

\tableofcontents

\newpage

\section{Introduction}

Being part of the standard model of particle physics, quantum chromodynamics
(QCD) is generally believed to be the correct theory of the strong interactions.
A particular feature of QCD is that its fundamental fermions, the quarks, cannot
be observed as free particles, but are always confined in composite particles,
the hadrons, such as the protons and neutrons. The vortex model~\cite{'tHooft:1977hy,
Vinciarelli:1978kp,Yoneya:1978dt,Cornwall:1979hz,Nielsen:1979xu}
assumes that the center of the gauge group is crucial for confinement. The center
degrees of freedom can be extracted from gauge field configurations by maximal
center gauge (MCG) and center projection\footnote{We use direct maximal center
gauge, which is equivalent to Landau gauge in the adjoint representation,
and which maximizes the squared trace of link variables.}~\cite{DelDebbio:1996mh}. These d.o.f. are dubbed P-vortices and can be viewed as two-dimensional surfaces on the
four-dimensional lattice. They are thought to approximate objects already
present in configurations before the extraction step. These latter objects are
called thick vortices, carry quantized magnetic center flux and are responsible
for confinement according to the vortex model.  The extracted P-vortex surfaces
are complicated, unorientable random surfaces percolating through the lattice. These
and other P-vortex properties are in good agreement with the requirements to
explain confinement, which was shown both in lattice Yang-Mills theory and
within a corresponding infrared effective model, see {\it
e.g.}~\cite{DelDebbio:1996mh,Langfeld:1997jx,DelDebbio:1997ke,Kovacs:1998xm,Engelhardt:1999wr,Engelhardt:2003wm,Hollwieser:2014lxa,Altarawneh:2015bya,Hollwieser:2015qea}.
The vortex model can be applied to other infrared features of QCD not
immediately related to confinement, such as the topological properties of gauge
fields. In particular, it was shown how the topological susceptibility present in
QCD can be calculated from center vortices~\cite{
Bertle:2001xd,Engelhardt:2000wc,Engelhardt:2010ft,Hollwieser:2010mj,
Hollwieser:2011uj,Schweigler:2012ae,Hollwieser:2012kb,Hollwieser:2014mxa} and vortices are also able to explain
chiral symmetry breaking~\cite{deForcrand:1999ms,Alexandrou:1999vx,Engelhardt:1999xw,
Engelhardt:2002qs,Leinweber:2006zq,Bornyakov:2007fz,Hollwieser:2008tq,Hollwieser:2009wka,Bowman:2010zr,Hollwieser:2013xja,Hollwieser:2014osa,Trewartha:2014ona,Trewartha:2015nna}.
This way, the vortex model provides a unified picture for the infrared, low
energy sector of QCD, explaining both confinement and the chiral and topological
features of the strong interaction. A recently published work~\cite{Greensite:2014gra} also favors the center vortex degrees of freedom to be the dominating fluctuations in the QCD vacuum. 

However, some of the properties of full QCD are obscured in the P-vortex
(vortex-only) configurations, especially when it comes to topological properties
in connection with fermions. In particular, we address the problem of reproducing a finite chiral condensate
in center projected (Z(2)) configurations, using overlap
(and chirally-improved) Dirac operators. Low-lying eigenmodes and
also zero modes are not found in these configurations; the spectra show a
large eigenvalue gap for vortex-only configurations. In~\cite{Hollwieser:2008tq}
the reason for the large gap in the vortex-only case was shown to be connected
to the lack of smoothness of center projected lattices, {\it i.e.}, maximally
nontrivial plaquettes - the vortex plaquettes. In that case, the exact symmetry
of the overlap operator is strongly field-dependent, and does not really
approximate the chiral symmetry of the continuum theory. It was further shown
that the overlap operator produces more reasonable spectra when applied to a
smoother version of the center projected lattice. The procedure applied however
requires knowledge of the original lattice. In the present work, we want to
explore another strategy: Starting from $Z(2)$ vortex configurations, we want to
embed the corresponding physics in full $SU(2)$ configurations by smoothing the
thin vortices\footnote{We also want to mention a similar attempt to smear
vortices in~\cite{Bruckmann:2003yd}, although for an explicit example of a vortex
configuration and with a different goal, {\it i.e.}, to understand topological charge contributions from vortex writhe and intersections. In the present investigation we want to develop a smearing method for random vortex configurations.}\cite{Bruckmann:2003yd}.
We speculate that the infrared aspects of the QCD vacuum can be understood in
terms of thick center vortices, which can be derived from thin vortex structures
by a new smearing method, introduced in the following. The idea and the
goal of this method can be summarized as follows: Remove maximally nontrivial
plaquettes without destroying the vortex structure and reproduce gluonic and
fermionic observables of the original $SU(2)$ configurations using the smoothed center
vortices. Section~\ref{sec:meth} presents the development of the new vortex
smearing method, including a brief summary of its relevant steps in
Sec.~\ref{sec:sum}. In section~\ref{sec:res} we apply the
vortex smearing method to 1000 $Z(2)$ vortex configurations, obtained from Monte
Carlo-generated full $SU(2)$ gauge fields after maximal center gauge and center
projection. We present results for various gluonic and fermionic observables
comparing the original (full) and vortex smeared lattice configurations.
Section~\ref{sec:class} gives more insight into the actual effect of the vortex
smearing by applying it to classical, {\it i.e.}, planar and spherical vortex
configurations. We finish with concluding remarks in section~\ref{sec:con}.

\section{Method}\label{sec:meth}

Center vortex gauge fields are generally not smooth enough to fulfill
e.g. the L\"uscher condition~\cite{Luscher:1981zq}, especially for thin vortex
(Z(2)) configurations. The problem, mentioned above, of the overlap Dirac operator 
with $Z(2)$ vortex configurations is caused by maximally nontrivial plaquettes, which are the
locations where P-vortex flux pierces lattice planes. The actual (closed) vortex
surface is located on the dual lattice\footnote{To each cell $c_n$ on the original
lattice corresponds a cell $c_{D−n}$ of the dual lattice, where $D$ is the
dimension of the lattice. The dual lattice is shifted with respect to the
original lattice half a lattice spacing $a/2$ in each direction. In four
dimensions, there is a one-to-one correspondence between original and dual
plaquettes, {\it i.e.}, $xy$-plaquettes correspond to $zt$-plaquettes etc. and
corresponding plaquettes share the same center point.}, but we call plaquettes
with center flux $-\mathbbm{1}$ dual vortex plaquettes or simply vortex
plaquettes in the following. The idea is to smooth out the thin vortices to
regain a finite thickness. This can be understood in two ways. One is to
distribute the center vortex flux of the vortex plaquettes, {\it i.e.}, Tr
$U_{\mu\nu}=-2$ to several (neighboring) plaquettes. On the other hand, we can
think in terms of link variables, applying a smooth link profile, {\it i.e.}, a
"slow" rotation of the links within several lattice spacings instead of the
sudden jump from $+\mathbbm{1}$ to $-\mathbbm{1}$ or the other way around. Both
ideas thicken the vortices in the sense that the center flux is not restricted
to a singular surface but spread out over a few lattice spacings.
We are going to discuss both approaches, which are related of course,
starting with a simple rotation smearing.

\subsection{Link rotation smearing}\label{sec:rot}
\begin{figure}[h]
\centering
\includegraphics[width=.9\linewidth]{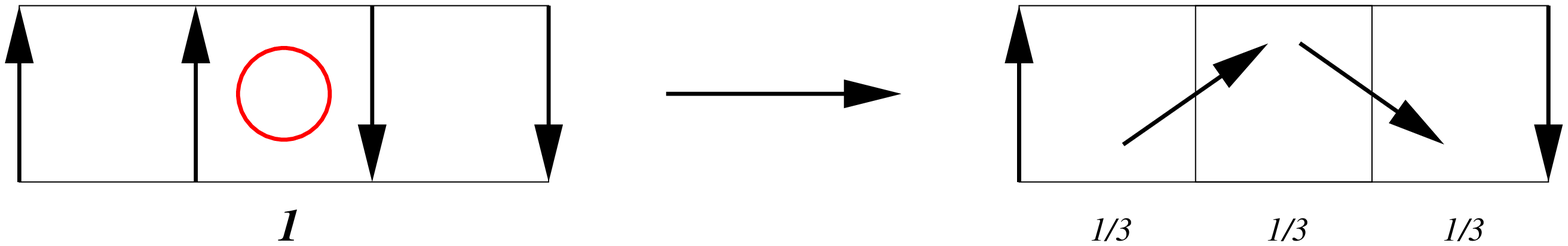}
\caption{Illustration of vortex smearing: the $Z(2)$ links of the thin
vortex plaquette (indicated by red circle) and the link rotation profile of the
smeared (thick) vortex. Using steady rotations of $\pi/3$ in the same $U(1)$
subgroup, the center vortex flux $-\mathbbm{1}$ of the thin (dual) vortex
plaquette is distributed uniformly within the original and neighboring
plaquettes. This is indicated by the numbers below the plaquettes as multiples
of $\pi$, indicating the composition of the total center vortex flux
$\exp{i\pi}\equiv-\mathbbm{1}$. The $U(1)$ subgroup for each vortex plaquette is
chosen such as to minimize the affected plaquettes.}\label{fig:links}
\end{figure}

We start with identifying the vortex plaquettes, {\it i.e.}, plaquettes with
Tr $U_{\mu\nu}=-2$ in a given $Z(2)$ configuration. The plaquette $U_{\mu\nu}$ is
given by the product of four links, {\it i.e.}, $U(\vec x)_{\mu\nu}=U(\vec
x)_{\mu}U(\vec x+\hat\mu)_{\nu}U^\dagger(\vec x+\hat\nu)_{\mu}U^\dagger(\vec
x)_{\nu}$. In fact, for $Z(2)$ gauge variables $U=\pm\mathbbm{1}$ and the ordering of
the product is irrelevant. Therefore
we next identify the pair of opposite links causing the overall $-\mathbbm{1}$
of the vortex plaquette, either $U(\vec x)_\mu U(\vec x+\hat\nu)_\mu$ or $U(\vec
x)_\nu U(\vec x+\hat\mu)_\nu$ gives $-\mathbbm{1}$. Then we smear these two
links as illustrated in Fig.~\ref{fig:links}, {\it i.e.}, the $\pm\mathbbm{1}$
links are rotated away from the center values in order to get
a smooth transition from $+\mathbbm{1}$ to $-\mathbbm{1}$ instead of an instant
jump between neighboring links. This of course removes the maximally nontrivial
(vortex) plaquette, indicated in Fig.~\ref{fig:links} with a (red) circle, and
spreads its center flux ($-\mathbbm{1}$) within its neighboring plaquettes. In
Fig.~\ref{fig:links} we use steady rotations of $\pi/3$, {\it i.e.}, the
two links are given by rotations of $\pi/3$ and $2\pi/3$, which distributes the
vortex flux uniformly to the three plaquettes, each carrying $1/3$ of the total
center vortex flux. Of course, the link rotations also lead to nontrivial
plaquettes in the directions orthogonal to the plotted plane, but with opposite
flux directions in forward and backward directions. These additional
contributions will not be treated individually since different vortex structures
would make the procedure very complex, but they will be taken into account by
the plaquette minimization technique discussed below.

\begin{figure}[h]
	\centering
	a)\includegraphics[width=.48\linewidth]{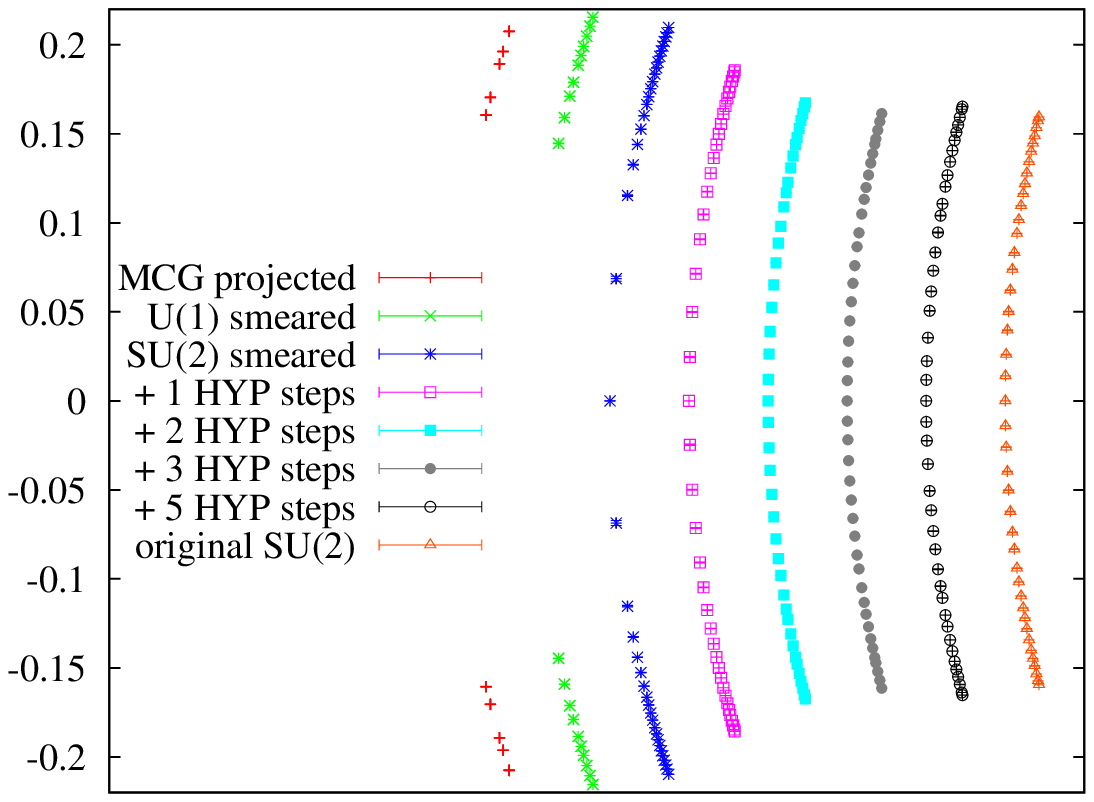}
	b)\includegraphics[width=.48\linewidth]{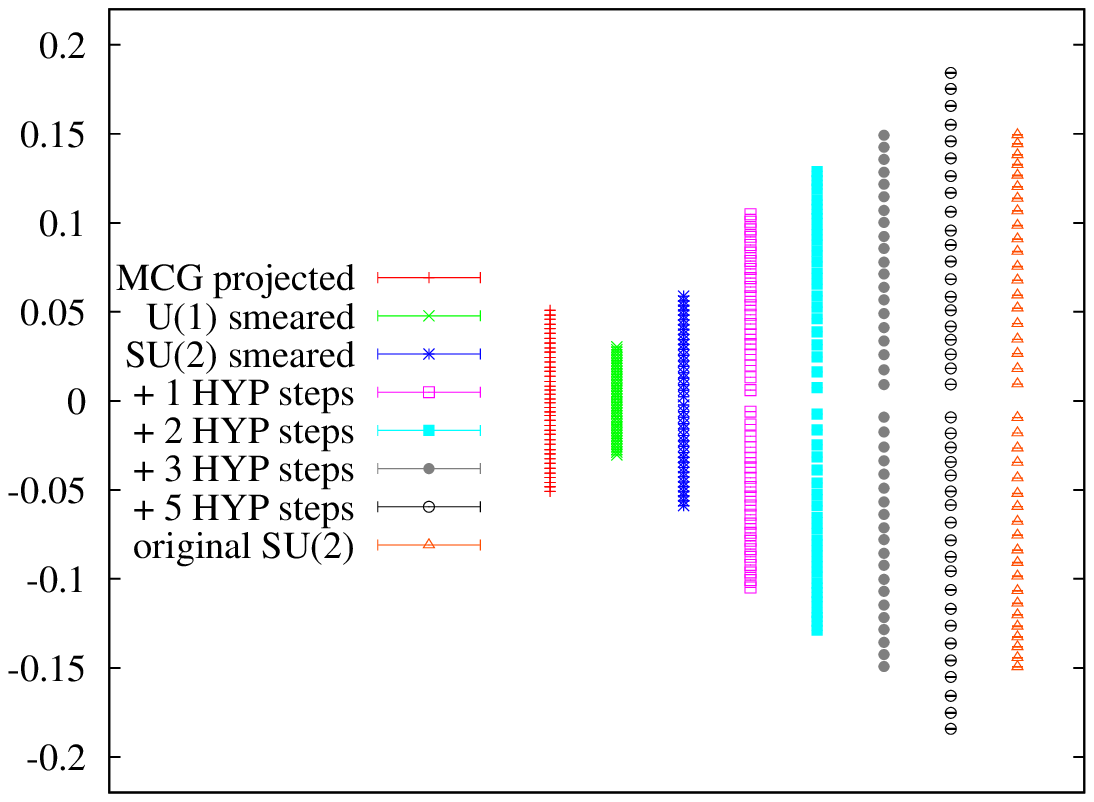}
	\caption{20 lowest a) overlap and b) asqtad staggered eigenvalues for
	original (full) $SU(2)$, Maximal Center Gauge projected $Z(2)$, $U(1)$ and $SU(2)$ vortex smeared configurations with different numbers of HYP smearing steps.}
	\label{fig:hyp}
\end{figure}

There are of course many ways to implement these rotations, therefore we perform a
systematic analysis to explore which method is best suited for our goal.
In order to reproduce the original vortex structure the smoothed $SU(2)$ links
should stay in the same hemisphere of the corresponding $S_3$ after MCG projection,
hence we apply only rotations smaller than $\pi/2$. A statistical analysis
shows that the maximally nontrivial plaquette reduces more effectively if the
two corresponding links rotate in the same $U(1)$ subgroup of $SU(2)$, however the
situation is nevertheless not trivial: It is interesting to observe that,
overall, the smallest plaquette values are observed for link rotations of $\pi/5$ or $\pi/6$ away from
their corresponding center elements, whereas for $\pi/3$ and $\pi/4$ we still
observe plaquettes with Tr $U_{\mu\nu}=-2$. These situations can occur at vortex
corners or intersections, where different rotations affect single plaquettes. In
fact, we can easily construct situations where multiples of $\pi/3$ or $\pi/4$
add up to $\pm\pi$, resulting in $\exp{\pm i \pi}=-\mathbbm{1}$, see
also~\cite{Hollwieser:2012kb}. For rotations up to $\pi/5$ MCG and center
projection also reproduce the original vortex structure very well. If we now
restrict all rotations to the same $U(1)$ subgroup, the smeared configurations
still show a gap in the overlap spectra, {\it i.e.}, no near-zero modes are
found, as it is the case for maximal Abelian projected configurations.
Therefore, we generalize the procedure; for each vortex plaquette we randomly choose a $U(1)$ subgroup to
perform the smearing rotation. This way, the eigenvalue gap closes and a finite
density of near-zero modes shows up. In order to improve the result, instead of
randomly choosing the $U(1)$ subgroups we try to minimize the affected plaquettes.
Now the maximally nontrivial plaquette reduces further, however the
eigenvalue spectra do not change significantly. Finally, we try various smearing
methods, {\it i.e.}, APE, EXP, LOG and their improved and HYP versions~\cite{Falcioni:1984ei,Teper:1987wt,Albanese:1987ds,Morningstar:2003gk,Hasenfratz:2007rf,Durr:2007cy}
, to make the smeared vortex configurations even smoother. Even though the average
plaquette now reduces further, intriguingly the maximally nontrivial
plaquette moves back towards $-\mathbbm{1}$. While the standard smearing
routines act too mildly, the improved ones, using bigger Wilson loops or the
hypercubic nesting trick, smooth the configurations enough within a few smearing
steps. In Fig.~\ref{fig:hyp} we show the overlap and asqtad staggered spectra
for original (full) $SU(2)$, MCG projected $Z(2)$ and various vortex and HYP smeared
configurations. We see that 2-3 HYP smearing steps seem to be appropriate to
reproduce the original Dirac spectra. However, even by systematically scanning
the parameter sets for the various smearing routines, we cannot avoid that the
vortex structure is deformed during the smearing and we are not able to
reproduce the initial vortex configuration after MCG projection. Therefore we
rule out standard smearing routines and try yet another strategy, {\it i.e.}
distributing the vortex flux of a single vortex plaquette (Tr $U_{\mu\nu}=-2$)
to several (neighboring) plaquettes.

\subsection{Center vortex flux distribution}
The first step is again to identify the vortex plaquettes, {\it i.e.}
plaquettes with Tr $U_{\mu\nu}=2\exp{\pm i\pi}=-2$ in a given $Z(2)$ configuration. The plaquettes and therefore also the vortex structure are by definition gauge invariant. The
$Z(2)$ links however are not and therefore the pair of links giving the overall
$-\mathbbm{1}$ identified in the link smearing procedure discussed above 
is somewhat arbitrary. For the link rotation smearing we
restricted ourselves to the direction of the jump from $+\mathbbm{1}$ to
$-\mathbbm{1}$ (or the other way around) to perform the smooth rotation since
this jump in fact defines the vortex in the $Z(2)$ configuration within its
specific gauge. Thinking in terms of vortex flux distribution, however, there is
no such preferred direction and we want to distribute the flux symmetrically
among neighboring plaquettes. Therefore we now smear all four links of the
vortex plaquette by individual link rotations in the same $U(1)$ subgroup in order
to guarantee uniform flux distributions. The $U(1)$ subgroup is either chosen
randomly for each vortex plaquette, or such as to reduce the plaquettes
orthogonal to the vortex plaquettes, affected by the individual link rotations. 

\begin{figure}[h]
\centering
\includegraphics[width=.9\linewidth]{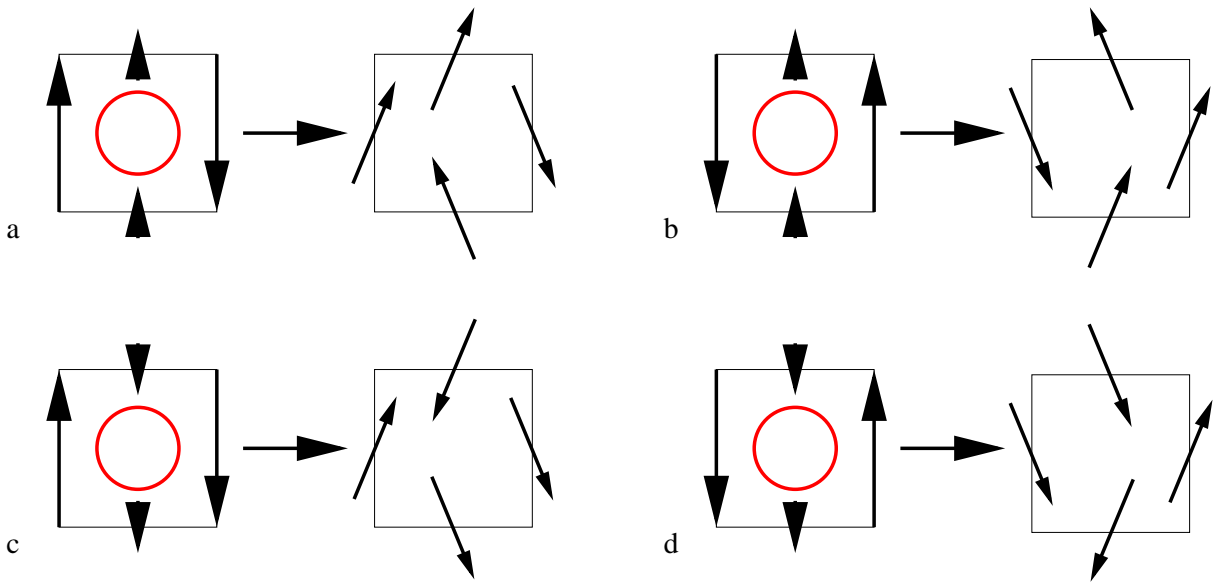}
\caption{Distribution of center vortex flux: the $Z(2)$ links of the initial
	vortex plaquette ($\exp{\pm i\pi}=-\mathbbm{1}$) and the corresponding
	smeared links using $\pm\pi/8$ rotations away from the initial links. We
	distinguish four cases according to the initial link configurations. The
	smeared links distribute the flux as shown in Fig.~\ref{fig:flx}, for
rotations in the same $U(1)$ subgroup.}\label{fig:smrflx}
\end{figure}

\begin{figure}[h]
	\centering
	\includegraphics[width=.9\linewidth]{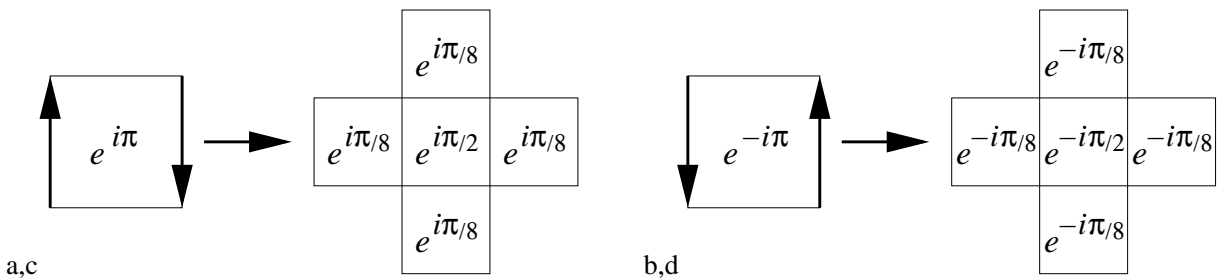}
	\caption{Vortex flux distributions for smeared vortex plaquettes with link
	rotations shown in Fig.~\ref{fig:smrflx}. They add to a total flux of
	$\pm\pi$, consisting of $\pm\pi/2$ at the original vortex and $\pm\pi/8$ at
	neighboring plaquettes. These vortex flux distributions among plaquettes are
	valid for link rotations in the same $U(1)$ subgroup only, and do not
	necessarily correspond to the final vortex smeared configurations, where the
	individual rotations are performed in different $U(1)$ subgroups, such as to
	minimize the affected (including orthogonal) plaquettes, see text.}
	\label{fig:flx}
\end{figure}

Fig.~\ref{fig:smrflx} shows an example of how to change the individual links using $\pm\pi/8$ rotations away from the original links. We have to distinguish four different cases according to the initial link configurations in order to get flux distributions of $\exp{\pm i\pi/2}$ and $\exp{\pm i\pi/8}$ at the original and four neighboring plaquettes summing up to the initial center element. The flux distributions for the individual cases are shown in Fig.~\ref{fig:flx}. There are of course many ways to distribute the center
vortex flux symmetrically and uniformly among various plaquettes, and even more
ways to realize these distributions by different link configurations. In order
to reproduce the initial vortex configuration however, we have to restrict the
individual rotations to $\pm\pi/8$ giving the maximally possible center flux
distribution shown in Fig.~\ref{fig:flx}. 

Smearing the $Z(2)$ configuration in this way (with individual rotations up to
$\pm\pi/8$) is not enough to close the gap in the overlap spectrum. Standard
smearing routines APE, EXP and LOG are again too mild to resolve this problem, whereas their improved HYP versions again destroy the vortex structure. Choosing the $U(1)$ subgroup for each of the four
rotations individually in order to minimize the affected plaquettes, instead of
applying the individual rotations to the four links in the same subgroup, not
only destroys the uniform flux distribution but also does not close the gap in
the overlap spectrum. Distributing the flux to more and more plaquettes
dissolves the vortex structure in a sense, especially when it comes to edges and
corners of the vortex structure. We therefore resort to yet a further technique:
resolving the vortex structure within a finer lattice in order to generate more
lattice spacings in which to smear it. 

\subsection{Vortex (lattice) refinement and blocking}\label{sec:ref}
\begin{figure}[h]
	\centering
	\includegraphics[width=.8\linewidth]{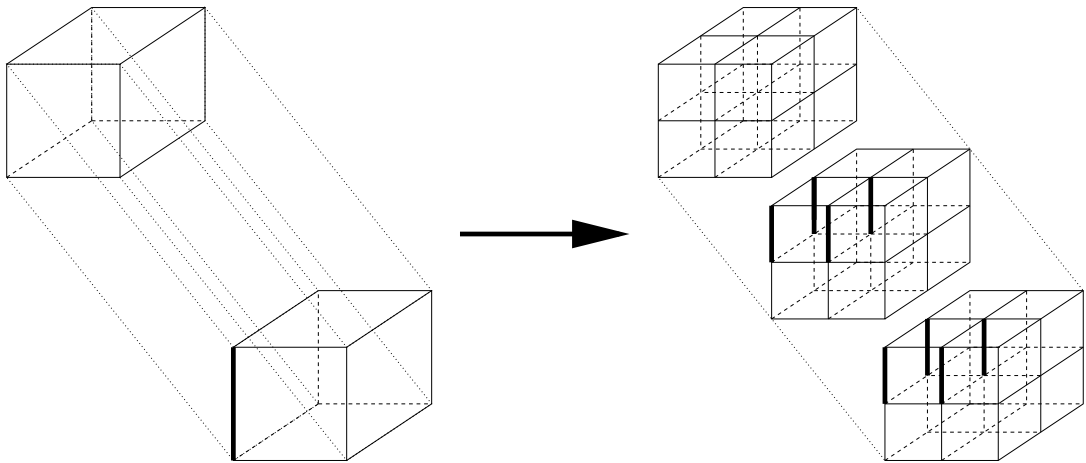}
	\caption{Refinement routine: Example of one $-\mathbbm{1}$ (fat z-) link giving
	eight $-\mathbbm{1}$ and eight $\mathbbm{1}$ links after refinement.} 
	\label{fig:ref}
\end{figure}

By smearing the vortex structure on the original lattice we quickly end up
destroying the vortex structure, since we cannot treat every single vortex
edge, corner, writhing or intersection point, etc. independently. However, we
can obtain more freedom in treating these structures by putting the vortex configuration on a finer lattice. 
For $Z(2)$ gauge links the refinement procedure can be defined straightforwardly
and yields exactly the same vortex structure but on a finer lattice. 
The refinement procedure is illustrated in Fig.~\ref{fig:ref}; we double the
number of links in each direction, hence the lattice volume increases by a
factor of $2^4=16$. If the initial link was $\mathbbm{1}$ we only insert
two $\mathbbm{1}$ links, however if the initial link has value $-\mathbbm{1}$, we
insert a $\mathbbm{1}$ and a $-\mathbbm{1}$ link in forward direction. The new
link pairs are copied forward by half the initial lattice spacing in all orthogonal
directions, {\it e.g.} an x-link $U_x(\vec x)=-\mathbbm{1}$ at $\vec
x=(x,y,z,t)$ gives $\tilde U_x=\mathbbm{1}$ at $(x,y,z,t)$, $(x,y+1/2,z,t)$,
$(x,y,z+1/2,t)$, $\ldots$, $(x,y+1/2,z+1/2,t+1/2)$ and $\tilde U_x=-\mathbbm{1}$ at $(x+1/2,y,z,t)$, $(x+1/2,y+1/2,z,t)$, $(x+1/2,y,z+1/2,t)$, $\ldots$, $(x+1/2,y+1/2,z+1/2,t+1/2)$ 
. 

\begin{figure}[h]
	\centering
	\includegraphics[width=.8\linewidth]{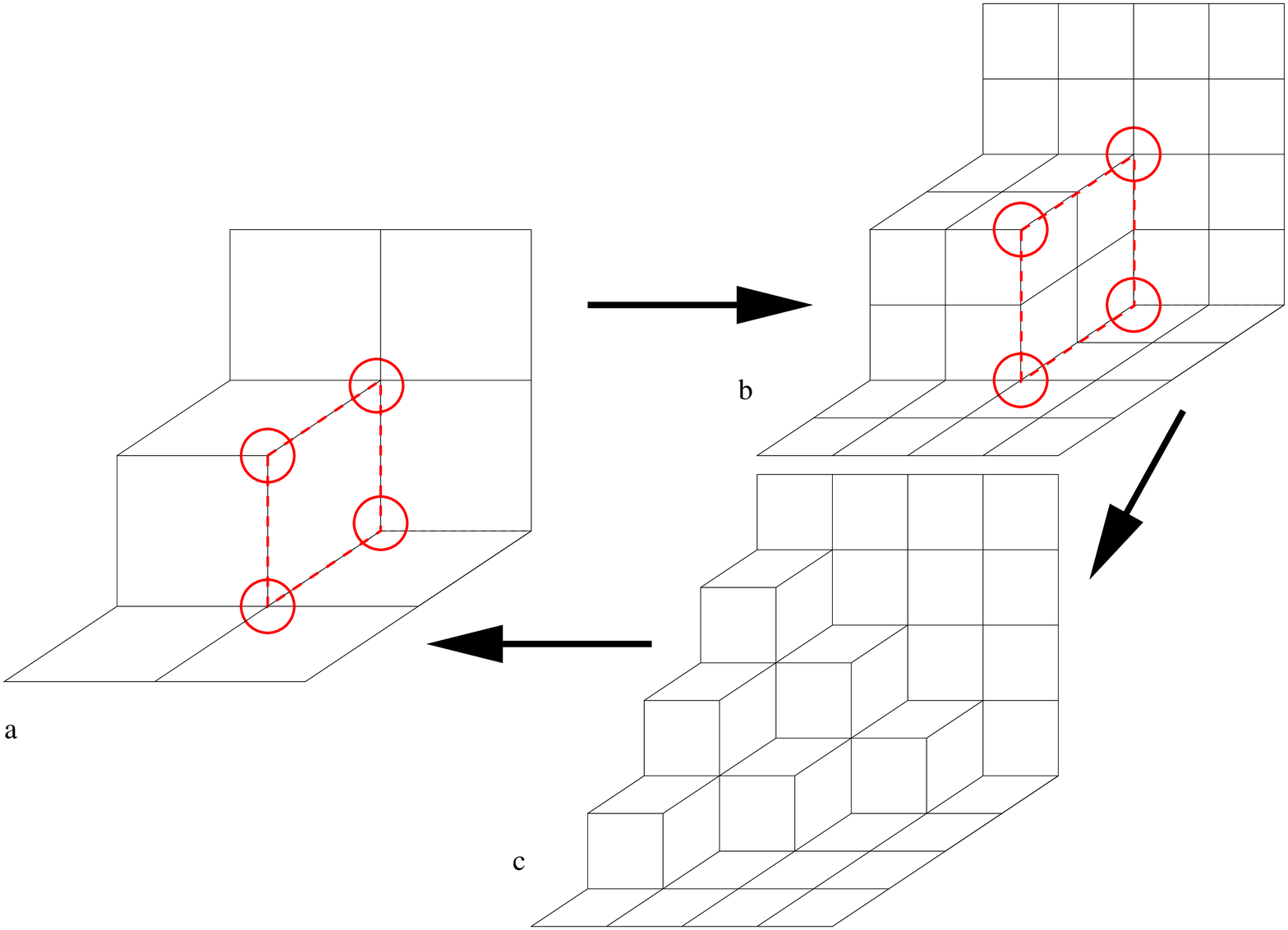}
	\caption{Vortex refinement (a$\rightarrow$b), smearing (b$\rightarrow$c) and
	blocking (c$\rightarrow$a): Example of a partial vortex surface on the dual
lattice including edges and corners (red circles). Smearing the central (red
dotted) plaquette in the coarse lattice (left) influences all other attached plaquettes since we smear
in the directions orthogonal to the plaquette ({\it i.e.}, its dual plaquette on
the original lattice) and the vortex corners (and edges) are deformed.
For the refined lattice on the right, smearing individual plaquettes only affects
its direct neighbors if they are connected via an edge and therefore plaquettes
attached to the vortex corners do not affect each other directly during
smearing. Further, deformations of the refined vortex surface during smearing
within the original lattice spacing (as indicated in c) will reproduce the
original vortex structure after blocking.}
	\label{fig:vcoref}
\end{figure}

Just as refinement gives the same vortex structure on the finer lattice, the
inverse procedure, blocking, again gives the original configuration. During
blocking, the copies between the coarse lattice planes are thrown away and the two refined links, {\it e.g.}
$\tilde U_x(\vec x=(x,y,z,t))$ and $\tilde U_x(\vec x=(x+1/2,y,z,t))$, are multiplied
to reproduce the original $U_x(\vec
x=(x,y,z,t))=\mathbbm{1}\cdot\pm\mathbbm{1}=\pm\mathbbm{1}$ link, see also
Sec.~\ref{sec:block} for more details. On the refined
lattice, however, one now has the advantage that deformations of the vortex surface within the original lattice
spacing still yield the correct vortex structure after blocking. Visualizing the actual
(closed) vortex surface on the dual lattice, refinement not only multiplies the
number of vortex plaquettes, yielding more (refined) plaquettes (and
therefore also links) at {\it e.g.} vortex edges or corners, as shown in
Fig.~\ref{fig:vcoref}, but also adds links (and plaquettes) in directions orthogonal to the vortex surface, which we may use to smear
our configurations. This way we may not make the vortex thicker in terms of the
original lattice, but we can make the configuration smoother by adding additional
rotations to the (refined) links or distributing the vortex flux to more (refined) plaquettes.

The refinement and blocking procedures are discussed more extensively
in~\cite{Engelhardt:2000wc,Bertle:2001xd} where they help to remove ambiguities
from vortex intersections, corners or writhing points, or, respectively, ultraviolet
artifacts during vortex topological charge calculation. In fact, during
vortex topological charge calculation the lattices are refined threefold, {\it
i.e.}, resulting in a lattice spacing $a/3$ in order to resolve intersection
lines and to make sure that neighboring vortex surfaces cannot interact with
each other. Even though smearing on finer and finer lattices might be more and
more efficient, we restrict ourselves to a twofold refinement to limit the
computational cost for the overlap Dirac operator evaluation. Nevertheless,
resolving vortex structure ambiguities via refinement seems also useful for the
present
problem of smearing the vortex surface, since such structures, {\it i.e.}, vortex
intersections, corner or writhing points, are more easily deformed 
during the smearing process.

Having cast a configuration on a finer lattice, for the smearing routines we
again start by identifying the (maximally nontrivial) center vortex plaquettes with Tr $U_{\mu\nu}=-2$. On the refined
lattice there are of course more center vortex plaquettes compared to the
original lattice. As mentioned above, the lattice volume, {\it i.e.}, the number
of lattice points and equally the number of links and plaquettes is multiplied by $2^4=16$. As can be seen in Fig.~\ref{fig:ref}
however, the number of negative links is only increased by a factor of eight,
since we also add eight positive links for an initially negative link. The
number of vortex plaquettes finally is increased by four, as can be checked in
Fig.~\ref{fig:ref} too, but can also be easily understood in terms of the dual lattice,
where the vortex surface forms a closed surface of dual plaquettes, which are
simply refined to four smaller plaquettes each (see also Fig.~\ref{fig:vcoref}). 
We should therefore note that the vortex density is reduced by a factor four on
the refined, original lattice; this ultimately was the initial goal of the refinement procedure, resolving vortex structure ambiguities by increasing the distance between close
vortex structures or neighboring surfaces which would otherwise interact after
thickening them during vortex smearing.

\subsection{Refined link rotation smearing}\label{sec:reflrs}
\begin{figure}[h]
	\centering
	\includegraphics[width=.9\linewidth]{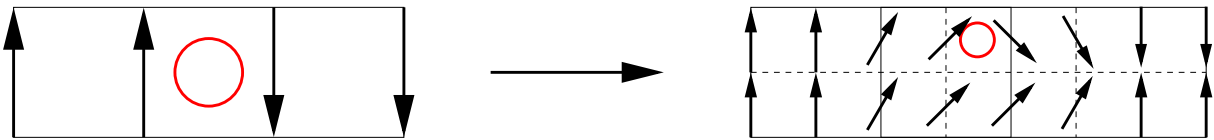}
	\caption{Refined link rotation smearing: The smearing rotation from
	$\mathbbm{1}$ to $-\mathbbm{1}$ can be extended to four links without
disturbing any other vortex plaquettes. Center vortex plaquettes of the initial
and refined (or MCG projected after smearing) configurations are again indicated
by red circles. Link rotations in the top/bottom row are given by $\pi/8, \pi/4,
3\pi/4, 7\pi/8$ and $\pi/8, \pi/4, \pi/4, \pi/8$ and the individual $U(1)$
subgroups are chosen such that the affected plaquettes are minimized.}
	\label{fig:refsmr}
\end{figure}

As in section~\ref{sec:rot}, we locate the opposite link pairs causing
(negative) vortex plaquettes now on the refined lattice and smooth out the jump
from $\mathbbm{1}$ to $-\mathbbm{1}$ or the other way around.
On the refined lattice we can extend the rotation to four links without disturbing any
neighboring vortex plaquettes and additionally apply rotations to the
neighboring links in link direction from the refinement procedure, see
Fig.~\ref{fig:refsmr}. We rotate the individual links either $\pi/8$ or $\pi/4$
away from their initial center elements and still reproduce the initial vortex configuration after blocking. 
At this stage, we made another interesting observation: Applying the Dirac
operators to the refined, smeared configurations gives spurious results,
the spectra show an even larger gap with individual eigenvalue bands, even for
the asqtad staggered fermions. The problem seems related to the refinement
procedure, since the asqtad (and the standard) staggered Dirac operator,
which identifies zero modes well on center projected configurations, gives
unphysical spectra
already for the simply refined (non-smeared) configurations. This observation is
insofar interesting as the refined lattices represent the same vortex
configurations, except that they of course are half as thick compared to the
original lattice due to the smaller lattice constant. The only difference to
vortex configurations on originally finer lattices seems to be the fact that
negative links only arise at every second (even) lattice slice of the corresponding
link direction ({\it e.g.} x-links in x-slices, {\it i.e.}, odd x-slices contain
only $\mathbbm{1}$ x-links). Even though this seems not very likely in Monte
Carlo generated vortex configurations, the fact that it causes a problem in
identifying Dirac operator zero modes is worth noting. While the staggered Dirac operator might fail because of its even/odd-lattice implementation, we cannot think of any plausible explanation for the failure of the overlap Dirac operator. However, we find in the next section that this problem is not as severe for the overlap compared to the
staggered fermions, since it is not present for the former in the case of flux
distribution smearing on refined lattices, and therefore we will not discuss it further for now.

\begin{figure}[h]
	\centering
	\includegraphics[width=.9\linewidth]{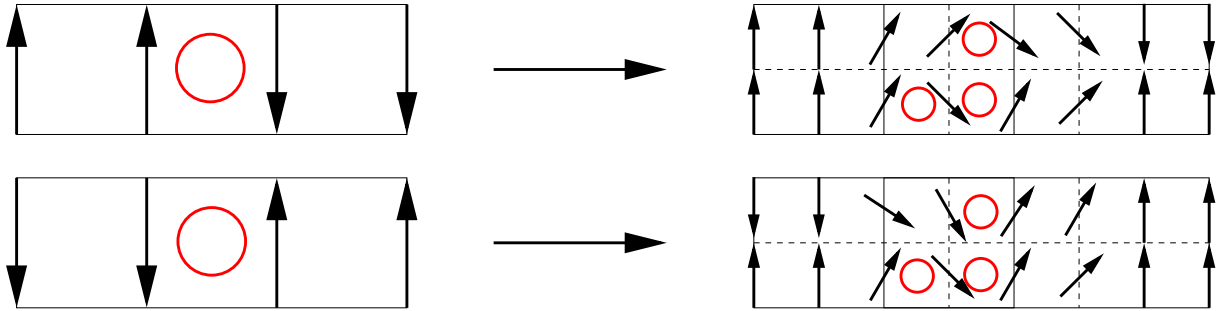}
	\caption{Refined link rotation smearing for $\mathbbm{1}\rightarrow-\mathbbm{1}$ 
		and $-\mathbbm{1}\rightarrow\mathbbm{1}$ link pairs. The individual
		$U(1)$ rotations of the links from left to right in the top and bottom
		rows of the two cases are given by $\pi/8, \pi/4, 5\pi/8, 3\pi/4; \pi/8,
		3\pi/4, \pi/8, \pi/4$ and $5\pi/8, 7\pi/8, \pi/8, \pi/8; \pi/8,
		3\pi/4, \pi/8, \pi/4$.
The odd (lower) lattice slices also contain a certain number of smeared links
close to $-\mathbbm{1}$, introducing additional vortex plaquettes (red circles)
which distort and therefore smear the vortex surface in link direction. 
The $U(1)$ subgroups are not the same for the individual rotations, but chosen such as to minimize the corresponding plaquettes.}
	\label{fig:refsmr1}
\end{figure}

\begin{figure}[h]
	\centering
	\includegraphics[width=.9\linewidth]{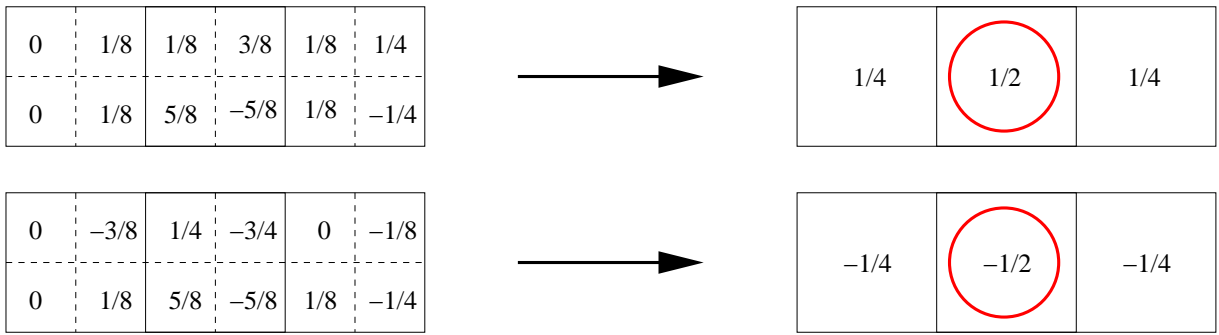}
	\caption{Vortex flux distributions for refined link rotation smeared vortex
	plaquettes with $\mathbbm{1}\rightarrow-\mathbbm{1}$ and
	$-\mathbbm{1}\rightarrow\mathbbm{1}$ link pairs in terms of fractions of
	$\pm\pi$ adding up to a total flux of $\pm\pi$, with $\pm\pi/2$ at the
	original vortex and $\pm\pi/4$ for neighboring plaquettes. These vortex flux
	distributions among plaquettes are valid for link rotations in the same U(1)
	subgroups only, and do not necessarily correspond to the final vortex
	smeared configurations, where the individual rotations are chosen such as to minimize the corresponding plaquettes.}
	\label{fig:refsmr2}
\end{figure}

In order to overcome this issue in the present case of link rotation smearing, we try to spread the vortex structure back to its
initial thickness, {\it i.e.}, two lattice constants of the refined lattice,
during the smearing process. Therefore we simply add another smeared link close
to $-\mathbbm{1}$ to the corresponding odd lattice slice of the refined
lattice. There are many different ways of doing the individual rotations;
finally, we settle with the smearing rotations shown in Fig.~\ref{fig:refsmr1},
which seem to give the best results. The figure shows all rotations in the same U(1)
subgroup, however, in practice, the subgroups for the individual rotations are
chosen such as to minimize the corresponding plaquettes, {\it i.e.}, reduce the
maximally nontrivial plaquette among the six plaquettes affected by the link
being rotated as much as possible. In order to check the vortex flux distribution among the refined and smeared plaquettes, we analyze
the idealized case of all rotations in one $U(1)$ subgroup, which of course is not
the optimal case for the overall smearing routine. We restrict
ourselves to the plaquettes in one plane only, cf. Fig.~\ref{fig:refsmr2}, noting however that the smearing
routine distributes vortex flux also to orthogonal plaquettes. The center vortex flux distribution is shown in Fig.~\ref{fig:refsmr2} in terms of fractions of $\pi$ for the individual refined and original plaquette values for the two cases plotted in
Fig.~\ref{fig:refsmr1}, {\it i.e.}, $\mathbbm{1}\rightarrow-\mathbbm{1}$ or
$-\mathbbm{1}\rightarrow\mathbbm{1}$. The individual contributions add up to
$\pm\pi$ respectively, giving a flux of $\pm\pi/2$ for the initial vortex
plaquette and $\pm\pi/4$ for the neighboring plaquettes. 

With this refined smearing procedure, a flat vortex surface is distorted
within the initial thickness it had before the refinement procedure, which seems to work just like a
smearing effect for the thick center vortices. In terms of the thin vortex
structure, {\it i.e.}, if we apply the MCG and project the smeared (thick)
vortices back to $Z(2)$, the thin vortex exhibits a rough instead of a
flat surface, since we introduced additional vortex plaquettes on the refined
lattice, within the original thickness of the vortex (see also
Fig.~\ref{fig:refsmr1}). 
The actual effect of this vortex surface distortion will be presented for
classical, {\it i.e.}, planar and spherical vortices in section~\ref{sec:class}.
These additional plaquettes are of course removed after blocking and we recover
the original vortex surface. They are also partially removed already on the
refined lattice by the smoothing procedure discussed in the next section.

\subsection{Refined vortex flux smearing}\label{sec:reflx}
\begin{figure}[h]
	\centering
	\includegraphics[width=.8\linewidth]{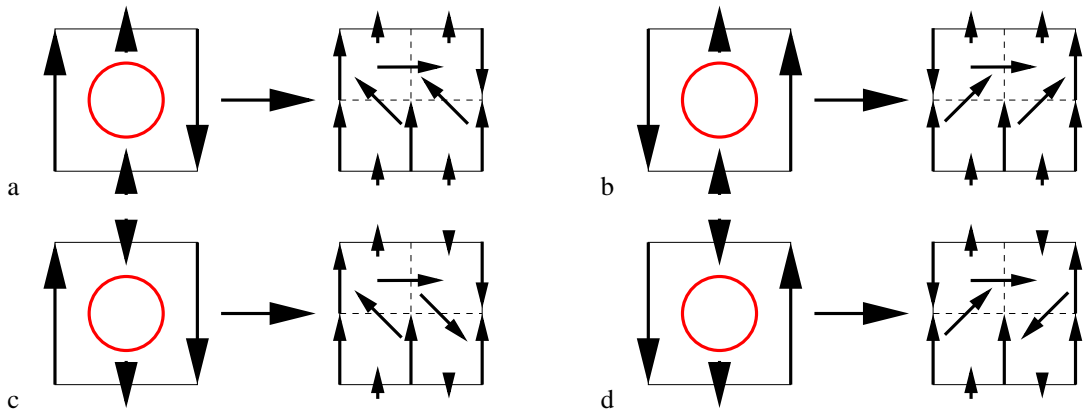}
	\caption{Examples of link configurations giving uniform center vortex flux
		distribution among the four refined plaquettes of the initial center
		vortex plaquette. The individual link rotations are given by $+\pi/2$,
		$\pm\pi/4$ and $\pm3\pi/4$ in the same $U(1)$ subgroup. Each refined
		plaquette carries one fourth of the initial center vortex flux; a),c)
	$\exp{i\pi}=-\mathbbm{1}$ and b),d) $\exp{-i\pi}=-\mathbbm{1}$. In order to
minimize the orthogonal plaquettes (the ones orthogonal to the displayed plane)
we vary the $U(1)$ subgroup by applying "2D gauge transformations" (see text) at
the central points of the plaquettes, indicated by the red circle.} 
	\label{fig:reflux}
\end{figure}

On the refined lattice, we have a straightforward way to distribute the
center vortex flux among the four refined plaquettes corresponding to each initial center vortex
plaquette without affecting neighboring plaquettes or even links. In
Fig.~\ref{fig:reflux} we show examples of link configurations to distribute the
center vortex flux $\exp{i\pi}=\exp{-i\pi}=-\mathbbm{1}$ uniformly among the
four refined plaquettes, each carrying one fourth of the initial center vortex
flux. The uniform distribution is of course only guaranteed if we apply all link
rotations of $\pm\pi/4$ and $\pi/2$ in the same $U(1)$ subgroup. 

Since we change only links at half the initial lattice spacings, {\it i.e.}, links
dividing the original plaquettes into four refined plaquettes, blocking
trivially (the mentioned links are thrown away) restores the initial $Z(2)$ link
configuration and therefore the original vortex structure. However, this method
also changes links orthogonal to the links giving the initial jump from $\mathbbm{1}$ to
$-\mathbbm{1}$ or vice versa. In fact these are the links in direction
of the jump (or, rotation, after smearing) plotted in Fig.~\ref{fig:reflux}
with rotations of $\pm\pi/4$. Depending on the vortex structure, the plotted
link configurations may still cause maximally nontrivial plaquettes in
directions orthogonal to the displayed plane; as mentioned above, we can easily
construct situations where $\pm\pi/4$ and $\pi/2$ links add up to $\pm\pi$, giving
$\exp{\pm i\pi}=-\mathbbm{1}$ for a plaquette. Therefore, we try to "2D gauge transform"
away from the simple examples in Fig.~\ref{fig:reflux}. In fact, applying a
gauge transformation at the central lattice site leads to arbitrary link
configurations without changing the plaquettes in the displayed plane. Since we only want to affect the
displayed links and not the ones orthogonal to the displayed (paper) plane, we do not
apply real (4D) gauge transformations, affecting all links at a certain point, but
restrict the transformations to the displayed links in the 2D plane. Using this
"2D gauge transformation" at the central points of the original vortex
plaquettes, using random $SU(2)$ vectors, we can minimize the affected plaquettes
orthogonal to the original or refined vortex plaquette. This way we eliminate
maximally nontrivial plaquettes and the overlap fermions seem to detect zero
modes properly. Staggered fermions, however, still show a gap, which might be related
to the problem discussed in the last section, {\it i.e.}, its even/odd lattice
implementation and the refinement procedure. In fact, the smeared configurations
shown in Fig.~\ref{fig:reflux} all have nontrivial links in the even (upper)
slice of the refined lattice. Since we apply the "2D gauge transformation" in
order to minimize the plaquettes, the link configuration we start with does not
matter. However, if we analyze the refined flux smeared configurations, we find
that the majority of links close to $-\mathbbm{1}$ is still found in the even lattice slices after the smearing
routine. This seems to be reasonable, since after refinement $-\mathbbm{1}$
links only appear in even lattice slices, see also Fig.~\ref{fig:ref}; links
close to $-\mathbbm{1}$ in odd lattice slices would obviously lead to plaquettes
close to $-\mathbbm{1}$ as well, which we try to avoid. By omitting the
minimizing "2D gauge transformation" and applying different link rotations in
order to reproduce the uniform flux distribution among the plaquettes we may
overcome the problem of the staggered Dirac operator in a similar way as in the
previous section. However, even though we examined many different combinations of
link rotations, we did not find a solution which gives equally good results for
both, overlap and staggered fermions. We also attempted extending the flux
distribution to the next neighboring plaquettes, and further also included "2D gauge
transformations" at the points next to the central point. Apart from not being
able to solve the initial problem this way, we furthermore lose the vortex
finding property. Therefore we settle on the link configurations shown in
Fig.~\ref{fig:reflux}, supplemented by the "2D gauge configuration" at the center of the original vortex
plaquette. This method gives us the best results towards our goal, approaching
continuum $SU(2)$ gauge dynamics, except for the staggered fermion spectra. These, however, can be improved with yet another, final step in our vortex smearing procedure,
described in the next section.

\subsection{Vortex smeared blocking}\label{sec:block}
A simple way to eliminate ultraviolet fluctuations of the
center projection vortices obtained in the maximal center gauge is to apply
blocking steps such as to transfer the vortex configurations onto new coarser
lattices, while always preserving their chromomagnetic flux content on length
scales larger than the new lattice spacing. 
Consider a new coarse lattice with $n$ times the spacing of an old fine lattice,
superimposed on the latter such that all sites of the coarse lattice coincide
with sites of the fine lattice. In this work we use $n=2$, but the blocking
procedure in principle is feasible for arbitrary $n$. The gauge phases associated with plaquettes on the coarse lattice then are defined to be equal to the $n\times n$ Wilson loops on the old fine lattice to which these plaquettes correspond. Equivalently, if an odd number of vortices pierces the $n\times n$ Wilson loop on the old fine lattice, then one vortex is defined to pierce the corresponding
plaquette on the new coarse lattice; if an even number of vortices pierces the
$n\times n$ Wilson loop on the fine lattice, then no vortex pierces the
corresponding plaquette on the coarse lattice. Note that, thinking in terms of thin center
vortices, {\it i.e.}, a $Z(2)$ lattice, this argumentation and the blocking simply
reduces to the multiplication of the $n^2$ plaquettes forming the $n\times n$
Wilson loop. Note also that blocking manifestly preserves the values of all
Wilson loops (as far as they can still be defined on the coarse lattice). Thus,
blocking leaves the string tension induced by a thin vortex ensemble invariant.
The only information that is lost during blocking are the original small
plaquettes, {\it i.e.}, ultraviolet fluctuations.

\begin{figure}[h]
	\centering
	\includegraphics[width=.9\linewidth]{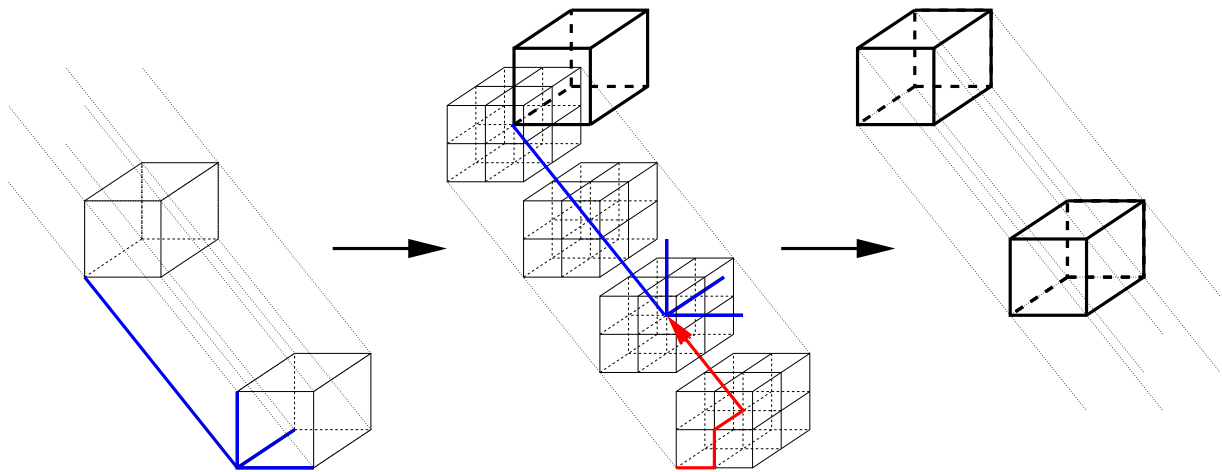}
	\caption{Refinement and smeared blocking procedure: After refinement the
	blocking is not performed starting at (1,1,1,1), the lower, left corner, but
	half an original ({\it i.e.}, one refined) lattice spacing forward in every
	space-time direction, {\it i.e.}, (2,2,2,2) on the refined lattice, indicated
	by the red arrow (the time direction is indicated by the fine lines
	connecting the space-like cubes). Without vortex smearing between steps 2
	and 3 the lattices before and after the whole procedure are actually the
	same; with vortex smearing the re-blocked lattice gives a smeared version of the original lattice.}
	\label{fig:reblk}
\end{figure}

Now, in terms of $Z(2)$ lattices, we have seen that blocking is the exact inverse
procedure to refinement. Hence, blocking a refined lattice exactly gives us
the same links and plaquettes present in the original lattice, and therefore also the
same vortex structure. For the following discussion let us identify original
(x,y,z,t) and refined lattice sites (2x-1,2y-1,2z-1,2t-1), and let us call the
latter "odd" lattice sites on the refined lattice, since all indices are odd.
Whenever one index of a refined lattice site is even, the lattice site is not
part of the original lattice and we may call it "half-even" or "even" if all
indices are even. It now is interesting to note that, due to the refinement
procedure we defined in section~\ref{sec:ref}, we can alternatively block the refined
lattices starting at (half-)even (refined) lattice points, {\it i.e.}, one refined (half an original) lattice spacing away from the original (odd refined) lattice sites in
any forward space-time direction and still get back the exact, original
configuration. Hence, instead of starting the blocking procedure at the (odd) refined
lattice point (1,1,1,1), which actually coincides with (1,1,1,1) on the original
lattice, we can also start blocking at, e.g., point (2,1,1,1) on the refined
lattice to reproduce the original configuration, or even at (2,2,2,2) as shown
in Fig.~\ref{fig:reblk}. Now, as mentioned before for the refined flux smearing
procedure, blocking the smeared lattice starting at odd refined lattice sites
recovers the original vortex configurations, as the links making up the original plaquettes are not changed during the smearing procedure. However, if we start the blocking at any (half-)even lattice site we have to multiply several smeared $SU(2)$ links instead of $\pm\mathbbm{1}$s and end up with a $SU(2)$ instead of a $Z(2)$ configuration.
This new, blocked configuration now represents some smeared version of the
original $Z(2)$ lattice, since the smeared links are derived from the original
lattice after refinement. In order to keep the smearing procedure symmetric,
{\it i.e.}, we do not want to favor any smearing direction, we start the
blocking procedure at the even lattice sites of the refined lattice, {\it i.e.}
one refined (half an original) lattice spacing forward in every space-time
direction, as indicated in Fig.~\ref{fig:reblk} by the red arrow. This procedure
we will call "smeared blocking" in the following.

\subsection{Vortex smoothing}\label{sec:smooth}
Since center projection vortices exhibit artificial ultraviolet fluctuations, we
also use the smoothing procedure first discussed in~\cite{Bertle:1999tw}. It
operates using elementary cube transformations on the lattice vortex surface
such that a net decrease in the number of vortex plaquettes is achieved. 

One point at which the smoothing procedure can be applied is before the
refinement process; note that, after refining, the vortex surface smoothing has no effect any more, since elementary cubes are made
up of eight smaller cubes after refinement. The initial idea was that smearing
smoothed vortex configurations might be easier since the number of small vortex
structures is reduced. However, we found that results for smeared original
and smoothed vortex configurations, even though they might differ for individual
configurations, agree within uncertainties in ensemble averages. In
fact, it was shown in~\cite{Bertle:1999tw} that smoothing does not change the
long-range physics of gauge configurations, since it only removes artificial
ultraviolet fluctuations of the vortices. 

On the other hand, smoothing does turn out to be useful at a different point;
namely, for our artificially distorted vortex surfaces after
refined smearing and MCG projection, since it removes those distortions which turn out to
be only elementary cube transformations of the refined vortex configurations.
However, the smoothing procedure does not necessarily reproduce the originally
refined vortex configuration, it only gives a smooth version of the vortex
surface within the original lattice spacing (two refined lattice spacings).
Again, see section~\ref{sec:class} for more details on the distortion and
smoothing effects on classical configurations.

\subsection{Summary}\label{sec:sum}
To conclude this section we briefly summarize the vortex smearing method for
$Z(2)$ configurations (the initial smoothing and final blocking are optional and
therefore put in parentheses): % We also refer to appendix~\ref{app:alg} for more specific details on the algorithms and individual rotation schemes.
\begin{itemize}
\item (smoothing of the thin vortex surface, see Sec.~\ref{sec:smooth} for
details)
\item refinement of the $Z(2)$ lattice configuration, see Fig.~\ref{fig:ref}
\item identification of vortex plaquettes, {\it i.e.}, plaquettes with Tr $U_{\mu\nu}=-2$ 
\item application of one of the two smearing routines:
\begin{itemize}
\item link rotation smearing:
\begin{itemize}
\item identification of opposite links causing the overall $-\mathbbm{1}$ of the plaquette, {\it i.e.}, $U_\mu(\vec x) U_\mu(\vec x+\hat\nu)$ or $U_\nu(\vec x) U_\nu(\vec x+\hat\mu)$ 
\item application of refined link rotation smearing as depicted in Fig.~\ref{fig:refsmr1},
	except for the $U(1)$ subgroups of the individual rotations not chosen
	uniformly, but such as to minimize 	the affected plaquettes
\end{itemize}
\item application of the refined vortex flux smearing, as depicted in
	Fig.~\ref{fig:reflux}, including "2D gauge transformations" in order to
	minimize the orthogonal plaquettes (see Sec.~\ref{sec:reflx} for details)
\end{itemize}
\item (vortex smeared blocking, see Fig.~\ref{fig:reblk})
\end{itemize}

\section{Results}\label{sec:res}
In order to test our method we use 500 thermalized L\"uscher-Weisz $SU(2)$ gauge
field configurations on $8^4$ lattices at coupling $\beta=3.3$ which gives a
lattice string tension $\sigma_{lat}=0.1112\pm0.0017$ corresponding to a lattice
spacing $a=0.1495\pm0.0012$fm. The locations of center vortices are identified
as usual by mapping the $SU(2)$ lattice to a $Z(2)$ lattice which contains, by
definition, only thin vortex  excitations. The mapping is carried out by fixing
the lattice to the direct maximal center gauge, which is equivalent to Landau
gauge in the adjoint representation, and which maximizes the squared trace of
the link variables. The gauge-fixing procedure is the over-relaxation method.
We also apply the above mentioned vortex smoothing and evaluate our results on
both original and smoothed vortex configurations after vortex smearing. As
mentioned above, the results for original and smoothed vortex configurations are
equal within uncertainties of ensemble averages. Hence, by combining the results we
can double our statistics - we may think of two different Gribov copies for each
Monte Carlo configuration, although the vortex structures are of course
correlated. In the following sections we present various observables for refined
link rotation smeared, refined vortex flux smeared and their vortex smeared
blocked configurations and discuss the individual advantages and drawbacks. 

\subsection{Fermionic Eigenvalues and Overlap zero modes}
\begin{figure}[h]
	\centering
	a)\includegraphics[width=.48\linewidth]{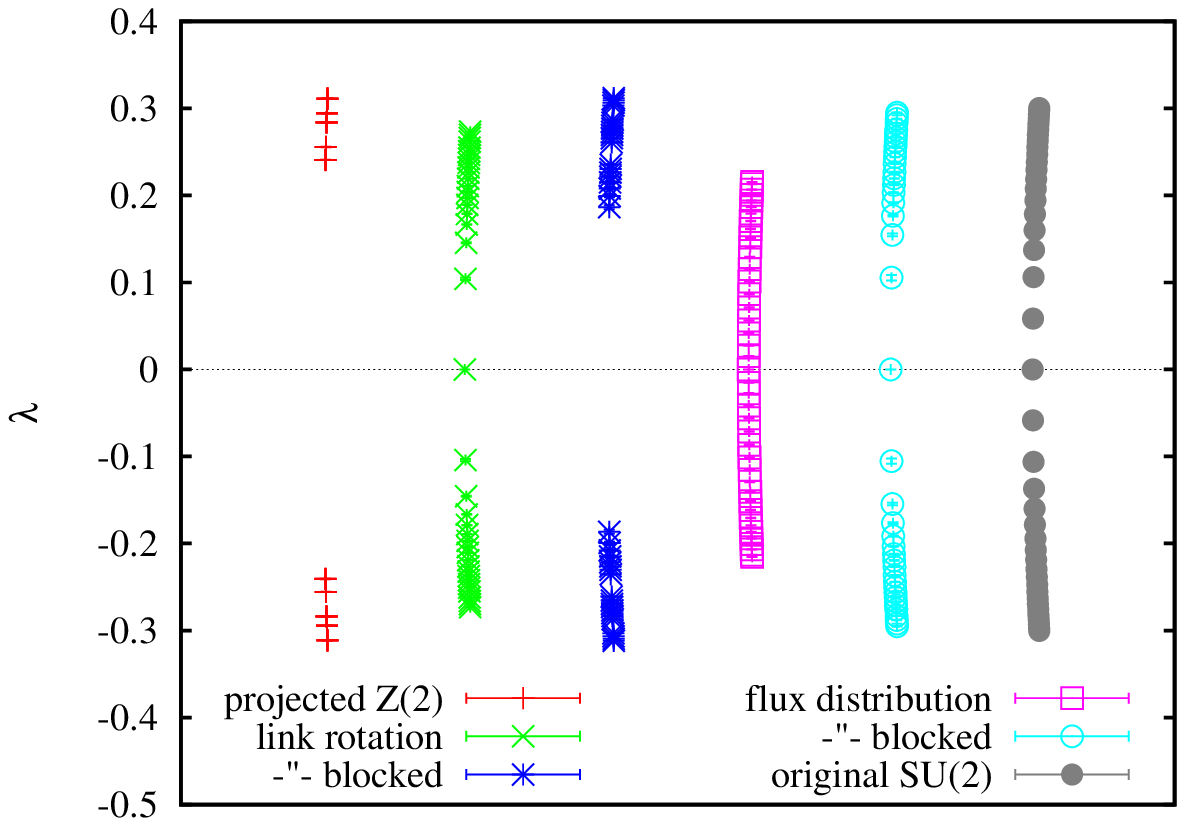}
	b)\includegraphics[width=.48\linewidth]{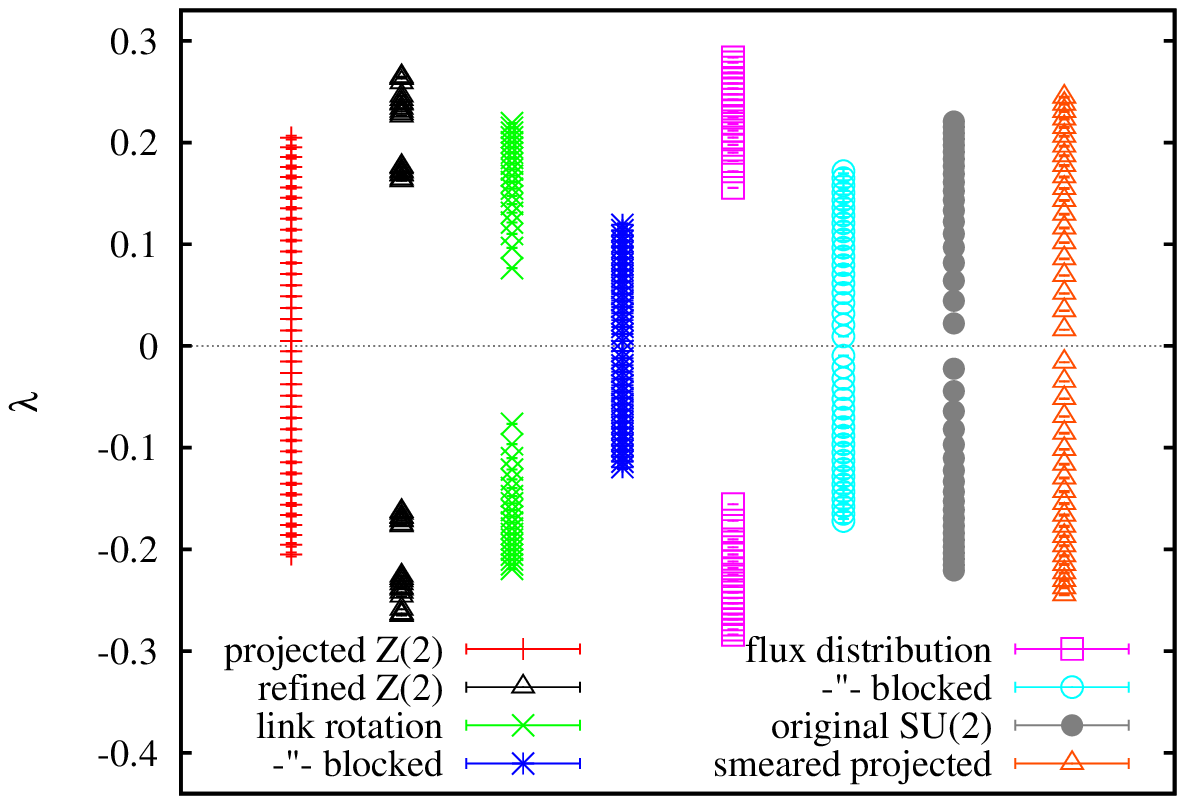}
	\caption{20 lowest a) overlap and b) asqtad staggered eigenvalues 
		for original (full) $SU(2)$, Maximal Center Gauge projected $Z(2)$ and
		vortex smeared configurations.}
	\label{fig:evs}
\end{figure}

Fermion eigenmodes are calculated by an implementation of the MILC~\cite{milc}
code at the Phoenix and Vienna Scientific Cluster (VSC) of the Vienna University
of Technology (VUT) and the Riddler Cluster at New Mexico State University
(NMSU). In Fig.\ \ref{fig:evs} we display the twenty lowest-lying complex
conjugate eigenvalue pairs of the overlap and asqtad staggered Dirac operators~ \cite{Narayanan:1994gw,Neuberger:1997fp},
for center projected, vortex smeared and original configurations. For the
spectra on the refined lattices, the eigenvalues are multiplied by a factor two,
to account for the refinement effect: For the free Dirac operator, using a plane
wave ansatz $\psi_{\alpha} (x) = u_{\alpha} \ \exp(i  p_{\mu} x_{\mu})$, the
eigenvalues are given by $\lambda\propto\sqrt{p_\mu p_\mu}$, see
e.g.~\cite{Hollwieser:2013xja}. The allowed values for $p_{\mu}$ are
\begin{equation*}
	p_i = \frac{2 n_i \pi}{a N_S}, \quad \quad p_4 = \begin{cases} \frac{2 n_4
	\pi}{a N_T} & \mbox{for periodic BC} \\ \frac{ (2 n_4 + 1) \pi}{a N_T} & \mbox{for anti-periodic BC} \end{cases},
\quad \quad n \in \mathbb{Z} \, ,
\end{equation*}
with $N_S$ the spatial and $N_T$ the temporal extent of the lattice. Hence,
even though there are $2^4=16$ times more eigenmodes on the refined lattice, the
eigenvalues scale linearly with $1/N_{S,T}$. The seeming mismatch is of course
compensated by much higher degeneracy of higher modes on the finer lattice.

\begin{figure}[h]
	\centering
	a)\includegraphics[width=.48\linewidth]{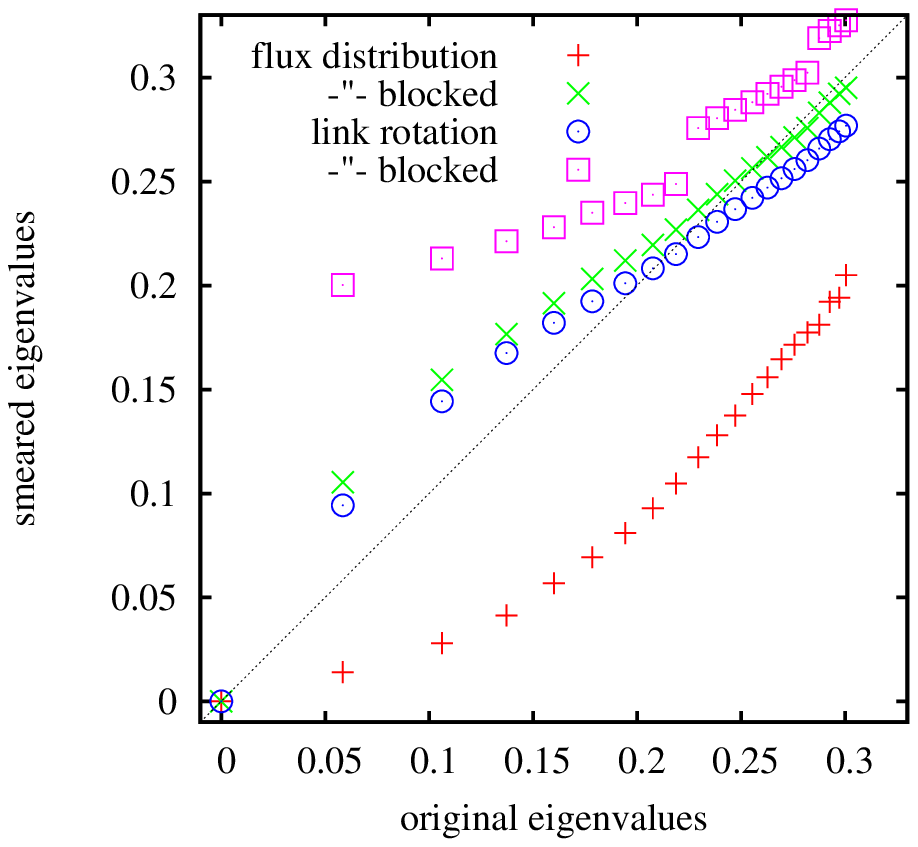}
	b)\includegraphics[width=.48\linewidth]{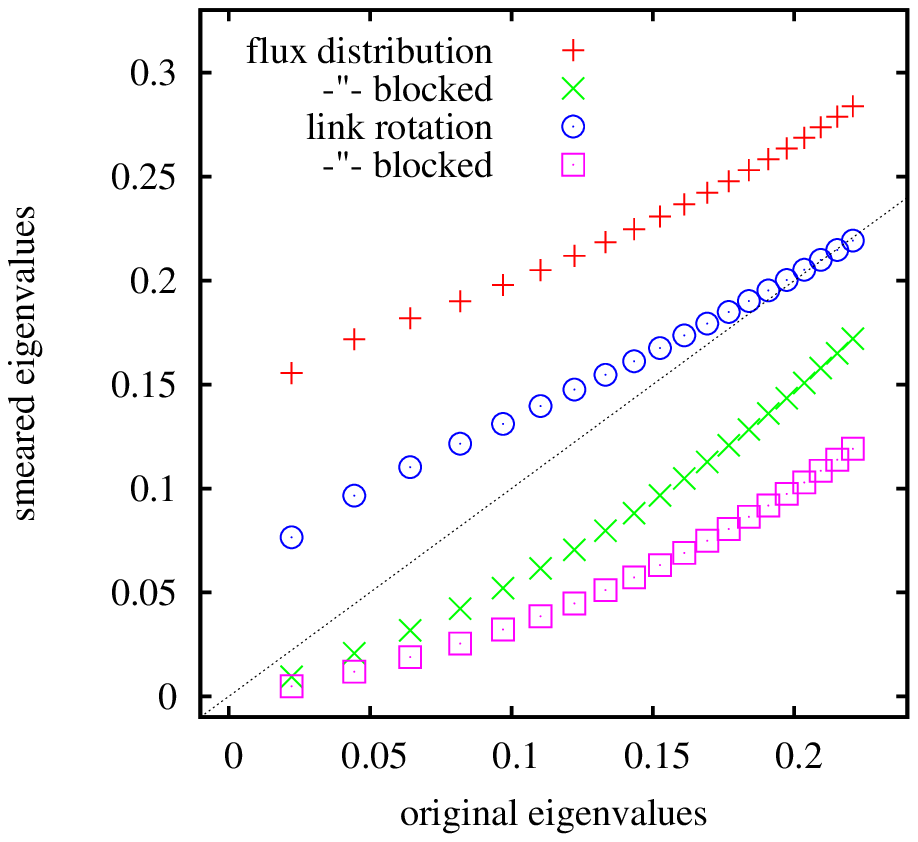}
	\caption{Scatter plot of a) overlap and b) asqtad staggered non-zero eigenvalues
		for	original (full) $SU(2)$ and vortex smeared configurations (ensemble mean eigenvalues).}
	\label{fig:vsc}
\end{figure}

\begin{figure}[h]
	\centering
	a)\includegraphics[width=.48\linewidth]{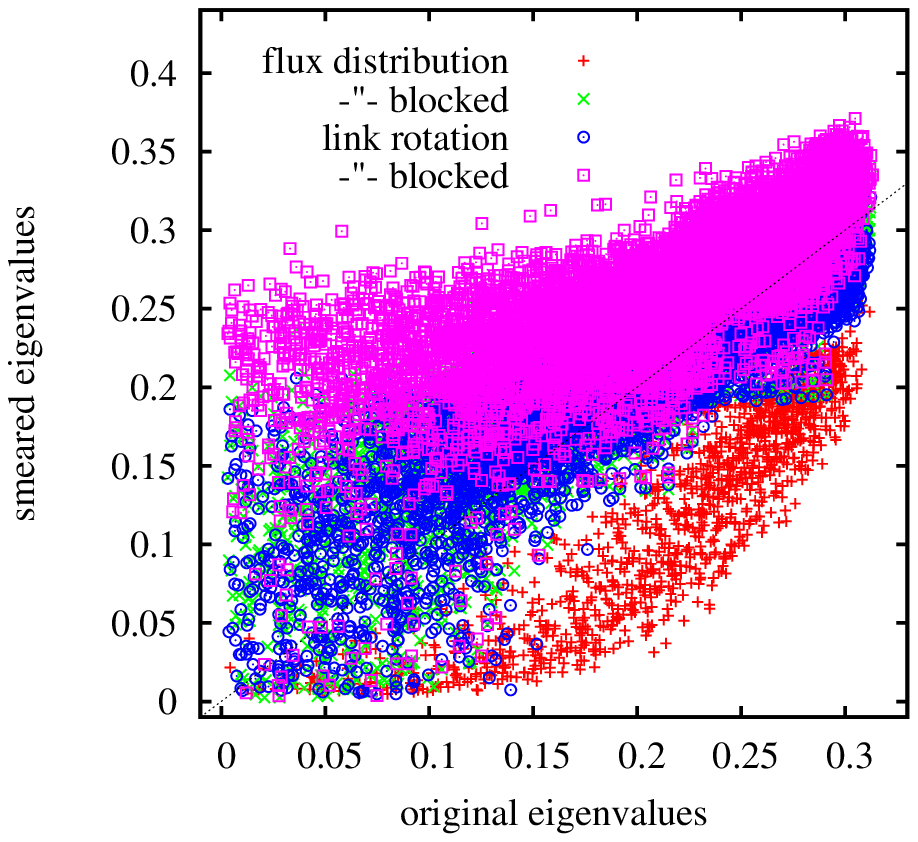}
	b)\includegraphics[width=.48\linewidth]{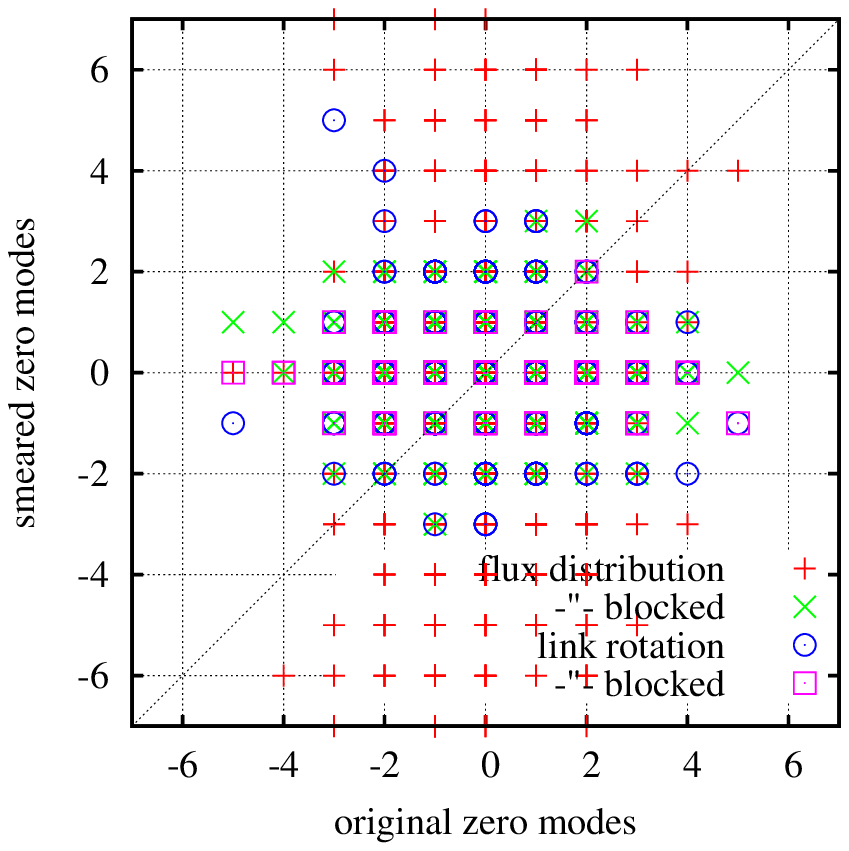}
	\caption{Scatter plot of overlap a) non-zero eigenvalues and b) number of
		zero modes for individual configurations.}
	\label{fig:zmc}
\end{figure}

Looking more closely at the overlap spectra, we see that there only appear to be five eigenvalue pairs (out of twenty) in the center projected case, indicating a four-fold degeneracy when the overlap operator is applied to $Z(2)$ lattice configurations.  This factor of four has the following origin:  In the first place, when link variables are simply plus or minus the $2\times 2$ identity matrix, the two colors decouple, and we have a factor of two degeneracy. Secondly, whenever the link variables are real and the Dirac operator has the Wilson or overlap (but not staggered) form, the eigenvalue equation $D \psi_n = \lambda_n \psi_n$ is invariant under charge conjugation.  Thus, if $\psi_n$ is an eigenstate with eigenvalue $\lambda_n$, then $C^{-1} \psi_n^*$ is also an eigenstate, with the same eigenvalue~\cite{Leutwyler:1992yt}.  This gives another factor of two, resulting in an overall four-fold degeneracy. 
For the vortex smeared configurations there is no such degeneracy and the spectra
approach the original (full) $SU(2)$ spectra. The actual correspondence of the
spectra can be seen in the scatter plots in Fig.~\ref{fig:vsc} for the ensemble
mean eigenvalues and in Fig.~\ref{fig:zmc}a) for the overlap eigenvalues on
individual configurations. The asqtad staggered results are not as good as for
the overlap Dirac operator in the sense that the gap is much larger than for the original
configurations, see Fig.~\ref{fig:evs}b. This large gap is caused by the
refinement procedure, as discussed already in Sec.~\ref{sec:reflrs}. Smearing the
refined $Z(2)$ configurations still shows large eigenvalue gaps which only go
away after smeared blocking. Since we focus on topological properties and therefore
especially on zero modes, the smearing routine was optimized to reproduce the
best overlap results. In Fig.~\ref{fig:evs}b) we also plot the asqtad
staggered spectra for simply refined and vortex smeared + MCG projected $Z(2)$
configurations; while the naive refinement process (without smearing) obviously
troubles the Dirac operators (the overlap response is similar), the latter
actually reproduces spectra which somewhat seem to interpolate between the original and projected cases.
In Fig.\ref{fig:zmc}b) we show a scatter plot of the number of zero modes for
original (full) $SU(2)$ and vortex smeared configurations. There is no one-to-one
correlation for the individual configurations; the reason for this will be
discussed in the next section where we analyze the influence of our method
on topological properties of the gauge field. 

Before that, we look at the distributions of the "unfolded" level spacing and
the lowest (non-zero) eigenvalues. Chiral random matrix theory (RMT) predicts
that these distributions are universal when they are classified according to
symmetry properties of the Dirac operator and the sector of fixed topological
charge under consideration~\cite{BerbenniBitsch:1997tx,Edwards:1999ra,
Verbaarschot:2000dy}. Fermion modes in the fundamental representation of gauge group
SU(N) of the overlap operator have the symmetry properties of the orthogonal
ensemble, {\it i.e.} their distribution is described by a Gaussian measure on
the space of real $N\times N$ symmetric matrices and is invariant under
orthogonal conjugation, whereas for the staggered operator they fall into the
symplectic ensemble, described by a Gaussian measure on the space of
quaternionic $N\times N$ Hermitian matrices and its distribution is invariant
under conjugation by the symplectic group.

\begin{figure}[h]
	\centering
	a)\includegraphics[width=.48\linewidth]{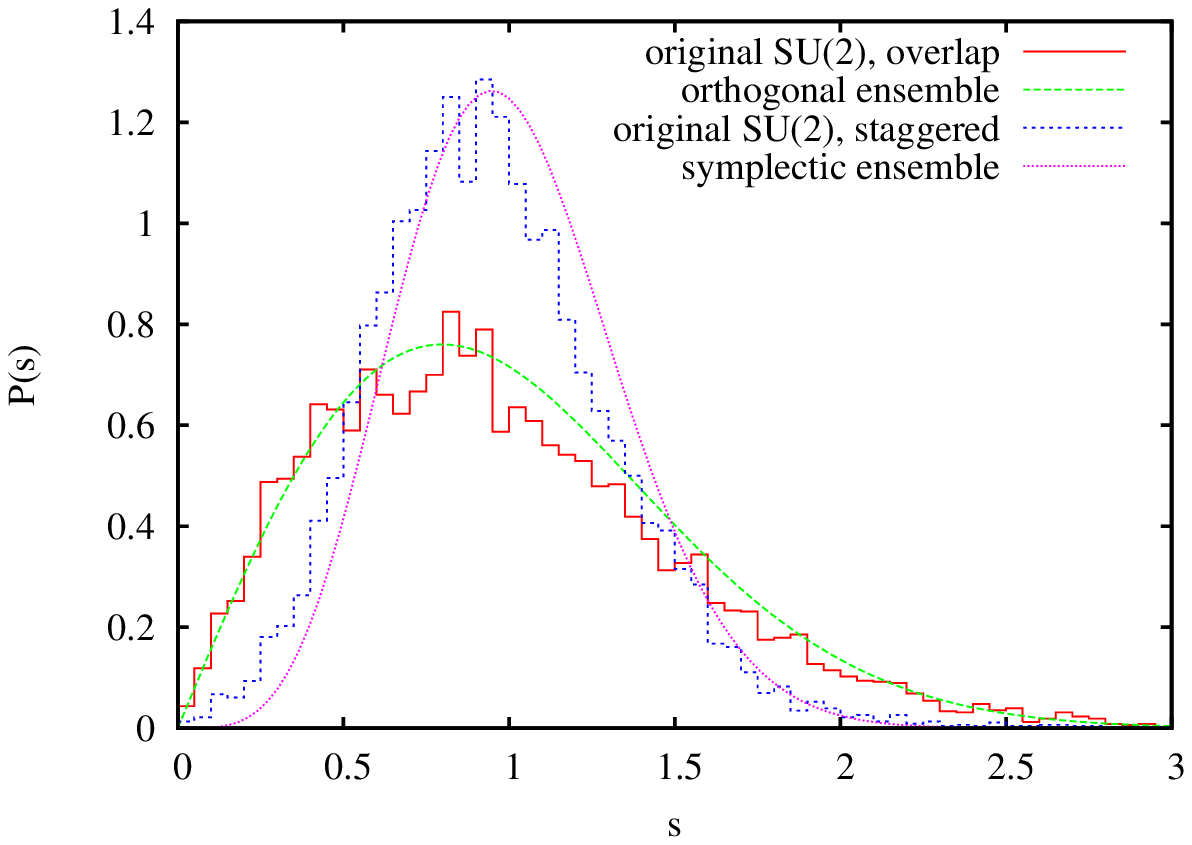}
	b)\includegraphics[width=.48\linewidth]{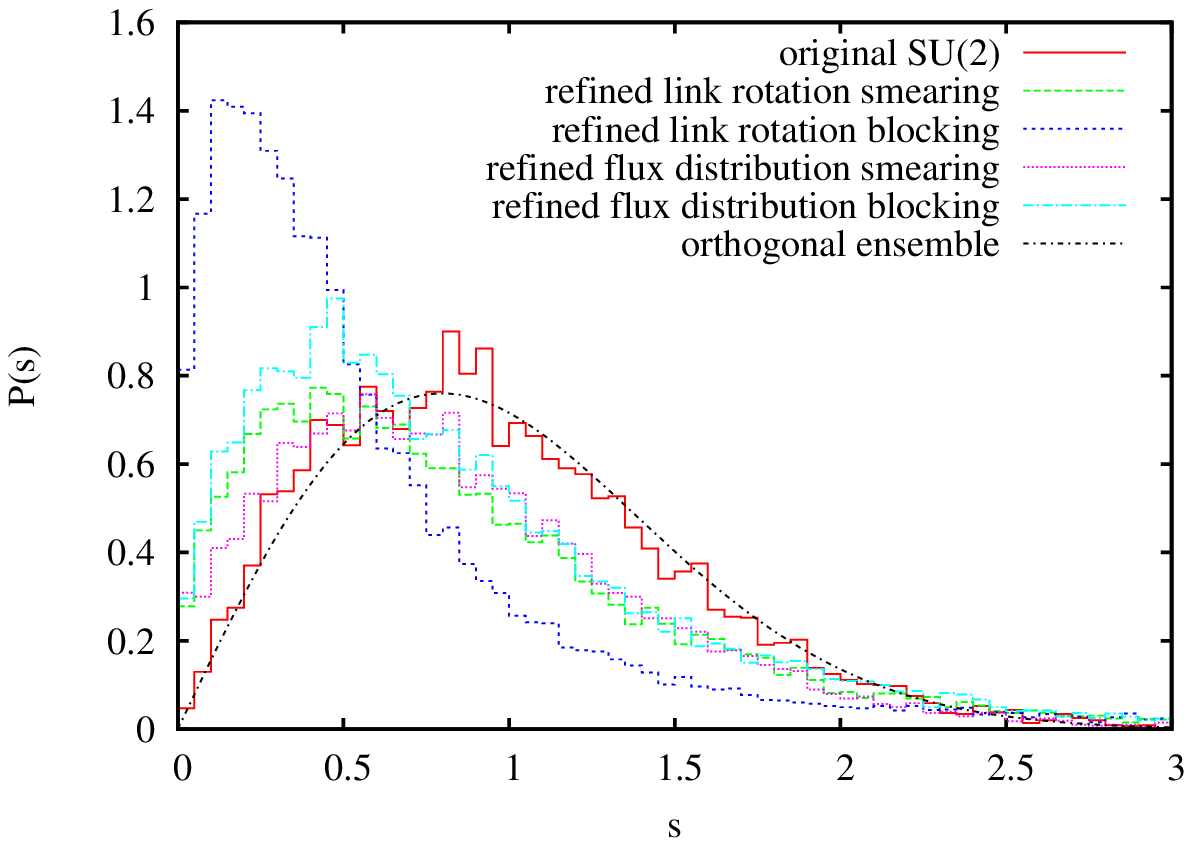}
	\caption{Distribution of the "unfolded" level spacing $s$ for 
		a) overlap and staggered eigenvalues on original configurations versus RMT
		universality predictions and b) overlap eigenvalues on original and smeared
		configurations. For better differentiation we attach the individual
	plots in Fig.~\ref{fig:zova} in Appendix A.}
	\label{fig:sov}
\end{figure}

\begin{figure}[h]
	\centering
	a)\includegraphics[width=.48\linewidth]{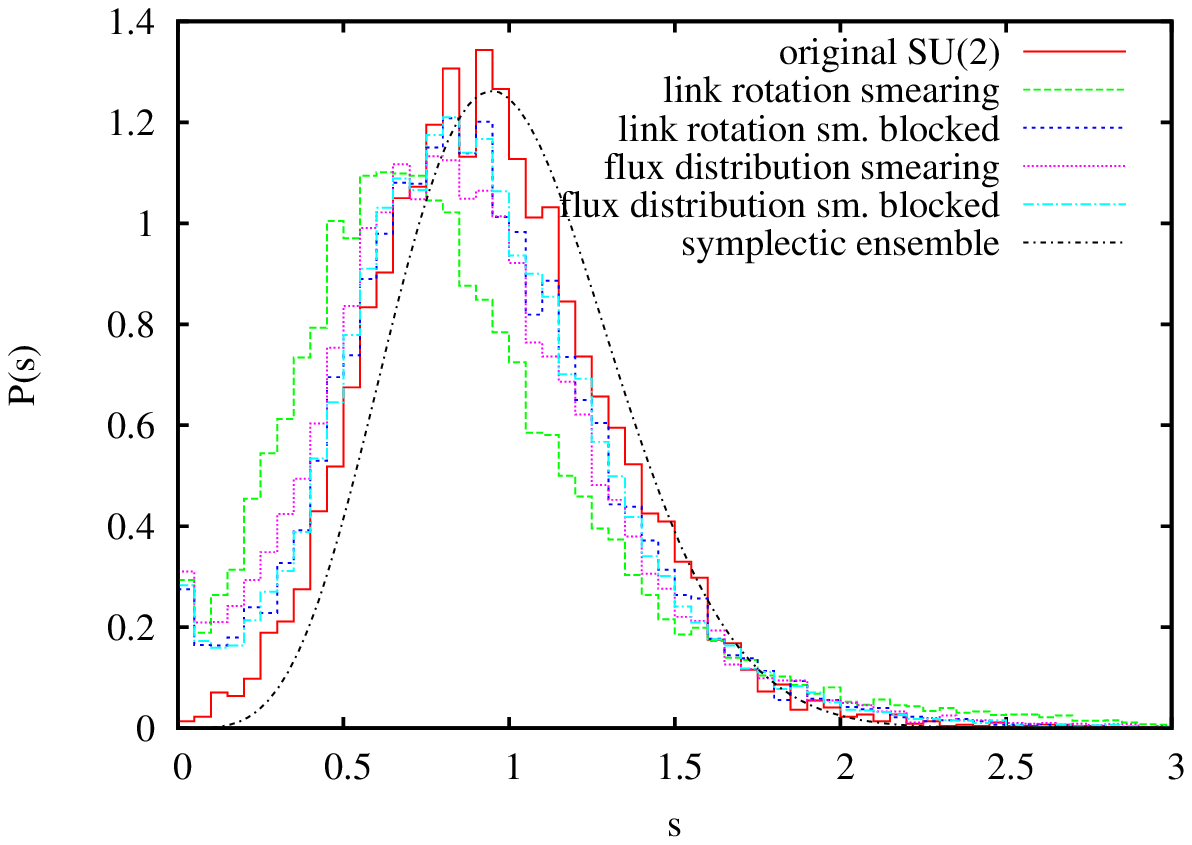}
	b)\includegraphics[width=.48\linewidth]{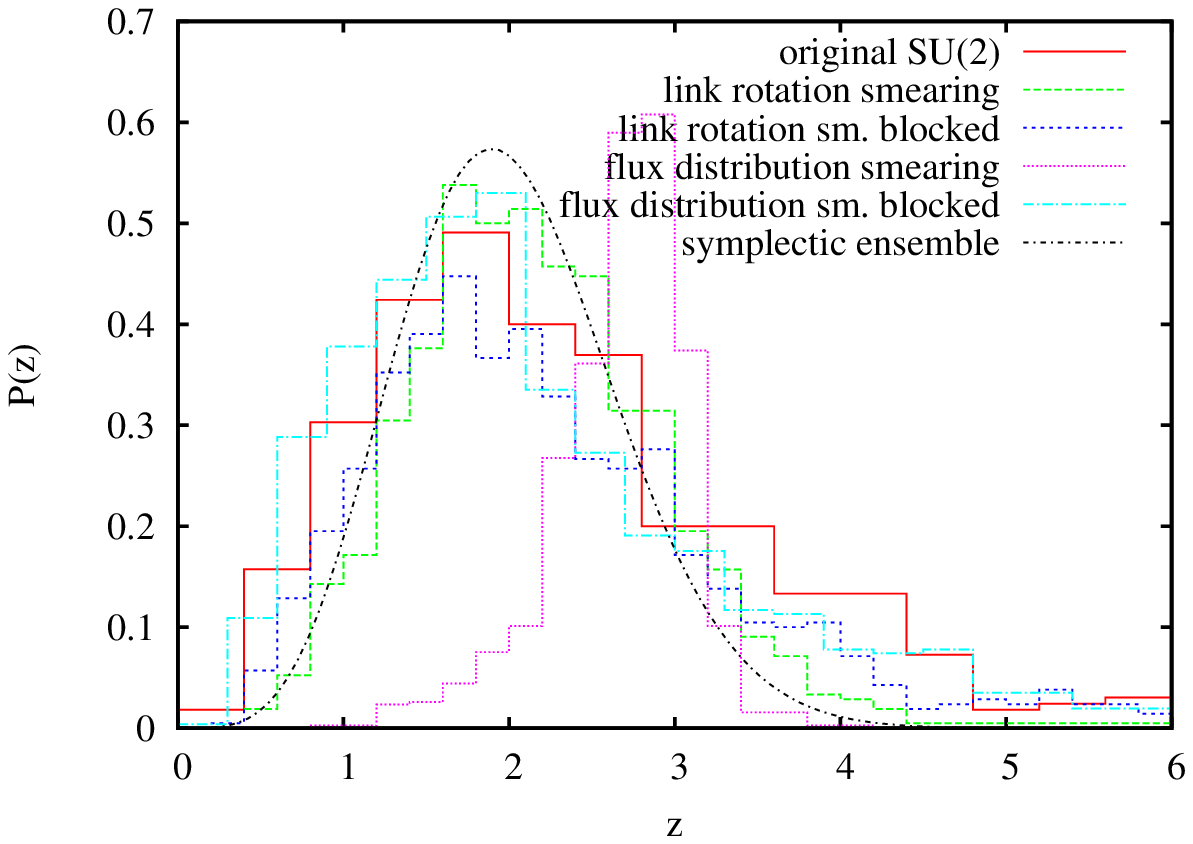}
	\caption{a) Distribution of the "unfolded" level spacing $s$ for staggered
		eigenvalues on original and smeared configurations. b) Distribution of
		the smallest staggered eigenvalue $\lambda_{min}$ and RMT universality
		prediction $P(z)$ versus the rescaled variable $z=\Sigma V\lambda_{min}$, 
		where $V$ is the volume and $\Sigma$ is the infinite volume value of the
		chiral condensate. Fitted values for $\Sigma$ are given in
		Table~\ref{tab:sig}. For better differentiation we attach the individual
	plots in Figs.~\ref{fig:zovb},\ref{fig:zovbb} in Appendix B.} 
	\label{fig:zst}
\end{figure}

The unfolding is done by first sorting all
non-zero positive eigenvalues $\lambda_i^n$ with $n$ labeling the configuration
number in ascending order. $N_i^n$ then gives the location of $\lambda_i^n$ in
the sorted list and is referred to as the unfolded spectrum. The level spacing
$s$ is simply given by $s=(N_{i+1}^n-N_i^n)/N_c$ where $N_c$ is the number of
configurations. The distributions of the unfolded level spacing $s$ in RMT are
well approximated by the various Wigner distributions~\cite{Halasz:1995vd}
\begin{equation}
P(s) = \begin{cases} {\pi\over 2} s {\rm e}^{-{\pi\over 4}s^2} & \mbox{orthogonal ensemble} \cr
%{32\over\pi^2} s^2 {\rm e}^{-{4\over\pi}s^2} & unitary ensemble \cr
{262144\over 729\pi^3} s^4 {\rm e}^{-{64\over 9\pi} s^2} & \mbox{symplectic  ensemble} . \end{cases}
\label{eq:level_space}
\end{equation}
In Fig.~\ref{fig:sov}a) we show the distribution of the unfolded level spacing
for overlap and staggered fermions on our original $SU(2)$ configurations. The
distributions are slightly shifted to lower values compared to RMT predictions, 
the reason could be a lack of statistics and our rather small lattice volume,
but more likely our choice of gauge coupling $\beta=3.3$, which is right at the
deconfinement phase transition where fluctuations can be expected.
Fig.~\ref{fig:sov}b) shows the results for overlap fermions on original and
smeared configurations and we observe that the smearing processes shift the
distributions further to the left, {\it i.e.} smaller level spacings. Only the
blocking procedure for link rotation smearing seems to strongly distort the level
spacing distributions. The results for staggered fermions are presented in
Fig.~\ref{fig:zst}, smearing again shifts the distributions slightly to
smaller level spacings, after blocking, however, smeared and original
distributions are not further apart than original and RMT predictions.

\begin{figure}[h]
	\centering
	a)\includegraphics[width=.48\linewidth]{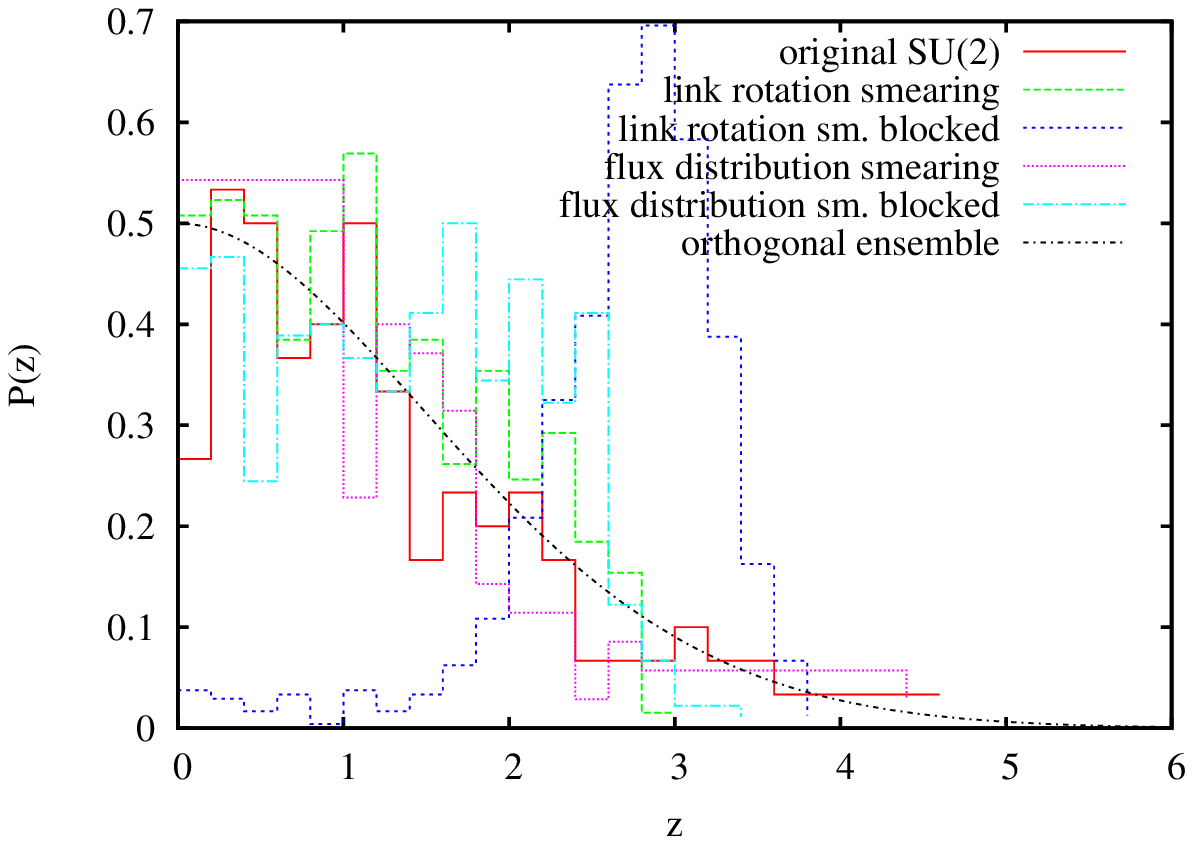}
	b)\includegraphics[width=.48\linewidth]{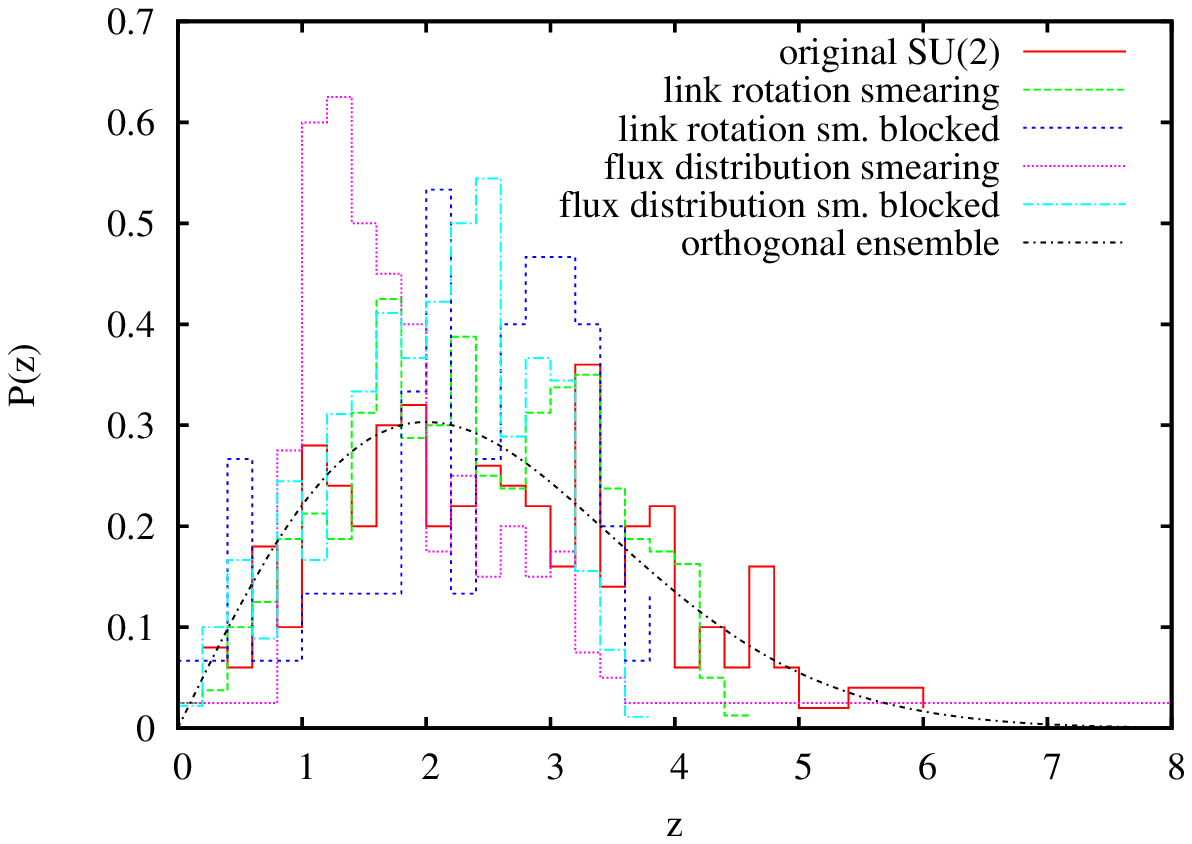}
	\caption{Distribution of the smallest overlap eigenvalue $\lambda_{min}$ and RMT
		universality prediction $P(z)$ for topological sectors a) $\nu=0$ and b)
		$\nu=1$ versus the rescaled variable $z=\Sigma V\lambda_{min}$, where
		$V$ is the volume and $\Sigma$ is the infinite volume value of the
		chiral condensate. Fitted values for $\Sigma$ are given in
	Table~\ref{tab:sig}. Find the individual
plots in Fig.~\ref{fig:zovc},\ref{fig:zovcc} in Appendix A.}
	\label{fig:zov}
\end{figure}

Next we look at the distribution of the lowest eigenvalue $\lambda_{\rm min}$
for the various ensembles. Chiral RMT predicts that these distributions are
universal when they are classified according to the number of exact zero modes
$\nu$ within each ensemble and then considered as functions of the rescaled
variable $z=\Sigma V \lambda_{\rm min}$. Here $V$ is the volume and $\Sigma$ is
the infinite volume value of the chiral condensate $\langle \bar \psi \psi
\rangle$. RMT gives for the distribution of the rescaled lowest eigenvalue for
the orthogonal ensemble, expected to apply to the overlap fermions, in the
$\nu=0$ and $\nu=1$ sector~\cite{Forrester:1993kf} 
\begin{eqnarray}
	P(z) = \begin{cases}  
		{2+z\over 4} {\rm e}^{-{z\over 2}-{z^2\over 8}} & \mbox{if $\nu=0$} \cr
{z\over 4} {\rm e}^{-{z^2\over 8}} & \mbox{if $\nu=1$} . \end{cases}
\label{eq:Pmin_OE}
\end{eqnarray}
For the symplectic ensemble, expected to apply to the staggered fermions, the
RMT prediction is~\cite{Forrester:1993kf,Nagao:1995np,Nagao:1998np}
\begin{eqnarray}
	P(z) = \begin{cases} 
\sqrt{\pi\over 2} z^{3/2} I_{3/2}(z) e^{-{z^2\over 2}} & \mbox{if $\nu=0$} \cr
{2\over {(2\nu+1)! (2\nu+3)!}} z^{4\nu+3} {\rm e}^{-z^2\over 2}
T_\nu(z^2) & \mbox{if $\nu > 0$} , \end{cases}
\label{eq:Pmin_SE}
\end{eqnarray}
where $I_{3/2}(z)$ is the modified Bessel function and $T_\nu(x)$ a rapidly
converging series based on partitions of integers, specified in the references.
Staggered fermions, however, do not have exact zero modes at finite lattice
spacing, even for topologically non-zero backgrounds, and thus seem to probe the
$\nu=0$ predictions of chiral random matrix theory only. We compare the RMT
predictions with our data in Fig.~\ref{fig:zst}b) for staggered and
Fig.~\ref{fig:zov} for overlap fermions. If one knows the value of the chiral
condensate in the infinite volume limit, $\Sigma$, the RMT predictions for
$P(z)$ are parameter free. On the rather small systems that we considered here,
we did not obtain direct estimates of $\Sigma$. Instead, we made one-parameter
fits of the measured distributions to the RMT predictions, with $\Sigma$ the
free parameter and results given in Table~\ref{tab:sig}. We note that the chiral
condensate is very small and results on the various ensembles vary, within
rather large uncertainties caused by small statistics and also the choice of gauge
coupling $\beta=3.3$ right at the deconfinement phase transition causing large
fluctuations. The distributions of the lowest
staggered eigenvalues in Fig.~\ref{fig:zst}b) are not consistent with RMT
predictions for the original configurations; after link rotation smearing, they
actually seem to agree with RMT predictions, but not for flux distribution
smearing. After blocking the original distributions are reproduced, for 
different fitting parameters, {\it i.e.} chiral condensates though. Finally, the
overlap results shown in Fig.~\ref{fig:zov} for $\nu=0$ and $\nu=1$ sectors are
broadly consistent with RMT predictions, with the exception
of link rotation smeared blocking in the $\nu=0$ case, which deviates
strongly from the other distributions, and a somewhat distorted distribution
for flux distribution smearing (without blocking) in the $\nu=1$ case.
For $\nu=0$ the distribution found in the original configurations differs at
zero eigenvalue from the RMT prediction, {\it i.e.}, we find less low-lying
eigenmodes. We conclude that the smeared ensembles roughly reproduce Dirac
spectra and distributions of the original configurations except for link
rotation smeared blocking with overlap and flux distribution smearing with
staggered fermions.

\begin{table}
\begin{tabular}{|l|c|c|c|}
\hline
Configuration & overlap, $\nu=0 $ & overlap, $\nu=1 $ & staggered \\
\hline
original $SU(2)$ & 0.018 (133) & 0.014 (255) & 0.016 (500) \\
link rotation smeared & 0.017 (341) & 0.011 (393) & 0.016 (1000) \\
-"- blocked & 0.011 (928) & 0.014 (66) & 0.015 (1000) \\
flux distribution smeared & 0.011 (87) & 0.010 (174) & 0.008 (1000) \\
-"- blocked & 0.013 (480) & 0.014 (444) & 0.017 (1000) \\
\hline
\end{tabular}
\caption{The chiral condensate $\Sigma$ from fits of the distribution
of the lowest eigenvalue to the RMT predictions and the number of
configurations in each topological sector in parenthesis.}
\label{tab:sig}
\end{table}

\subsection{Gluonic and Fermionic Topological Charge Correlation}

\begin{figure}[h]
	\centering
	a)\includegraphics[width=.48\linewidth]{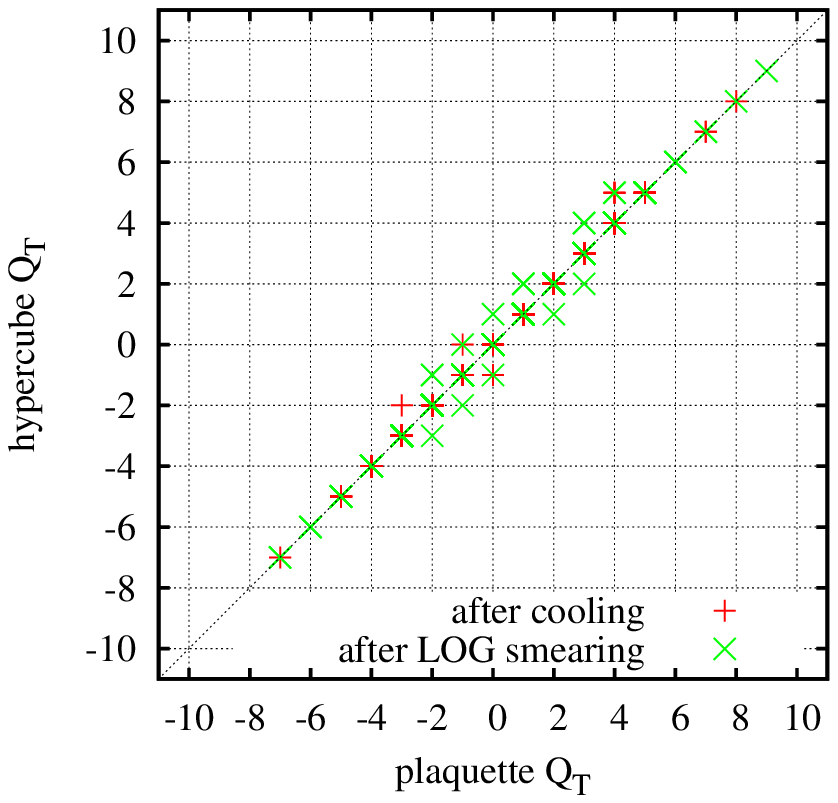}
	b)\includegraphics[width=.48\linewidth]{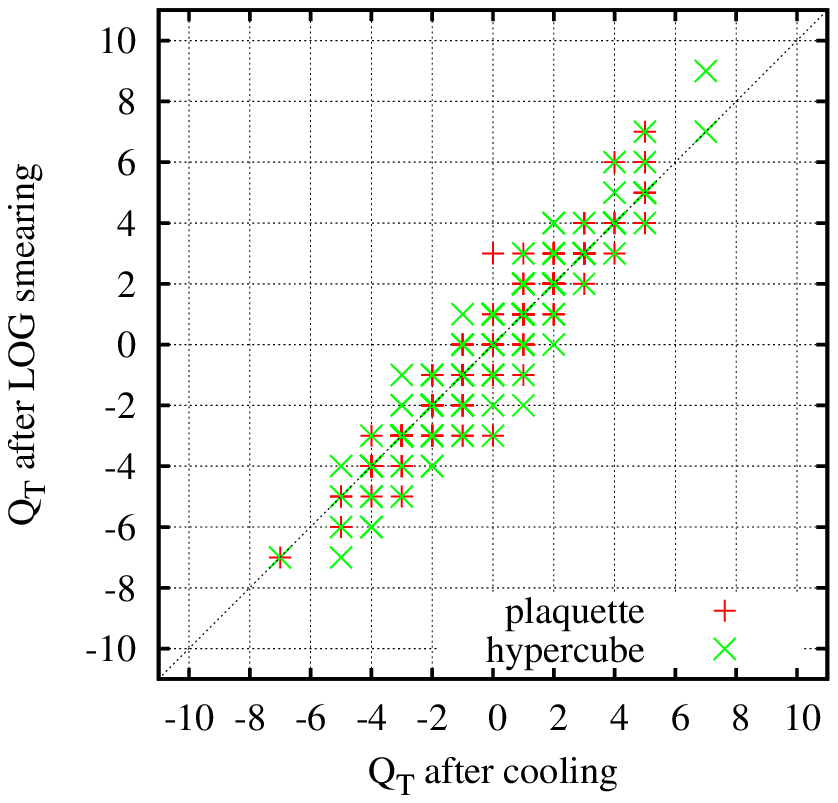}
	\caption{a) Scatter plot of plaquette and hypercubic definition of
	topological charge $Q_T$ after cooling and LOG smearing. b) Scatter plot of
	$Q_T$ after cooling and LOG smearing for plaquette and hypercube
	definitions. Both plots are for vortex smeared configurations, for original
	(full) configurations they look very similar.}
	\label{fig:qcs}
\end{figure}

If we want to recover the topological structure of the original (full) SU(2)
configurations from the vortex smeared configurations we face the problem that
standard gluonic definitions of topological charge via plaquette or hypercube
constructions need some smearing or cooling procedure in order to guarantee a
smooth gauge background which fulfills the L\"uscher condition~\cite{Luscher:1981zq}. However, smoothing and especially cooling destroys the relevant vortex structures. In a word, gluonic topological charge definitions are not very reliable in the background of center vortices, this was also discussed in~\cite{Hollwieser:2012kb}. 
For the gluonic topological charge $Q_T$ we use the integral (sum, on the lattice) of the
gluonic charge density $q(x) = \frac{1}{16\pi^2} \, \mathrm{tr} (\mathcal
F_{\mu\nu} \tilde{\mathcal F}_{\mu\nu})$ in the ``plaquette'' and/or
``hypercube'' definitions on the lattice~\cite{DiVecchia:1981qi,DiVecchia:1981hh},
which in fact give almost the same results, see Fig.~\ref{fig:qcs}a). 
In Fig.~\ref{fig:qcs}b) we show that the gluonic topological charge after cooling
or LOG smearing correlates very well on both, original and vortex smeared
configurations.

\begin{figure}[h]
	\centering
	a)\includegraphics[width=.48\linewidth]{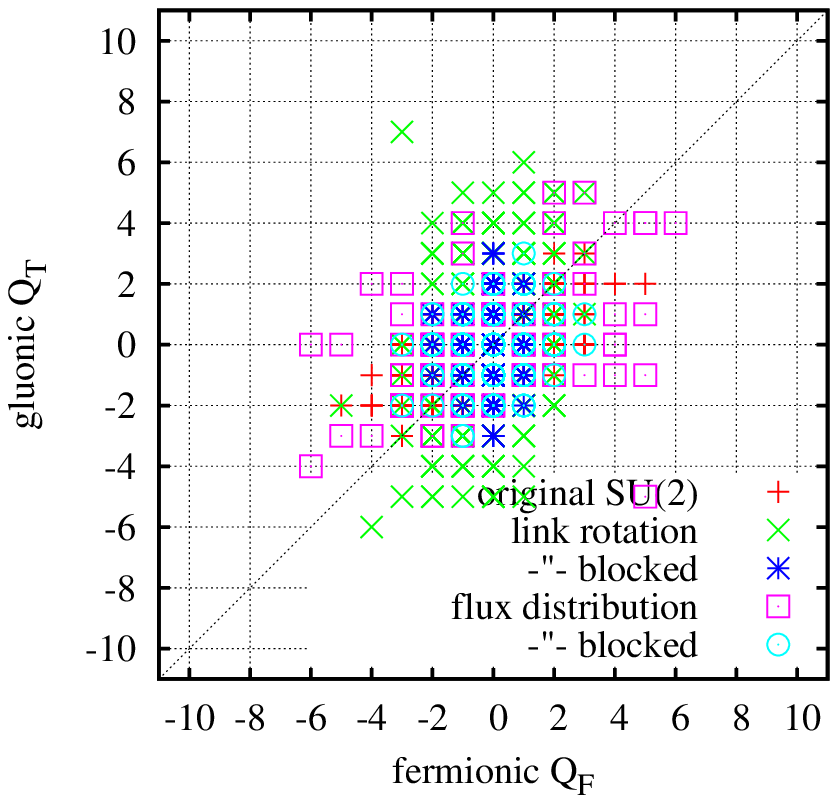}
	b)\includegraphics[width=.48\linewidth]{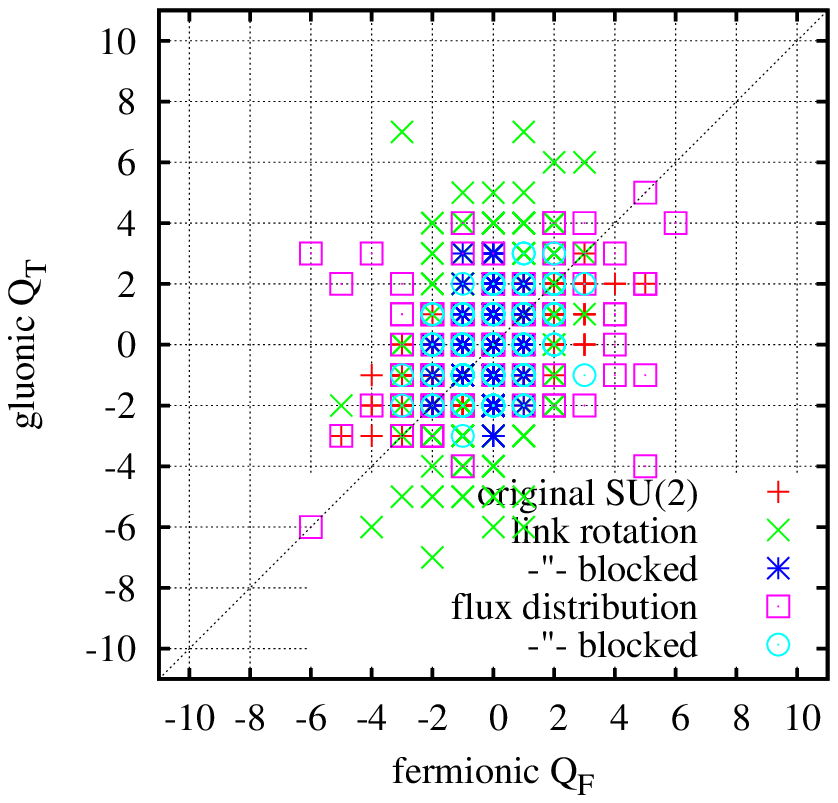}
	c)\includegraphics[width=.48\linewidth]{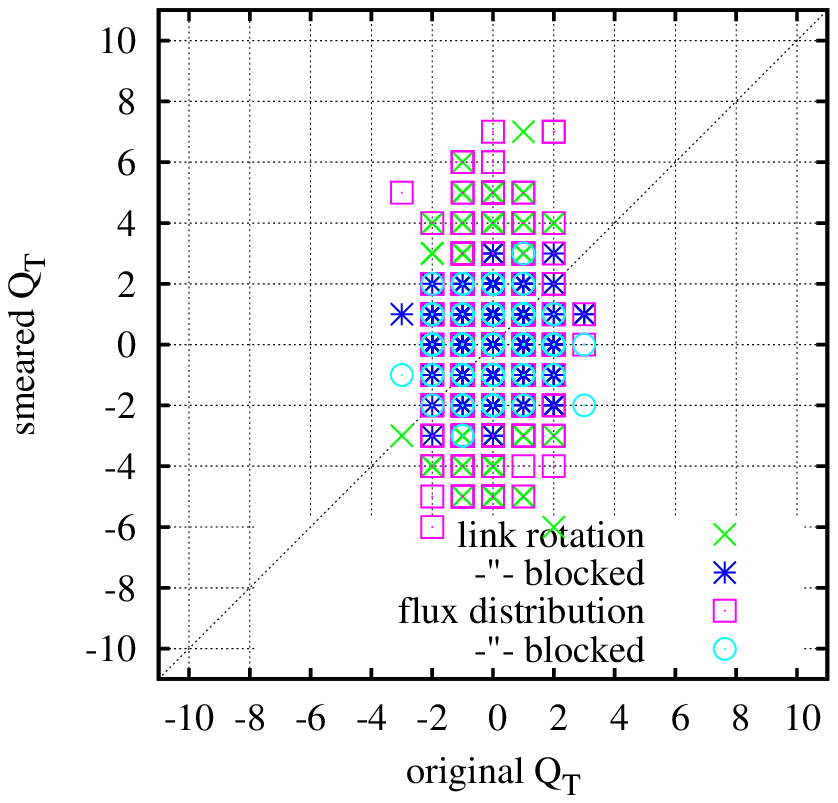}
	d)\includegraphics[width=.48\linewidth]{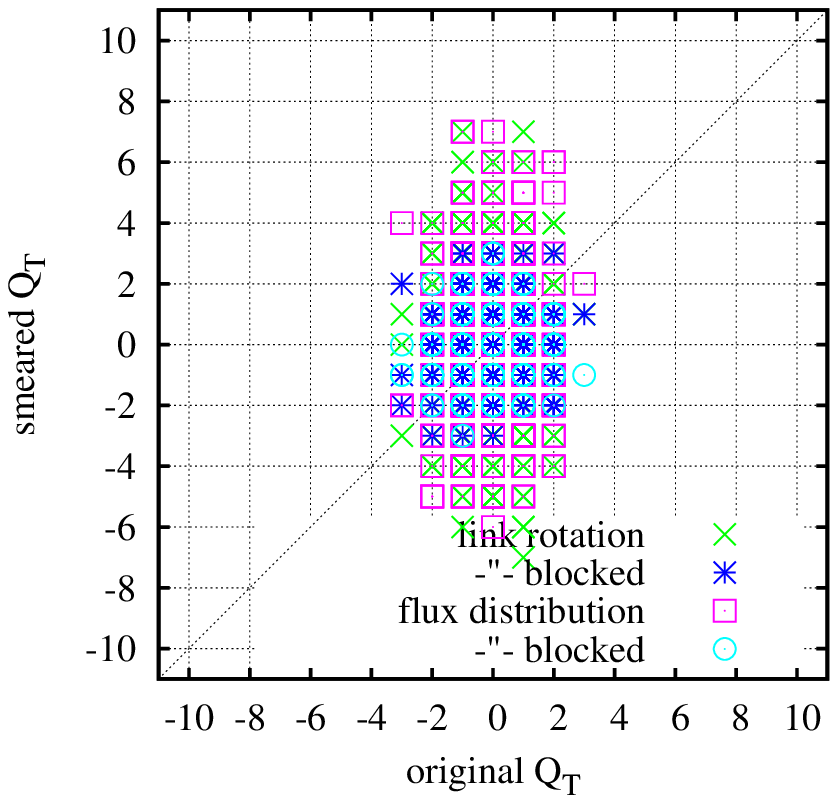}
	\caption{Scatter plot of fermionic $Q_F$ and gluonic topological charge
		$Q_T$ after a) cooling or b) (LOG) smearing on original (full) $SU(2)$ and
	various vortex smeared configurations ($Q_{F/T}$ on same configs);
	Scatter plot of gluonic topological charge $Q_T$ after c) cooling
	and d) (LOG) smearing for original (full) vs. vortex smeared configurations.
In the blocked vortex flux distribution smeared case one can actually see a
positive correlation, see also Fig.~\ref{fig:ggzz}.}
	\label{fig:QtQf}
\end{figure}

\begin{figure}[h]
	\centering
	a)\includegraphics[width=.48\linewidth]{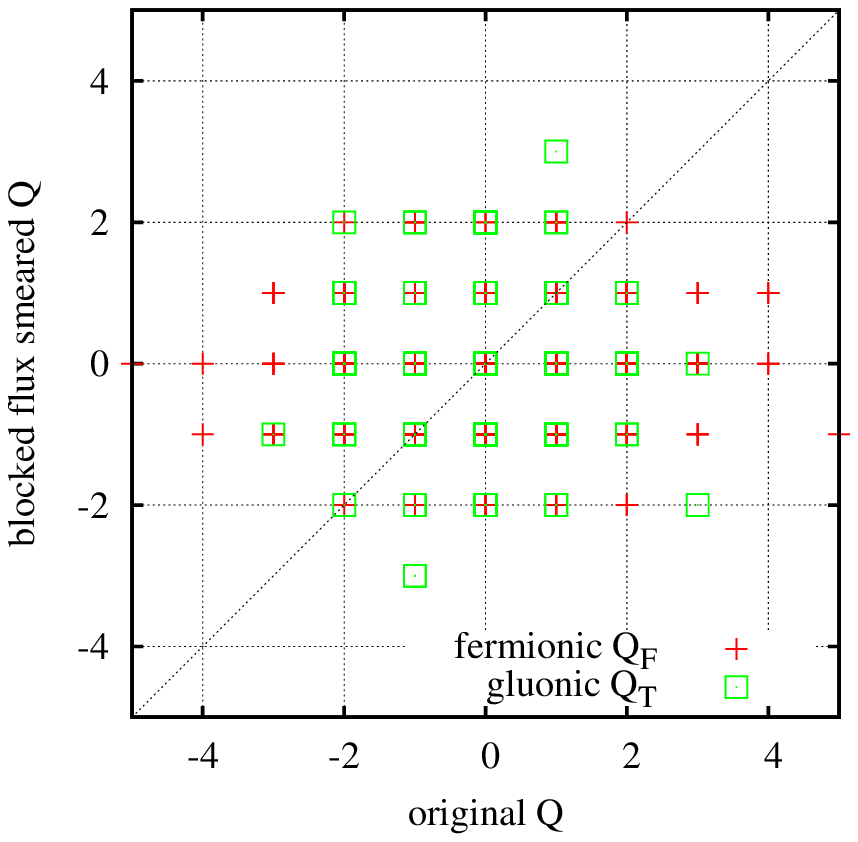}
	b)\includegraphics[width=.48\linewidth]{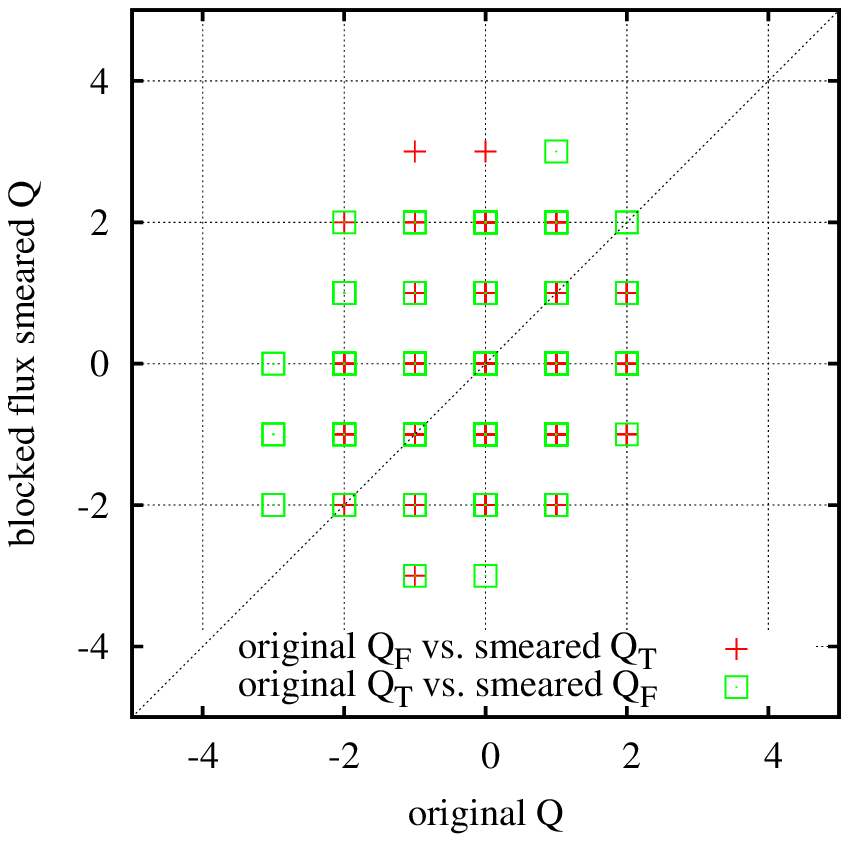}
	\caption{Scatter plot of fermionic and gluonic topological charge
		correlations and cross-correlations between original (full) vs. blocked
		vortex flux distribution smeared configurations.}
	\label{fig:ggzz}
\end{figure}

The Atiyah-Singer index theorem relates the number of exact fermionic zero modes
of a configuration and the topological charge $Q_F = \mbox{Tr} (\gamma_5 D_{ov})
= n_- - n_+ = \text{ind} D_{ov}$~\cite{Atiyah:1971rm}, which we call the
fermionic topological charge. The relation $Q_T\approx Q_F$ is not exact on either
original or vortex smeared configurations, see Fig.~\ref{fig:QtQf} for
topological charge correlation between the two definitions on individual
configurations. These results are not affected by additional cooling or LOG smearing. 
In view of the above concerns, it is not surprising that the correlation
of either fermionic $Q_F$ (as seen in Fig.~\ref{fig:zmc}b) or gluonic
topological charge $Q_T$ of the vortex smeared configurations with the original
topological charge is not very good. In Fig.~\ref{fig:QtQf} we see that the
refined vortex smeared configurations overestimate the original topological charge. 
For the smeared blocked configurations the net topological charge $Q_T$ is
comparable to the original one; actually, in the blocked vortex flux
distribution smeared case one can see a slightly positive correlation, see
Fig.~\ref{fig:ggzz}b. However the correlation between individual
configurations is not very good. Again, this is not very surprising since
cooling or smearing in order to evaluate the gluonic topological charge on the
lattice destroys its center vortex content. However, in~\cite{Bertle:2001xd} it
was shown that the vortex topological charge defined there based directly on the
structure of the vortex configurations gives a good estimate of the
topological susceptibility of the gauge field ensemble. Therefore we analyze the
vortex topological charge and the topological susceptibility next.

\subsection{Vortex Topological Charge and Topological Susceptibility}

Center vortices give rise to topological charge at intersection and writhing
points~\cite{Engelhardt:2000wc,Engelhardt:2002qs}. It is known
from~\cite{Bertle:2001xd} that center vortices reproduce the topological
susceptibility of the original gauge fields. We want to estimate how well our
vortex smearing procedure recovers this effect and further, how accurately the
vortex topological charge reveals the topological content of individual gauge
fields, {\it i.e.}, we are interested in the correlation of the vortex
topological charge with the index and gluonic topological charge definitions. 
The concepts of blocking and smoothing were introduced in Sec.~\ref{sec:smooth}
and~\ref{sec:block}, for more details on the use of these methods during vortex
topological charge calculation see~\cite{Bertle:2001xd}, where it was also 
shown that the string tension is rather independent of vortex smoothing;
however, smoothing reduces vortex topological charge and susceptibility by
removing short range fluctuations of the vortex
structure. On our original $8^4$ lattices with lattice spacing $a\approx0.15$fm
1-2 blocking steps seem appropriate during the vortex topological charge calculation
in order to get in the range of a physical vortex thickness of
$~0.4$fm~\cite{DelDebbio:1996mh,DelDebbio:1998uu,Engelhardt:1998wu}. The refined
lattices should accordingly be blocked 2-3 times.

\begin{figure}[h]
\centering
a)\includegraphics[width=.48\linewidth]{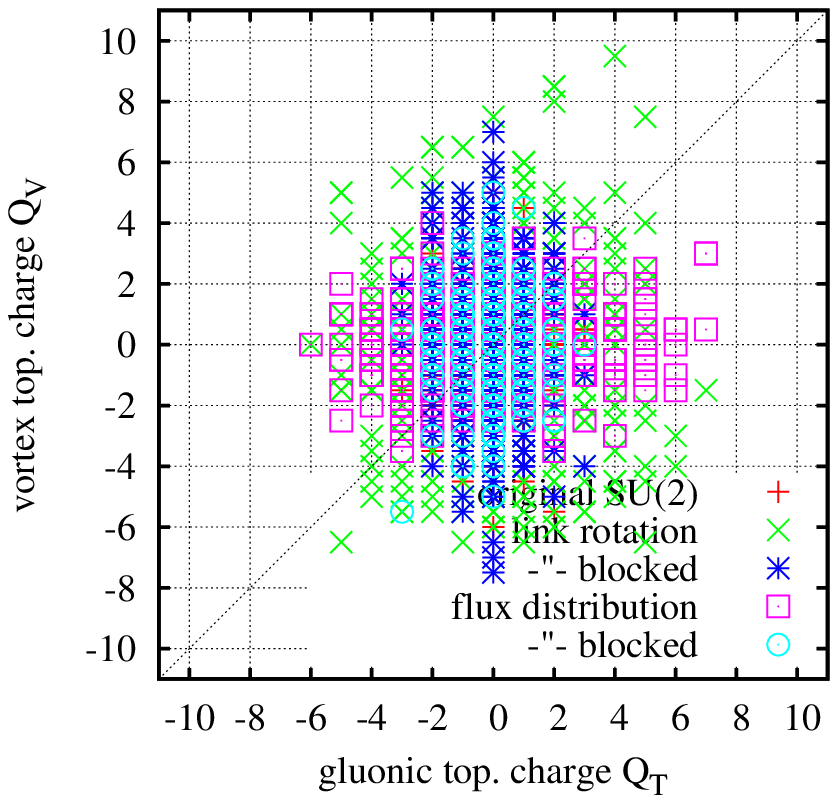}
b)\includegraphics[width=.48\linewidth]{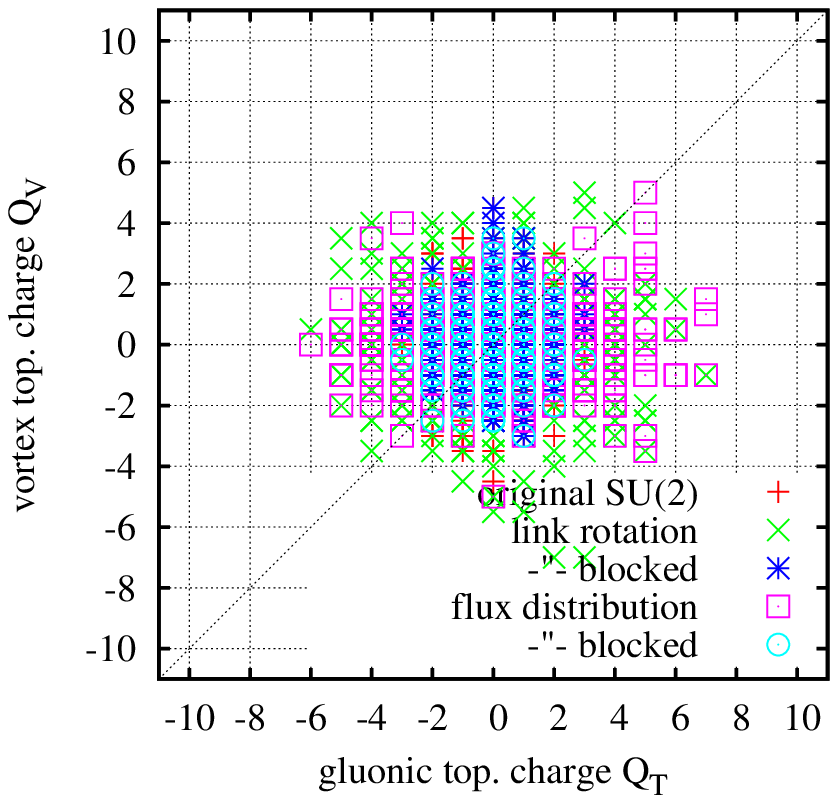}
c)\includegraphics[width=.48\linewidth]{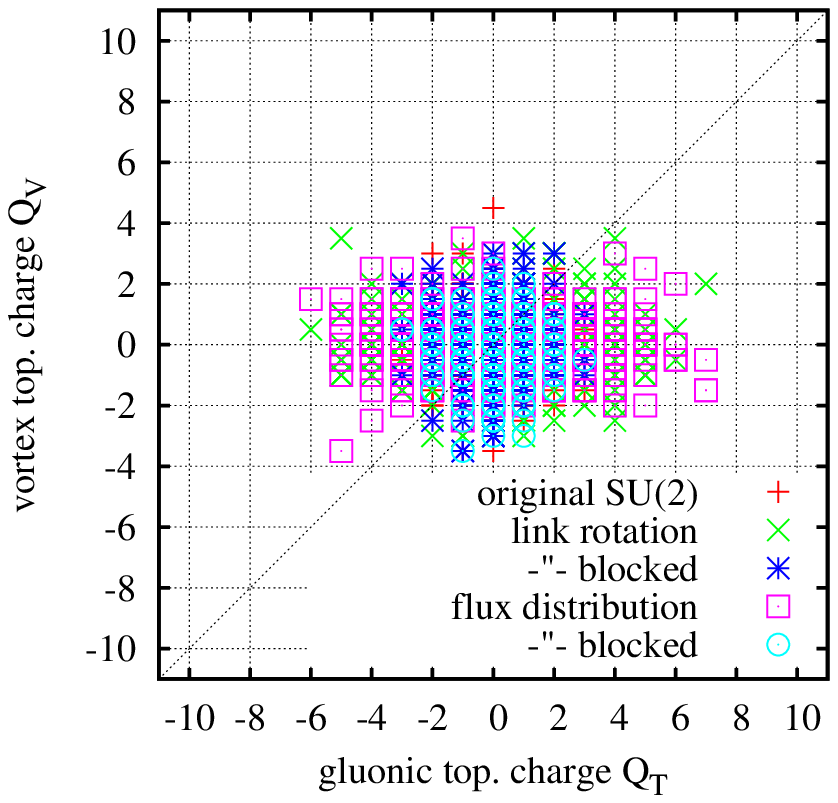}
d)\includegraphics[width=.48\linewidth]{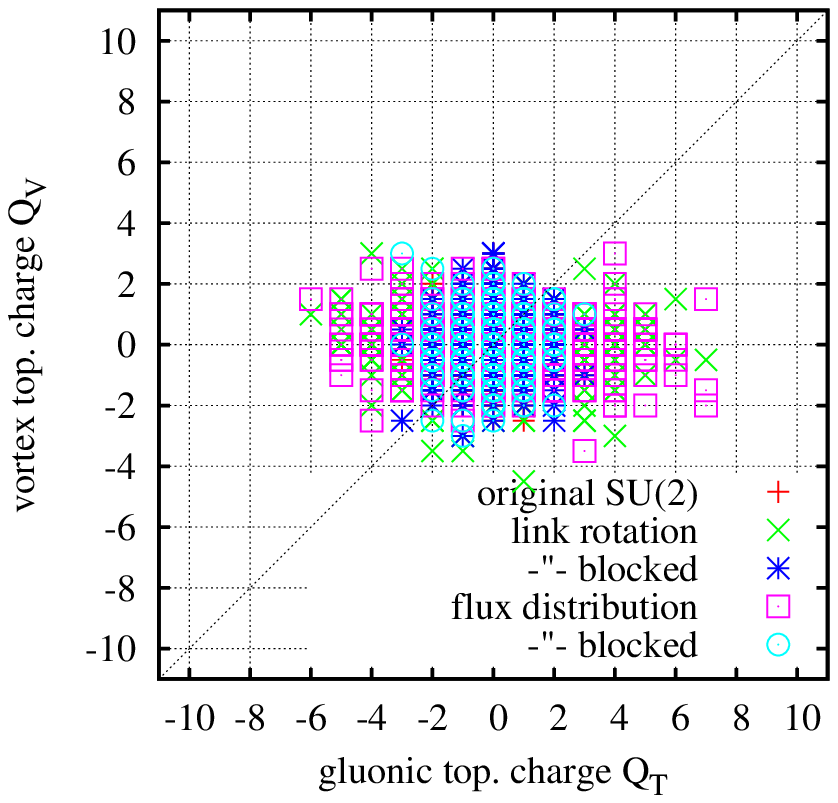}
e)\includegraphics[width=.48\linewidth]{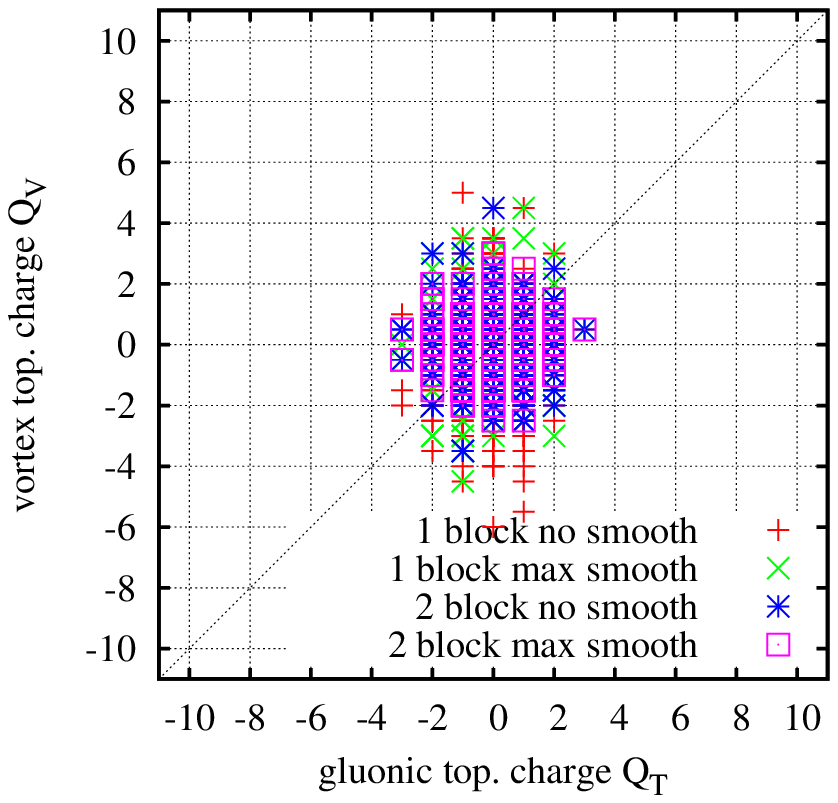}
f)\includegraphics[width=.48\linewidth]{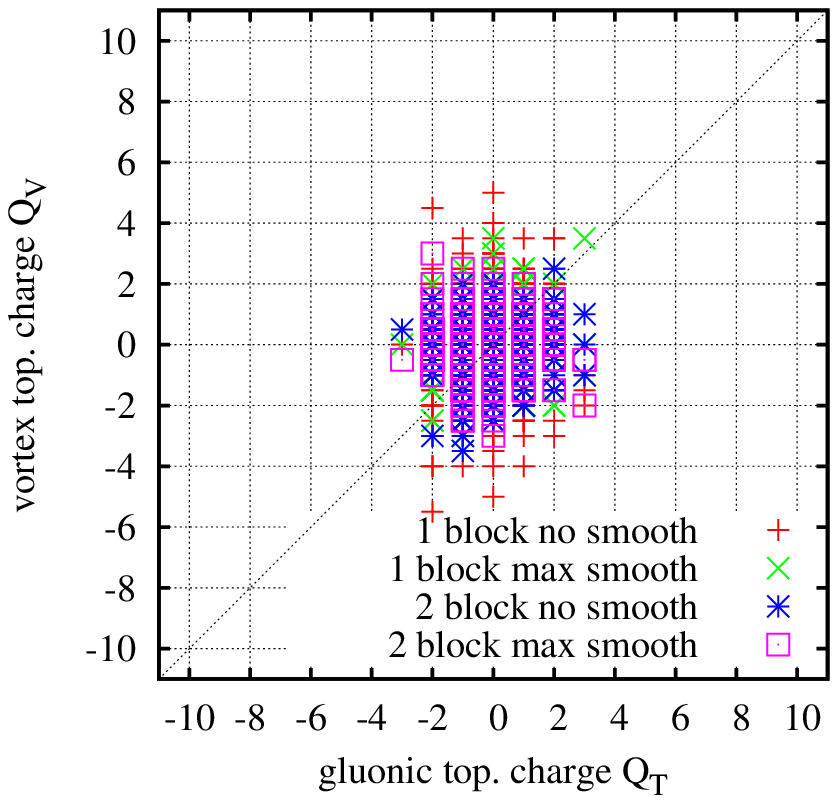}
\caption{Scatter plot of the vortex vs. gluonic topological charge 
	for original (full) $SU(2)$ and vortex smeared configurations. Vortex
	topological charge determination after a),b) one and c),d) two blocking
steps and a),c) no and b),d) maximal smoothing. 
In e/f we show the combined results (a-d) for blocked flux distribution smeared configurations again.
}\label{fig:qgv}
\end{figure}

\begin{figure}[h]
\centering
a)\includegraphics[width=.48\linewidth]{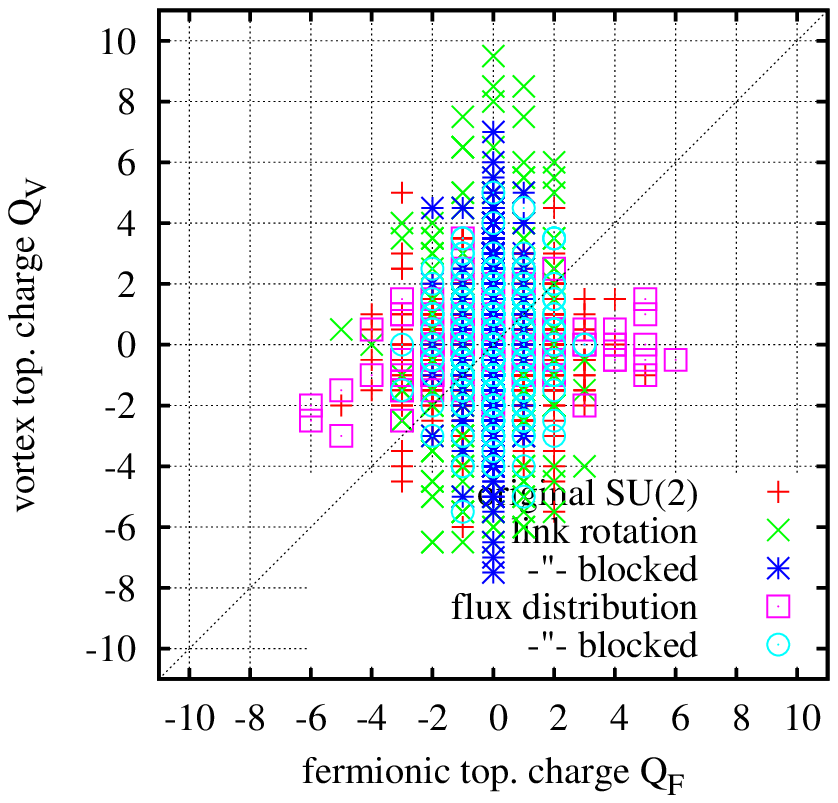}
b)\includegraphics[width=.48\linewidth]{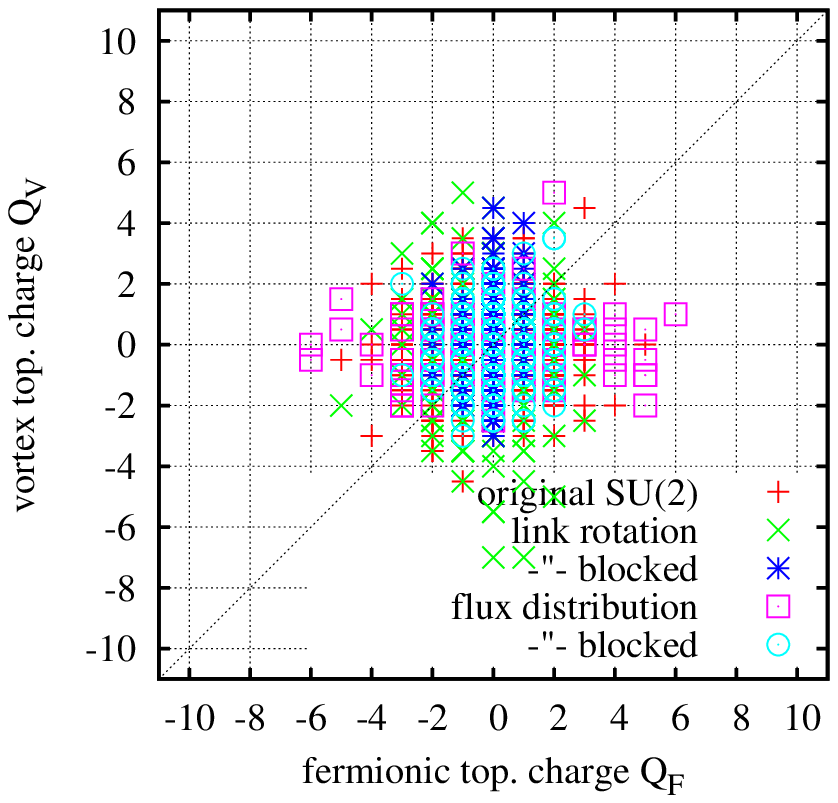}
c)\includegraphics[width=.48\linewidth]{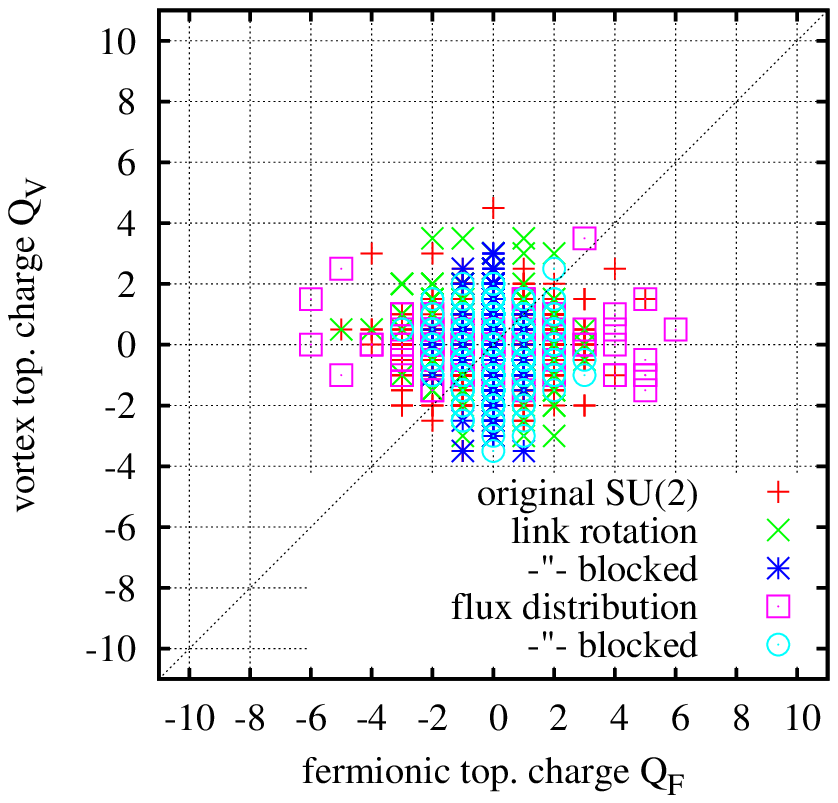}
d)\includegraphics[width=.48\linewidth]{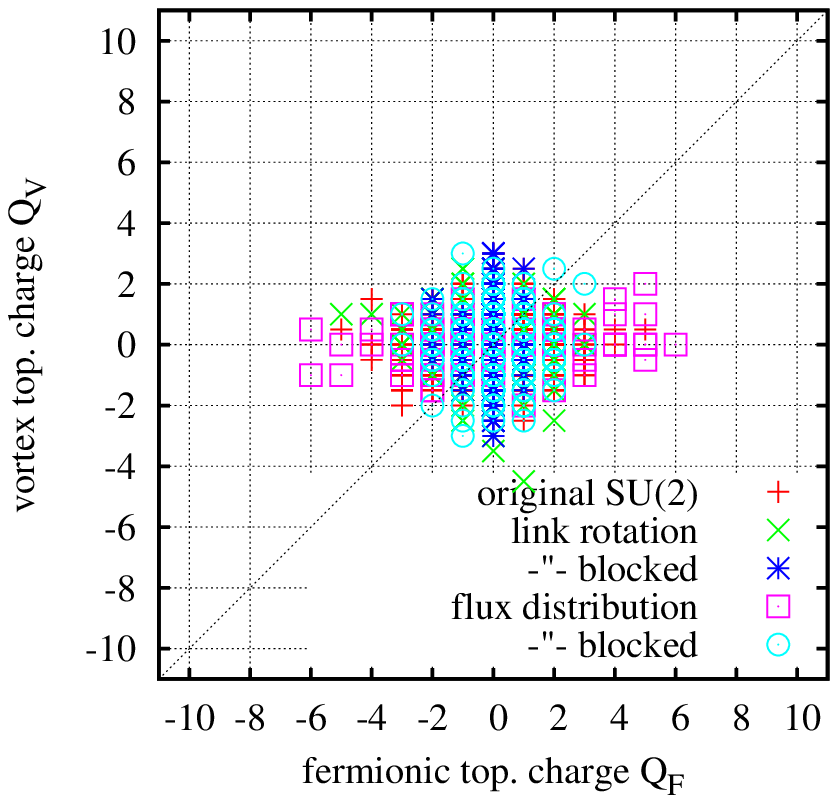}
e)\includegraphics[width=.48\linewidth]{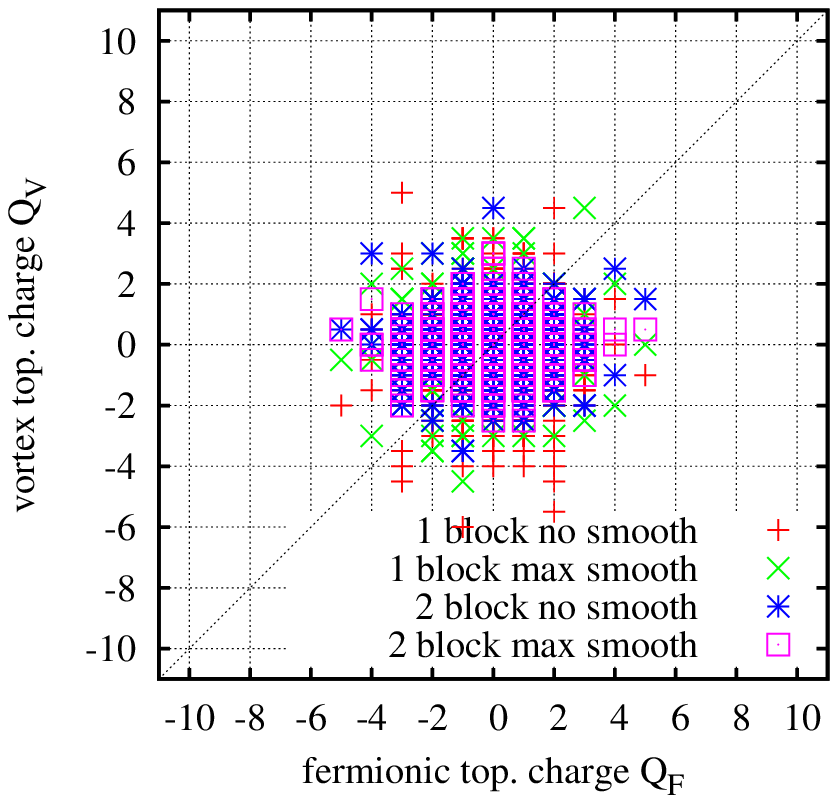}
f)\includegraphics[width=.48\linewidth]{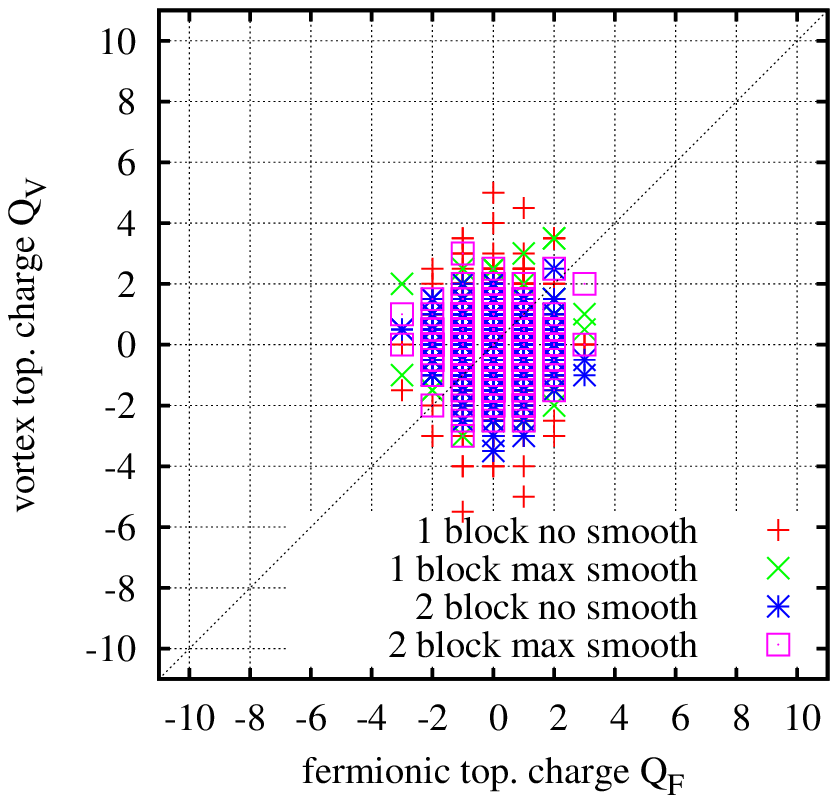}
\caption{Scatter plot of the vortex vs. fermionic topological charge for
original (full) $SU(2)$ and vortex smeared configurations. Vortex topological
charge determination after a),b) one and c),d) two blocking steps and a),c) no
and b),d) maximal smoothing. 
In e/f we show the combined results (a-d) for blocked flux distribution smeared configurations again.
}\label{fig:qzv}
\end{figure}

The vortex topological charge is not necessarily correlated to the index of the
Dirac operator, since the vortex configurations do not represent a topological
torus, as there are monopoles and Dirac strings present. Thus, the basic index
theorem is not valid and extra terms appear which are reflected in the
difference of the vortex topological charge and the index. During cooling or
smearing, monopoles and Dirac strings are smoothed out or fall through the
lattice and the toroidal topology is restored, hence $F\tilde F$ approaches the index
topological charge. However, the vortex finding property is lost during
smearing and the vortex topological charge quickly vanishes for full
configurations. These aspects were discussed in more detail in~\cite{Hollwieser:2012kb}.
Vortex topological charge depends on the orientation of the (thick) vortex
surfaces. The (thin) $Z(2)$ vortices lack any information of orientation and in
order to calculate the vortex topological charge, orientation is applied randomly
to the vortex surfaces. Similarly, during vortex smearing, by replacing the
"$Z(2)$ jump" with a smooth, random rotation in the $SU(2)$ space, we automatically
give the vortex surfaces a random orientation in the color space, which
influences the gluonic topological charge. Since these two processes are
independent, we can not expect that the smeared vortex configurations or the
vortex topological charge in general give comparable results for individual
configurations concerning topological properties. However, as stated at the
beginning of this section, the topological susceptibility of original gauge
fields is mirrored by the vortex topological charge. Hence, we want to verify if
this is also true for the vortex smeared configurations and if we can reproduce
the topological susceptibility of the original $SU(2)$ gauge fields. 
The vortex topological charge for original and smeared configurations is
essentially the same, since we deal with identical vortex structures. However,
the random application of vortex orientation ruins a one-to-one correlation,
unless we use the same random generator every single time. Even though this
random step has no influence on the rather dominant writhing point contribution to
topological charge, the random contribution of a few intersection points is
enough to destroy a one-to-one correlation between individual configurations.
Taking together all arguments from
the previous and this section concerning the different approaches to topological
charge determination, from center vortices, gluonic or fermionic definitions, it
is not surprising that the first of these does not exactly reproduce the latter ones for
individual configurations. In Figs.~\ref{fig:qgv} and~\ref{fig:qzv} we show
correlations between gluonic resp. fermionic and vortex topological charge - there is
no one-to-one correlation.

\begin{figure}[h]
	\centering
	a)\includegraphics[width=.48\linewidth]{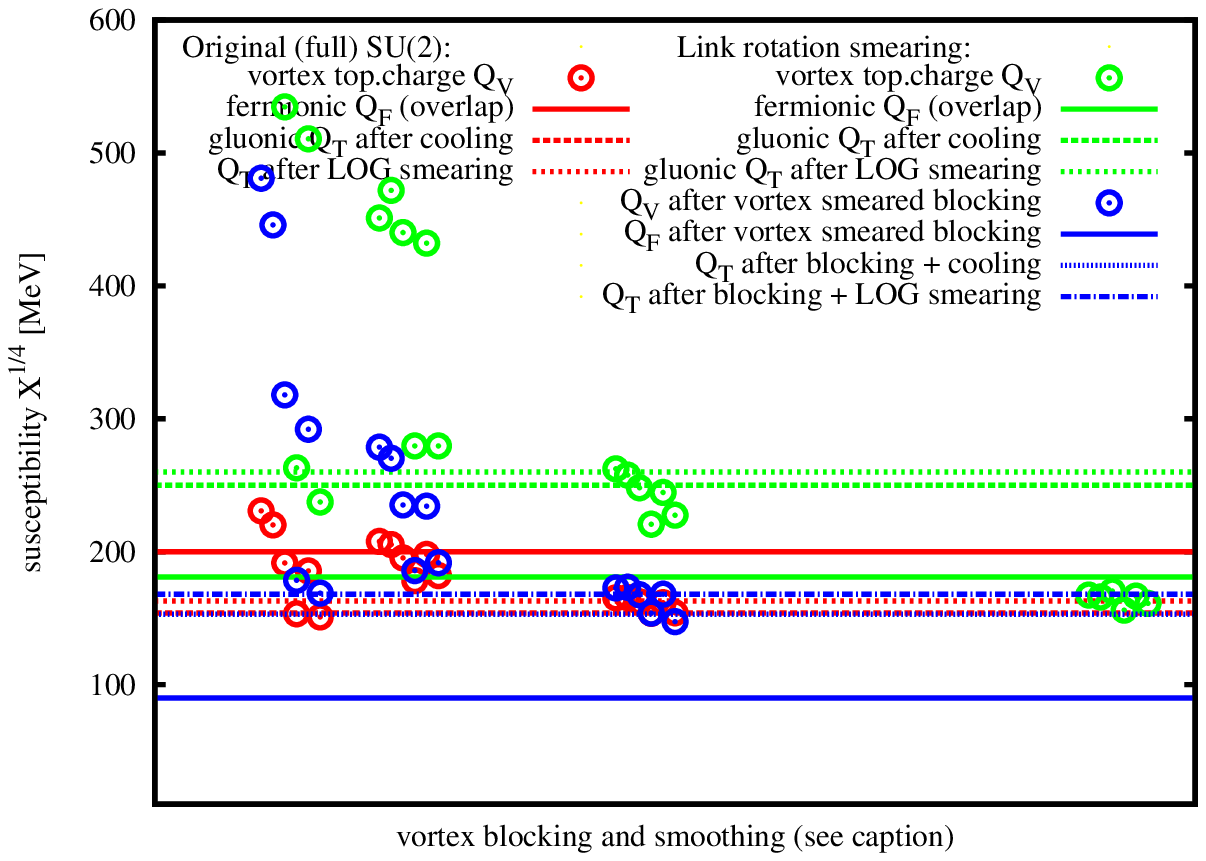}
	b)\includegraphics[width=.48\linewidth]{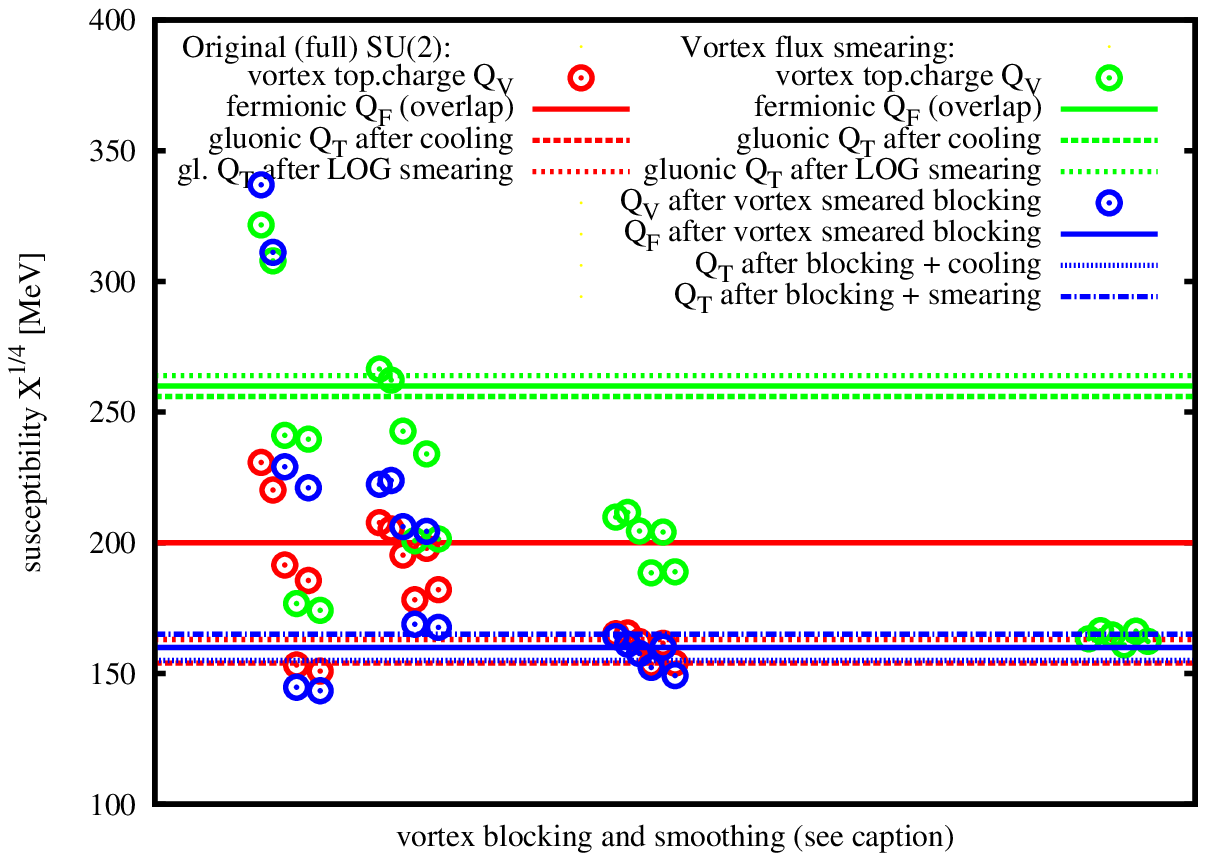}
	c)\includegraphics[width=.48\linewidth]{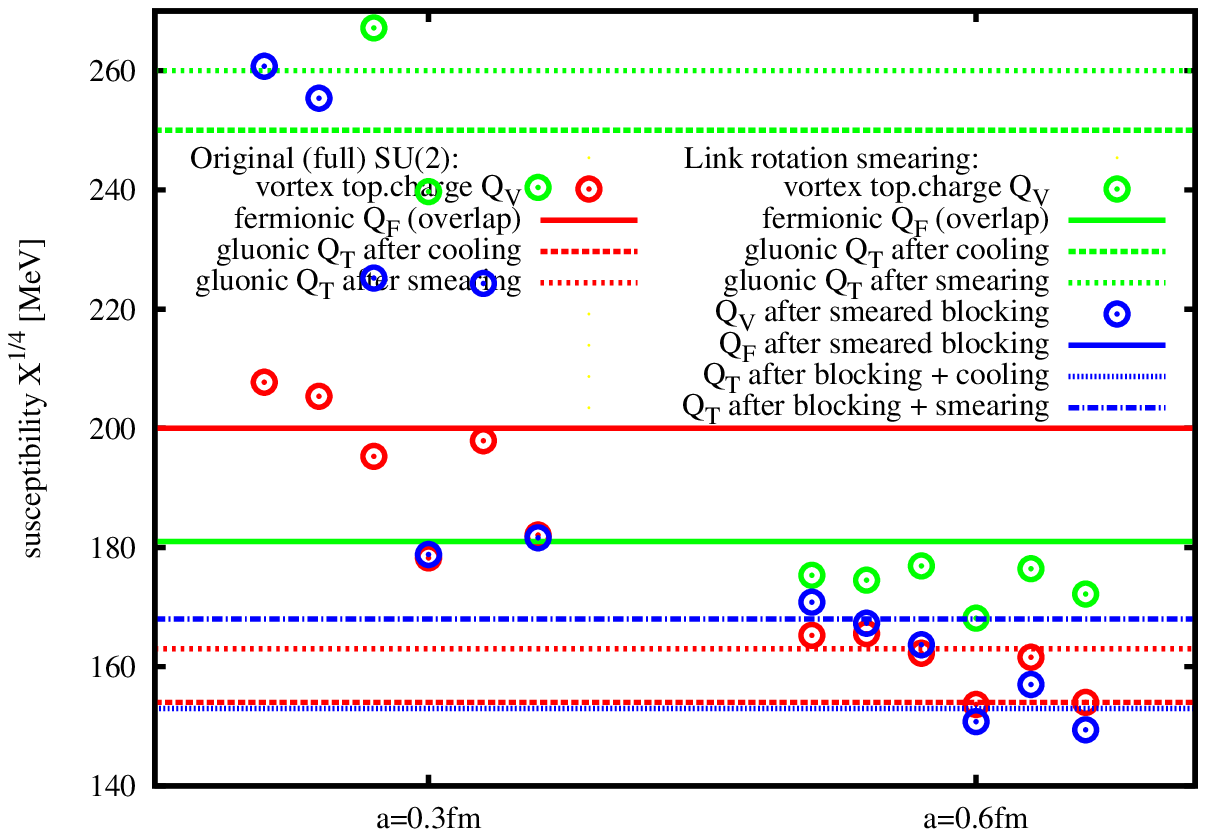}
	d)\includegraphics[width=.48\linewidth]{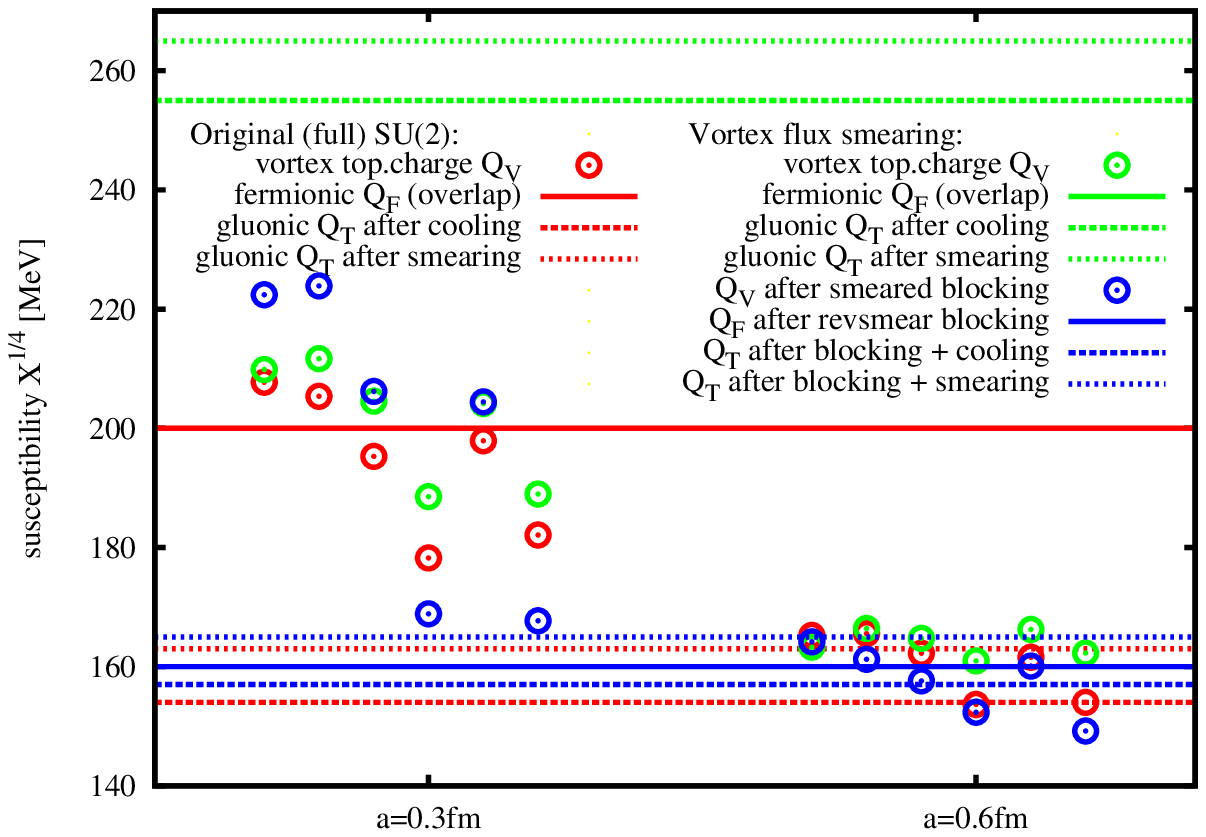}
	\caption{Topological susceptibility from fermionic $Q_F$, gluonic $Q_T$ and
	vortex topological charge $Q_V$ for original (full) $SU(2)$ (red) and refined
	a),c) link rotation (see Sec.~\ref{sec:reflrs}) or b),d) vortex flux
	distribution (see Sec.~\ref{sec:reflx}) smeared (green) and blocked (blue,
	see	Sec.~\ref{sec:block}) configurations. In a) and b) we show all blocking
	and smearing steps for $Q_V$, {\it i.e.}, we see four vertical groups of
	data points according to no, one, two and three blocking steps from left to
	right. Within these groups we plot the data points for zero to five smoothing 
	steps from left to right again. In c) and d) we show 1-2 blocking steps for
	$8^4$ lattices, {\it i.e.}, original and smeared blocked configurations, and
	2-3 blocking steps for refined ($16^4$) lattices, resulting in lattice
	spacings $a\approx0.3-0.6$fm respectively. We also zoom into the interesting
	susceptibility region and therefore miss a few data points from vortex
	topological charge $Q_V$ (green dots in a and c) and the blue line for
	topological susceptibility 	from fermionic $Q_F$ for blocked link rotation 
	smearing (blue solid line) in c. Note that for original (full) $SU(2)$,
the topological susceptibilities from fermionic (red solid line) and gluonic
topological charge definitions (red dashed lines) are not consistent, $\langle
Q_F^2\rangle/V=(200$MeV$)^4$ and $\langle Q_T^2\rangle/V\approx(160$MeV$)^4$
(averaging data from cooling and LOG smearing $Q_T$), presumably caused by our small original
lattice volume of about $(1.2$fm$)^4$. Vortex topological charge $Q_V$ (red
dots) reproduces the two values after one resp. two blocking steps in c) or d).
Refined smeared configurations (green) show very high susceptibilities for
gluonic and vortex topological charge, caused by artificial vortex fluctuations
introduced in the refined smearing methods. Blocking removes these fluctuations
and we observe good agreement between the different topological charge
definitions on original and vortex smeared configurations, especially for the
vortex flux distribution smearing - see d) at $a=0.6$fm ({\it i.e.}, two blocking
steps for $Q_V$).}\label{fig:susc}
\end{figure}

In Fig.~\ref{fig:susc} we show the topological susceptibilities for
original (full) $SU(2)$ and a) link rotation or b) flux distribution smeared configurations. 
The first thing we note is that for our original $SU(2)$ gauge ensemble,
the topological susceptibilities from fermionic and gluonic topological charge
definitions are not consistent, $\langle Q_F^2\rangle/V=(200$MeV$)^4$ and $\langle
Q_T^2\rangle/V=(160$MeV$)^4$ (averaging cooling and smearing $Q_T$), presumably caused
by our small original lattice volume of about $(1.2$fm$)^4$. It is very
interesting, however, that the vortex topological susceptibility reproduces these
values with one, respectively two blocking steps, averaging over the
corresponding smoothing steps, see red dots/lines in Fig.~\ref{fig:susc}. 
Next, we see that for the refined smeared configurations the gluonic and vortex topological charges (green dots and dashed lines) lead to much higher 
susceptibilities, caused by the artificial vortex fluctuations introduced during
the refined smearing process giving many (extra) contributions to $\mathcal
F_{\mu\nu} \tilde{\mathcal F}_{\mu\nu}$. This effect is larger for link rotation
(Fig.~\ref{fig:susc}a) compared to flux distribution smearing
(Fig.~\ref{fig:susc}b). After blocking, however,the original results are
reproduced, shown in the lower plots in each case, {\it i.e.}, compare red and
blue dots/lines in Fig.~\ref{fig:susc}c and d. The gluonic topological
susceptibilities after cooling and (LOG) smearing (blue dashed lines) agree with
the original values (red dashed lines) and vortex topological charge also
matches the original averages (blue and red dots). Concerning
fermionic topological susceptibility (solid lines), the refined link rotation smearing
reproduces a value of $\langle Q_F^2\rangle/V=(180$MeV$)^4$; after smeared
blocking, the value drops to $(95$MeV$)^4$, however. The vortex flux distribution
smearing method gives a more reasonable result, perfectly consistent with
gluonic and vortex topological susceptibilities. The refined version of flux
distribution smearing gives a topological susceptibility from fermionic $Q_F$ of
$(260$MeV$)^4$, which lies exactly between the gluonic values after cooling or
LOG smearing for the refined flux distribution smeared configurations. After blocking 
this value drops to $(160$MeV$)^4$, consistent with gluonic topological charge
susceptibilities from original (full) SU(2) and vortex flux smeared and blocked
configurations. For our rather small original lattices of about $(1.2$fm$)^4$, a topological
susceptibility of roughly $(160$MeV$)^4$ seems reasonable and we find that all
definitions of topological charge agree on this value after appropriate smearing
and blocking, as the lower plots at $a=0.6$fm show. The results confirm that vortices are
indeed able to reproduce the topological susceptibility of full QCD, either via
vortex topological charge or gluonic and fermionic definitions after vortex smearing.

\clearpage

\subsection{Center Vortex and Dirac Eigenmode Correlations}

We analyze the correlation of the overlap Dirac zero mode and first asqtad
staggered Dirac eigenmode of the vortex smeared configurations to the original
vortex structure. We use the correlator $C_\lambda(N_{\rm v}) =
\frac{\sum_{p_i}\sum_{x\in H}(V\rho_\lambda(x)-1)}{\sum_{p_i}\sum_{x\in
H}1}$~\cite{Kovalenko:2005rz},
where the sum is over sites $p_i$ on the dual lattice which belong to $N_{\rm v}$
plaquettes on the vortex surface (as identified from center projection),
$\rho_\lambda(x)$ is the eigenmode density and $V$ is the lattice volume.  At each
such vortex site on the dual lattice there is a second sum ($x\in H$) over sites
in a hypercube on the original lattice surrounding $p_i$. This correlator gives
the relative enhancement of the eigenmode density at the vortex surface. A
similar quantity $C_\lambda(q_{\rm v})$ can be formulated for vortex
topological charge density $q_{\rm v}$, with $p_i$ the sites on the dual lattice
carrying vortex topological charge $q_{\rm v}$. 

\begin{figure}[h]
	\centering
	a)\includegraphics[width=.48\linewidth]{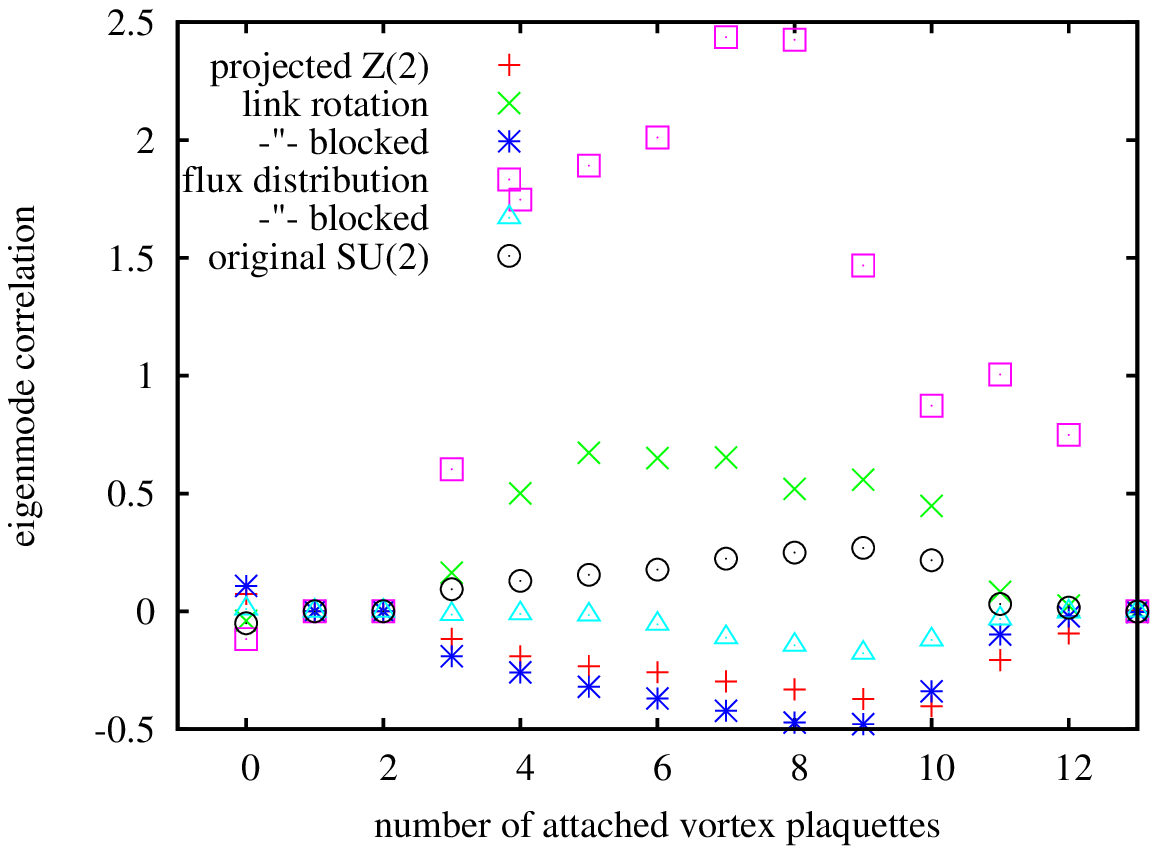}
	b)\includegraphics[width=.48\linewidth]{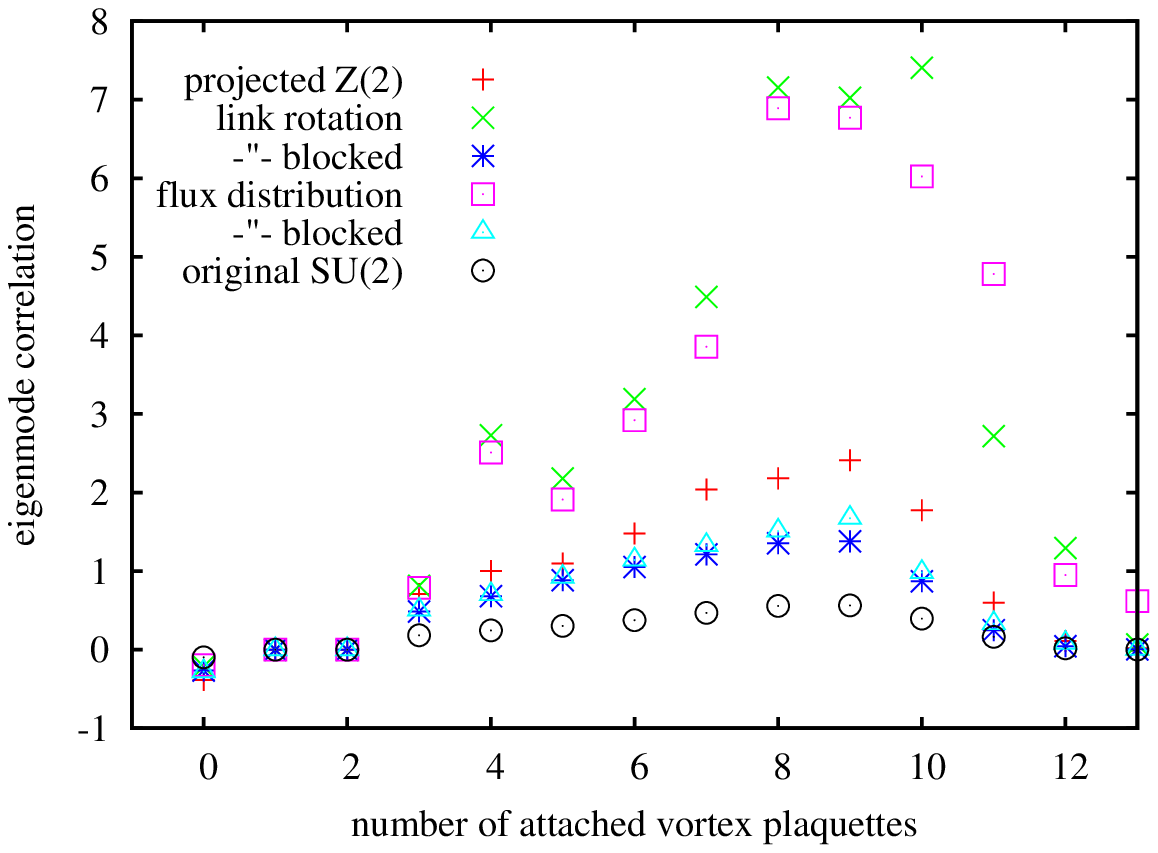}
	\caption{Vortex correlation of a) overlap and b) asqtad staggered eigenmodes 
		for	original (full) $SU(2)$, Maximal Center Gauge projected $Z(2)$ and vortex smeared configurations.}
	\label{fig:vcs}
\end{figure}

\begin{figure}[h]
	\centering
	a)\includegraphics[width=.48\linewidth]{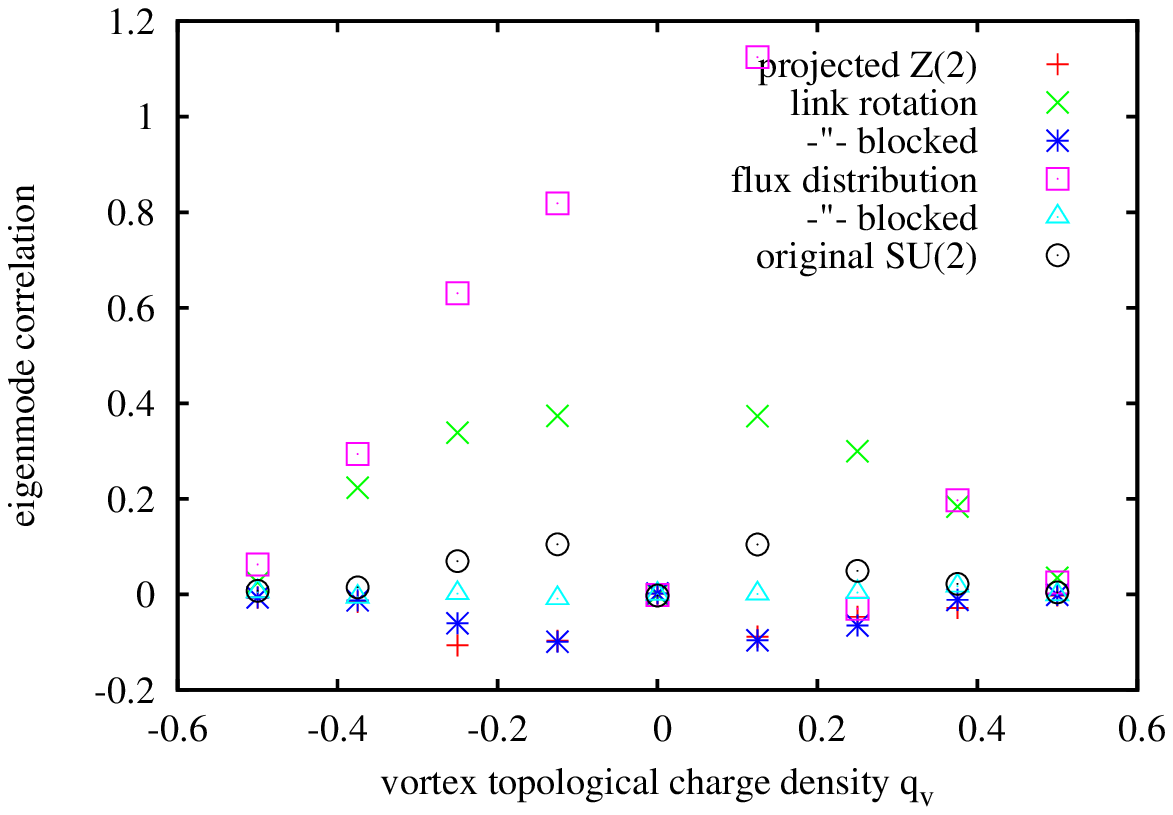}
	b)\includegraphics[width=.48\linewidth]{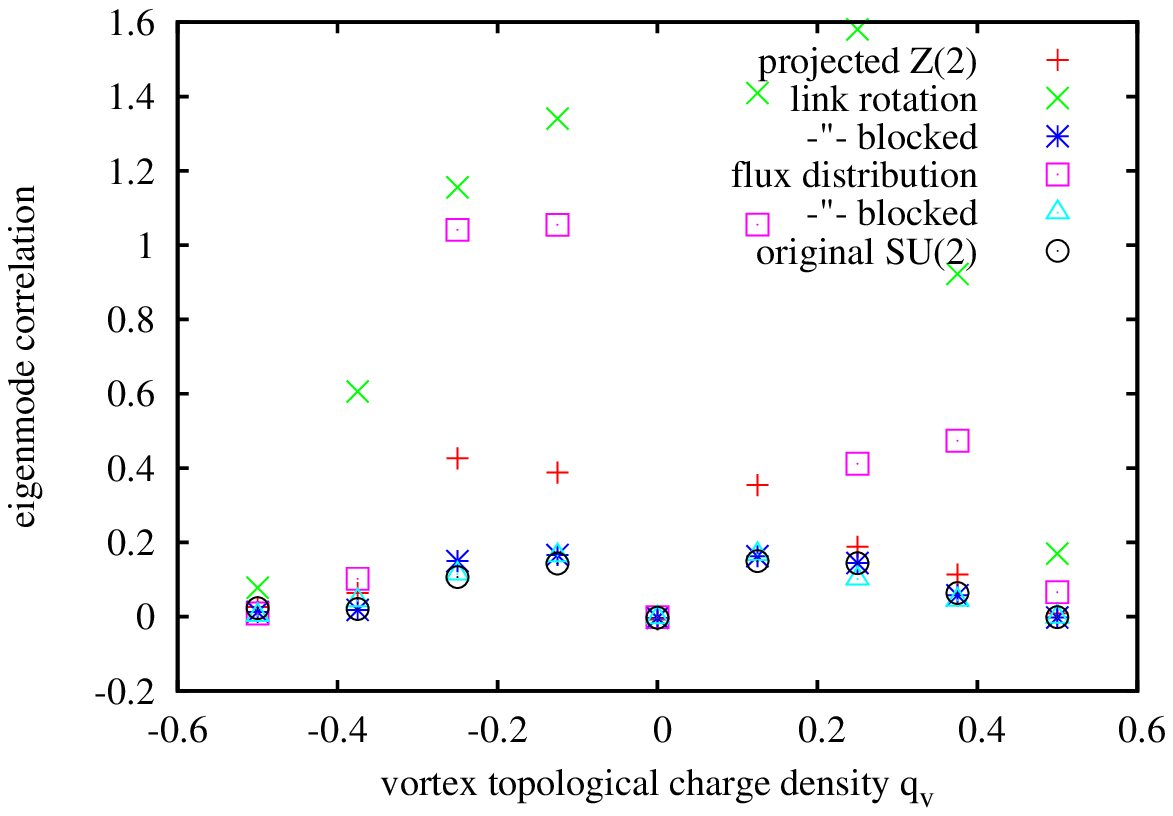}
	\caption{Correlation of vortex topological charge density with a) overlap zero mode
	and b) lowest asqtad staggered eigenmode densities for original (full) $SU(2)$,
center projected $Z(2)$ and vortex smeared configurations.}
	\label{fig:vcq}
\end{figure}

In Figs.~\ref{fig:vcs} and~\ref{fig:vcq} we display the data for
$C_\lambda(N_{\rm v})$ resp. $C_\lambda(q_{\rm v})$ computed for overlap and
staggered eigenmodes on the original (full), center projected $Z(2)$ and various
smeared configurations. We see that for refined vortex smeared configurations
the eigenmodes are strongly correlated to the vortex surface and topological
charge. The anti-correlation of overlap eigenmodes for the $Z(2)$ (center
projected) configurations is completely removed after refined smearing, however
after smeared blocking we lose the correlation to the original vortex
structure again. We tried to extend the second sum ($\sum_{x\in H}$ in
$C_\lambda$) to next-to-nearest neighbors but then the signal is lost in the
background noise, {\it i.e.}, we practically sum over all
sites where the correlator gives zero per definition. 
In the case of asqtad staggered eigenmodes we get good correlations
for all cases. The refined smeared configurations show drastically enhanced
correlations whereas blocked smeared results lie between original SU(2)
and center projected $Z(2)$ correlations.

\subsection{Wilson Loops and Vortex Limited Wilson Loops}

\begin{figure}[h]
	\centering
	a)\includegraphics[width=.48\linewidth]{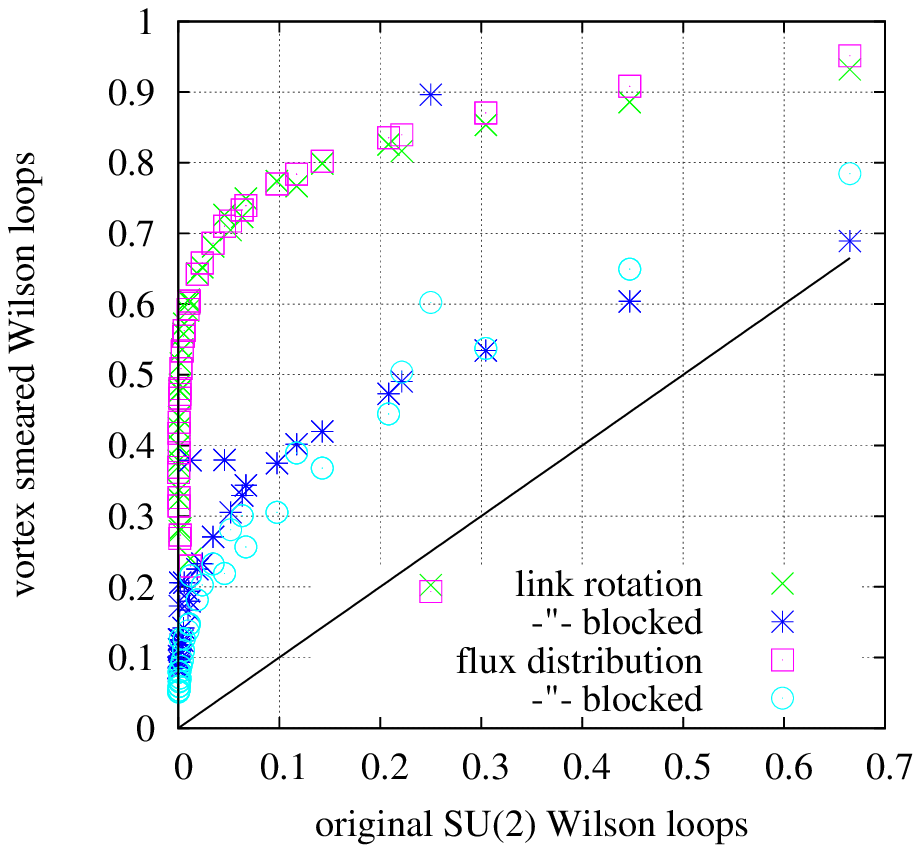}
	b)\includegraphics[width=.48\linewidth]{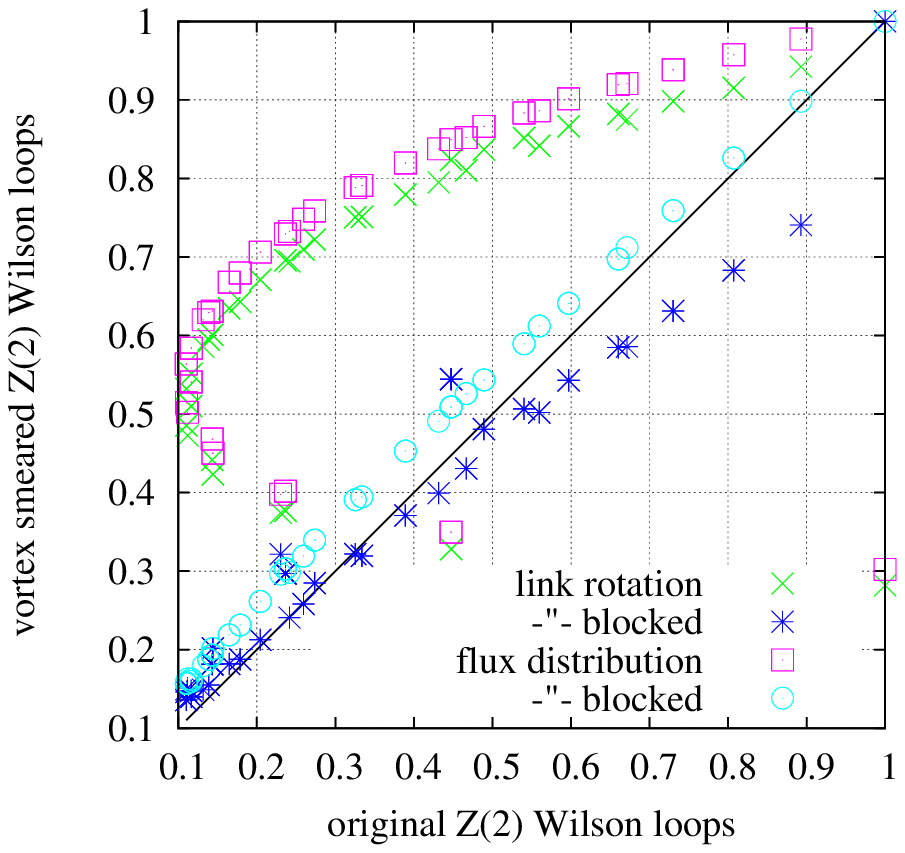}
	\caption{Scatter plot of a) $SU(2)$ and b) $Z(2)$ Wilson loops during smearing for original
	           and various vortex smeared configurations.}
	\label{fig:wil}
\end{figure}

In Fig.~\ref{fig:wil} we show the standard and center projected Wilson loops of
original $SU(2)$ vs. vortex smeared configurations. Plaquettes are systematically
minimized during smearing, hence small smeared Wilson loops tend to be much closer
to $\mathbbm1$. The $Z(2)$ Wilson loops, {\it i.e.}, Wilson loops after MCG and
center projection, for smeared blocked configurations however seem to reproduce the original $Z(2)$ Wilson loops.

\begin{figure}[h]
	\centering
	a)\includegraphics[width=.48\linewidth]{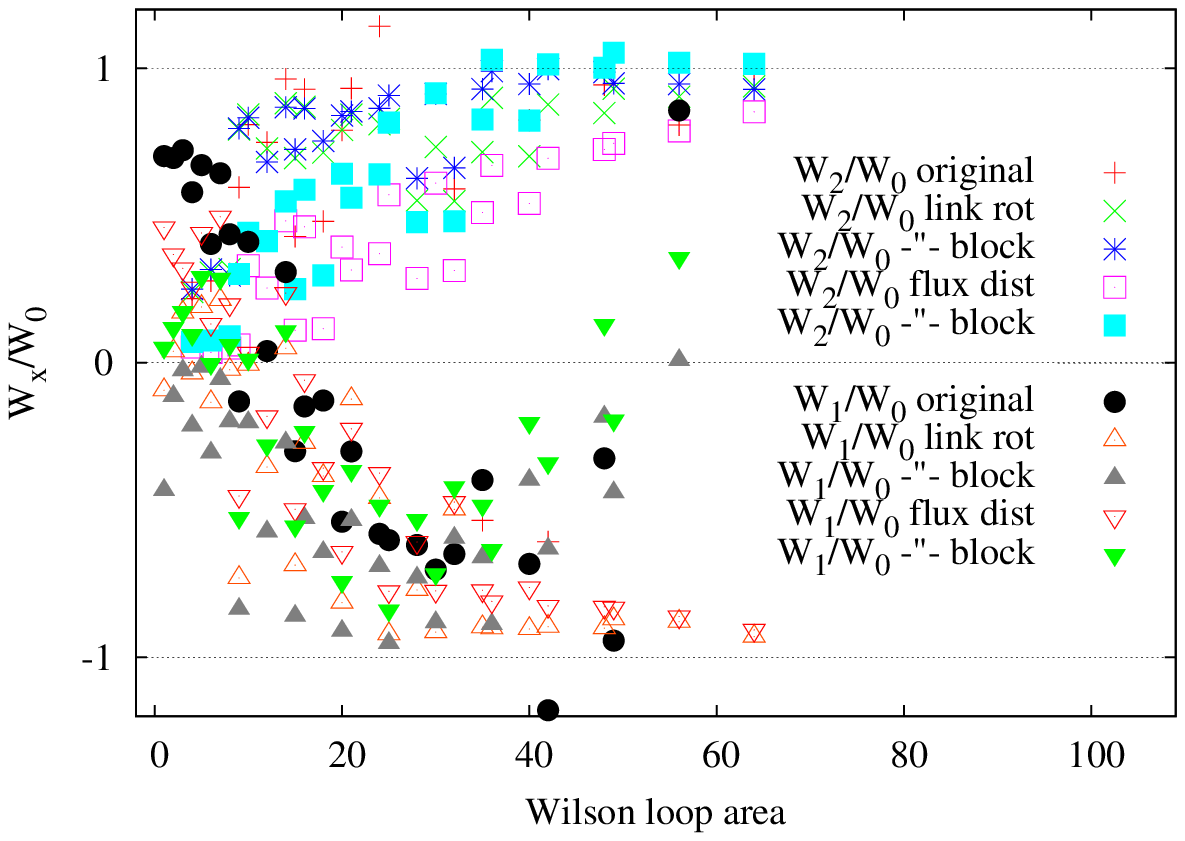}
	b)\includegraphics[width=.48\linewidth]{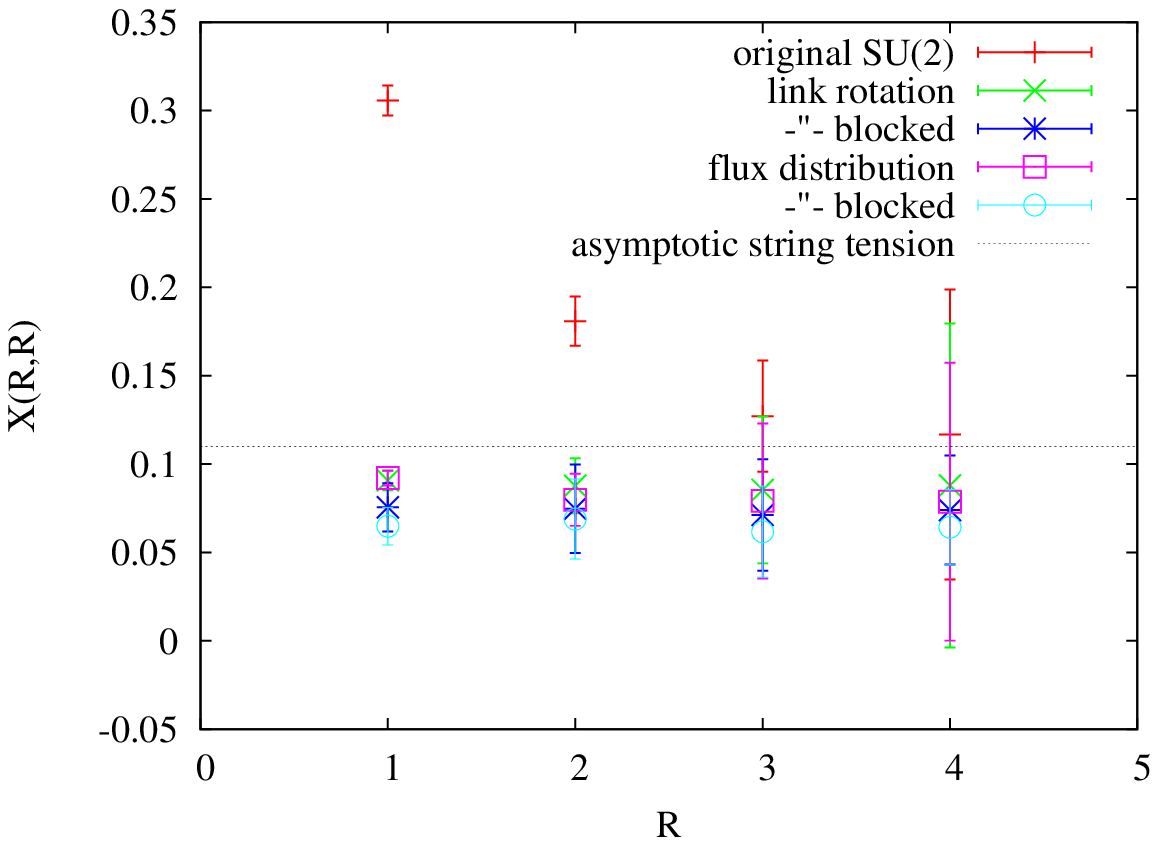}
	\caption{a) Vortex limited Wilson loops and b) Creutz ratios of original (full) $SU(2)$ and various vortex smeared configurations.}
	\label{fig:wvl}
\end{figure}

In Fig.~\ref{fig:wvl}a we show the ratios of vortex limited Wilson loops, {\it
i.e.}, Wilson loop averages $W_i$ evaluated on subsets of loops with $i$ original vortex piercings. The results show
that the vortex structure is preserved, only for large Wilson loops on blocked
smeared lattices the signal becomes weak. Finally, in Fig.~\ref{fig:wvl}b we plot
the Creutz ratios $\chi(R,R)$ which give for $R\rightarrow\infty$ the asymptotic
string tension. We find that the vortex smeared configurations reproduce the
original string tension, but not the Coulomb interaction for small $R$.

\clearpage

\section{Vortex Smearing and Classical Configurations}\label{sec:class}
Finally, we also tested the vortex smearing method on classical vortex
configurations, namely planar and spherical vortices. The figures give an
overview of the effects during vortex smearing and nicely illustrate the individual
steps. 

\subsection{Planar Vortex Pairs}
Plane vortices are constructed as presented in~\cite{Jordan:2007ff,Hollwieser:2011uj}. 
Due to the periodic boundary conditions of our lattice the vortex planes always
appear in pairs and we analyze two vortex pairs in perpendicular directions,
{\it i.e.}, $xy$- and $zt$-vortices. They intersect at four space-time points,
which contribute to the topological charge with contributions $Q=\pm1/2$
depending on the orientation of the intersecting vortex sheets. After vortex
smearing we may get total topological charge $Q=-2,-1,0,1$ or $2$, however the
smearing routine seems to prefer the cases of even $Q$ ($-2,0$ and $2$). In
fact, $Q=|1|$ requires orientation changes of single vortex sheets, which would
introduce magnetic monopoles and therefore additional gauge singularities which
seem to be suppressed by the minimizing of the plaquettes.

\begin{figure}[h]
	\centering
	a)\includegraphics[width=.3\linewidth]{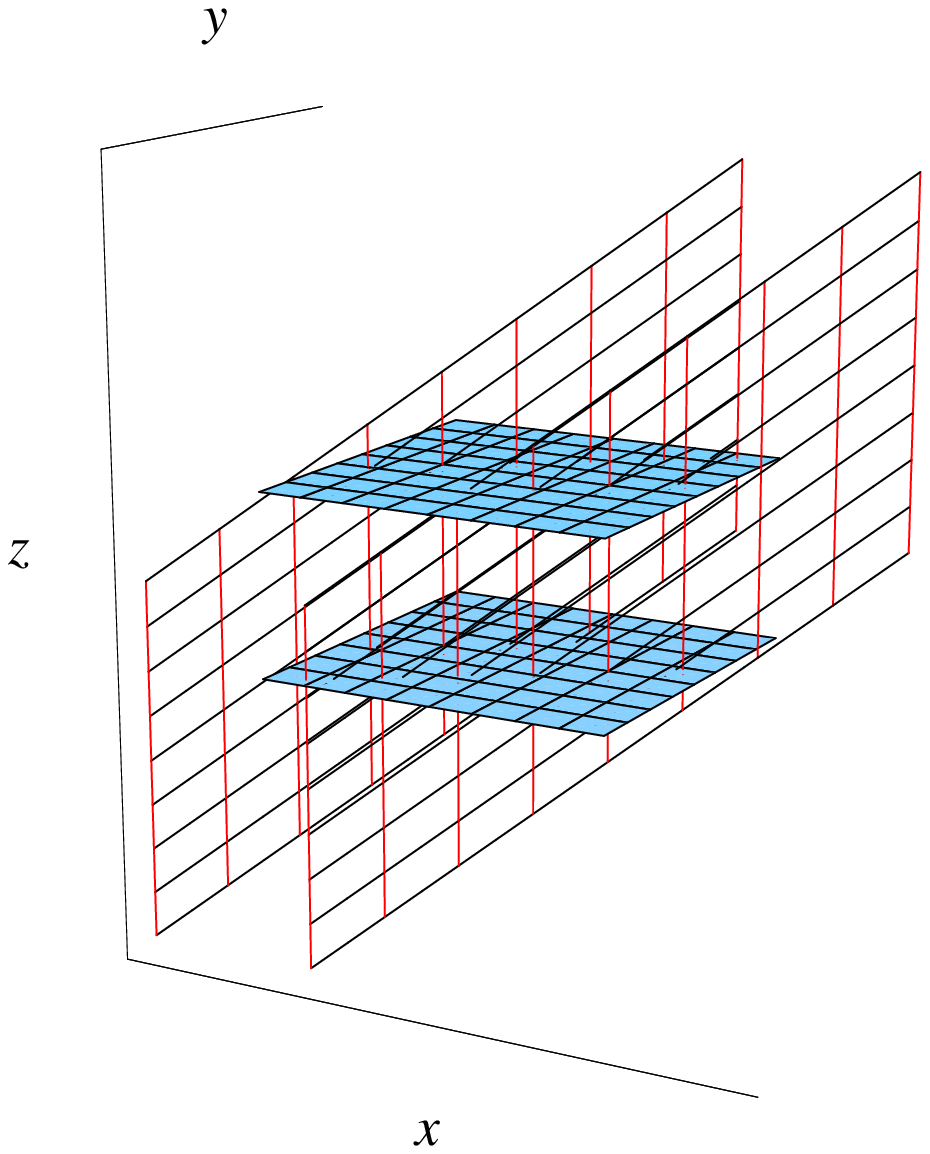}
	b)\includegraphics[width=.3\linewidth]{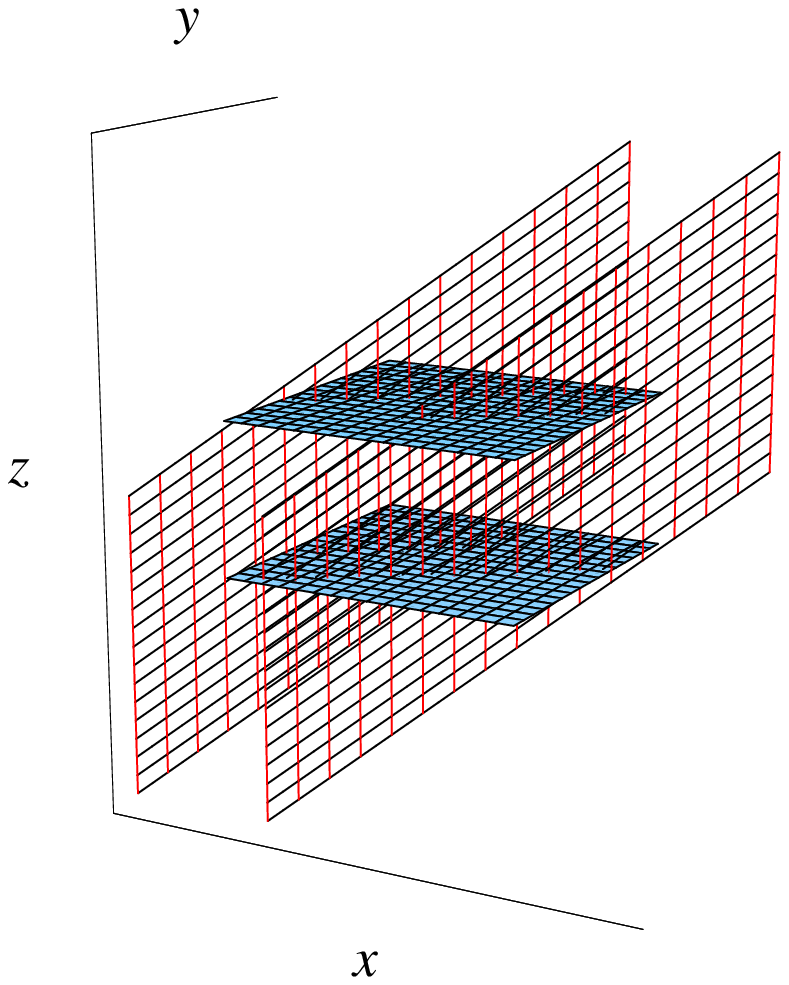}
	c)\includegraphics[width=.3\linewidth]{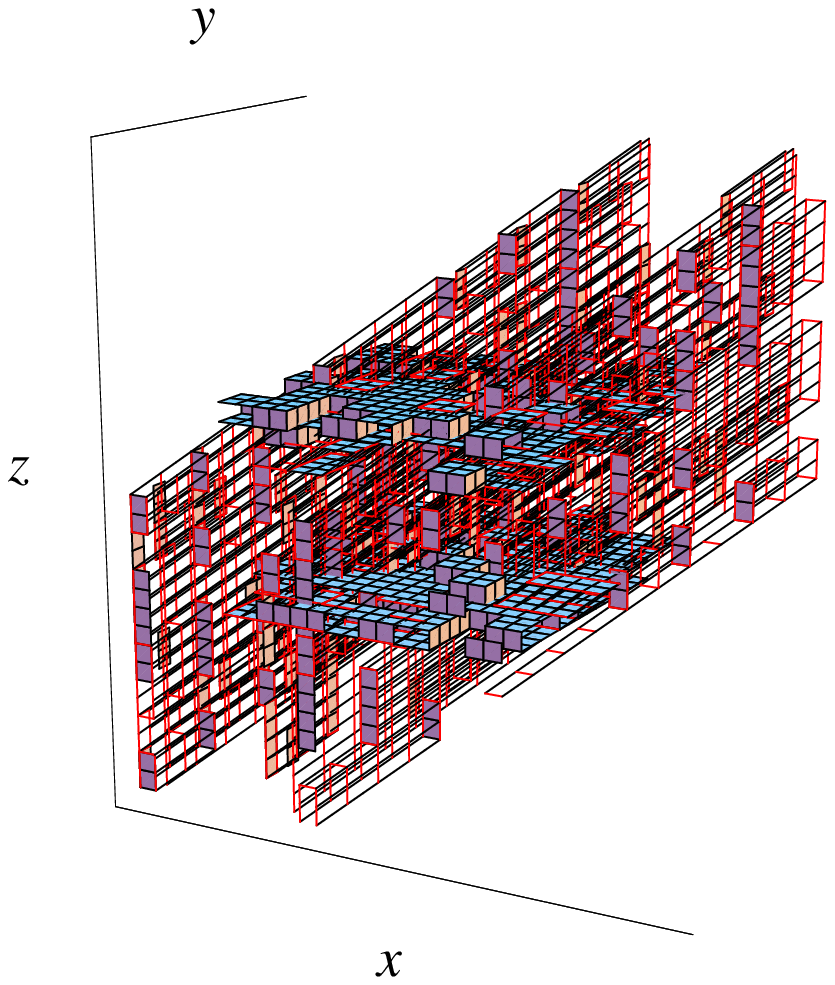}
	\caption{Two orthogonal pairs of planar vortices a) on a $8^4$ lattice and
		after b) refinement and c) vortex smearing on $16^4$ lattices. For a more
		detailed view of the configuration in c) see Fig.~\ref{fig:plan2}.
	Smoothing c) gives b) and blocking c) (or b) gives a). We plot the dual
vortex plaquettes, which actually represent the closed vortex surfaces.} 
	\label{fig:plan}
\end{figure}

Fig.~\ref{fig:plan} shows the vortex configuration on the initial $8^4$ lattice,
after refinement, vortex smearing and smoothing on the refined $16^4$ lattice.
As stated above,
refinement gives exactly the same vortex configuration on a finer lattice,
whereas the smearing routine seems to distort the plane vortex sheets.
However, after smoothing the smeared vortex surface as defined
in~\cite{Bertle:1999tw}, all distortions are removed in the case of plane vortex
sheets. After blocking even the unsmoothed smeared vortex configuration 
reveals the initial configuration. 

\begin{figure}[h]
	\centering
	\includegraphics[width=.22\linewidth]{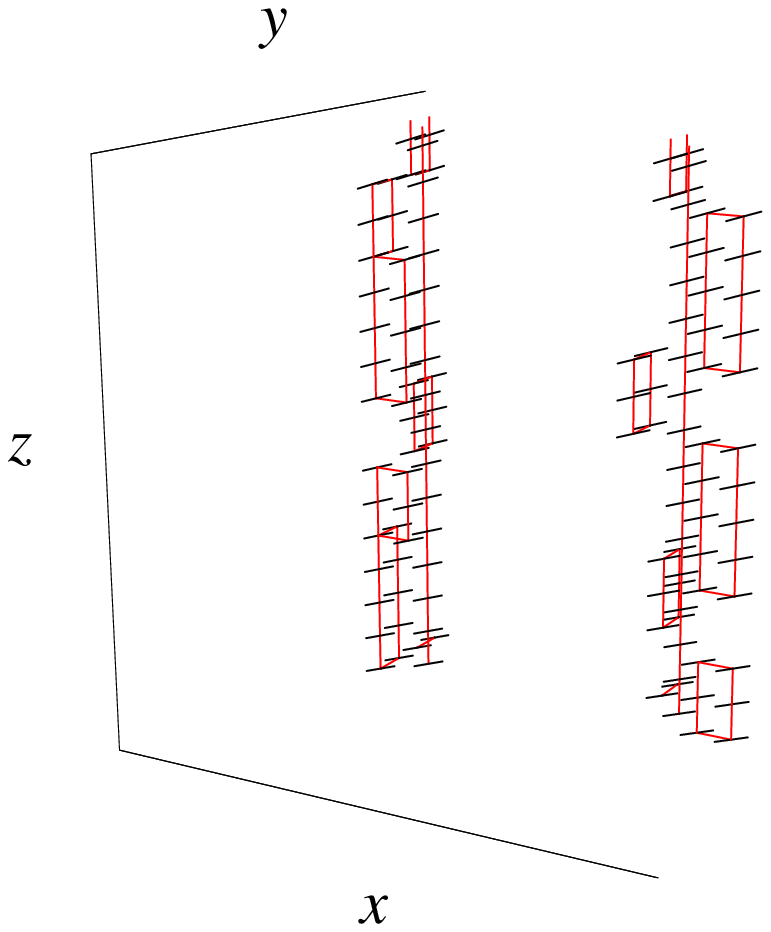}
	\includegraphics[width=.22\linewidth]{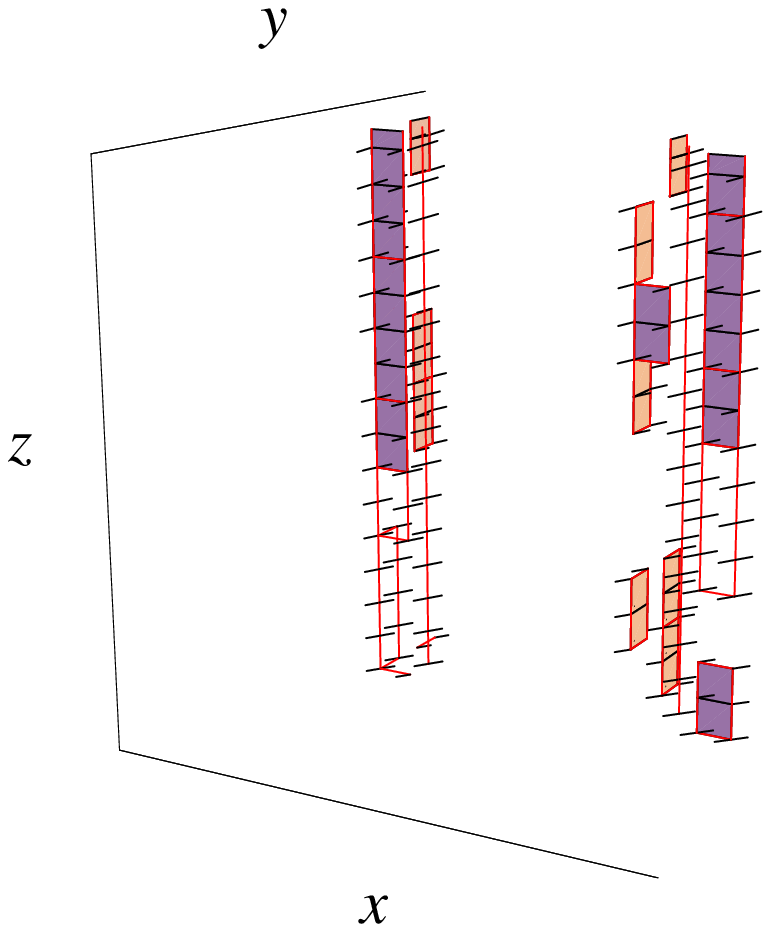}
	\includegraphics[width=.22\linewidth]{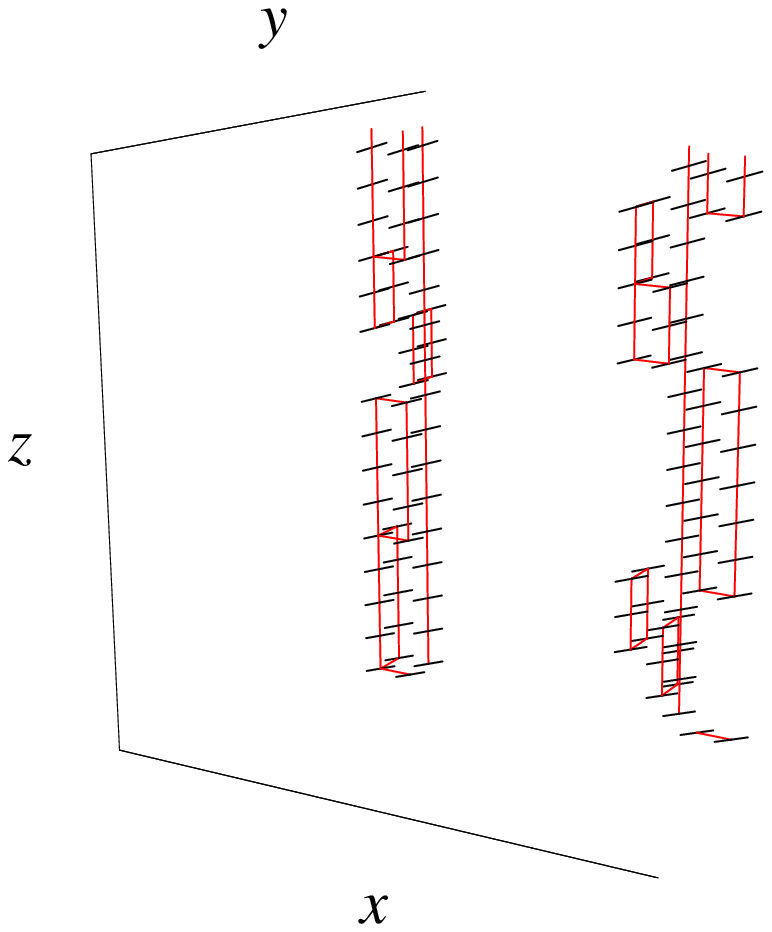}
	\includegraphics[width=.22\linewidth]{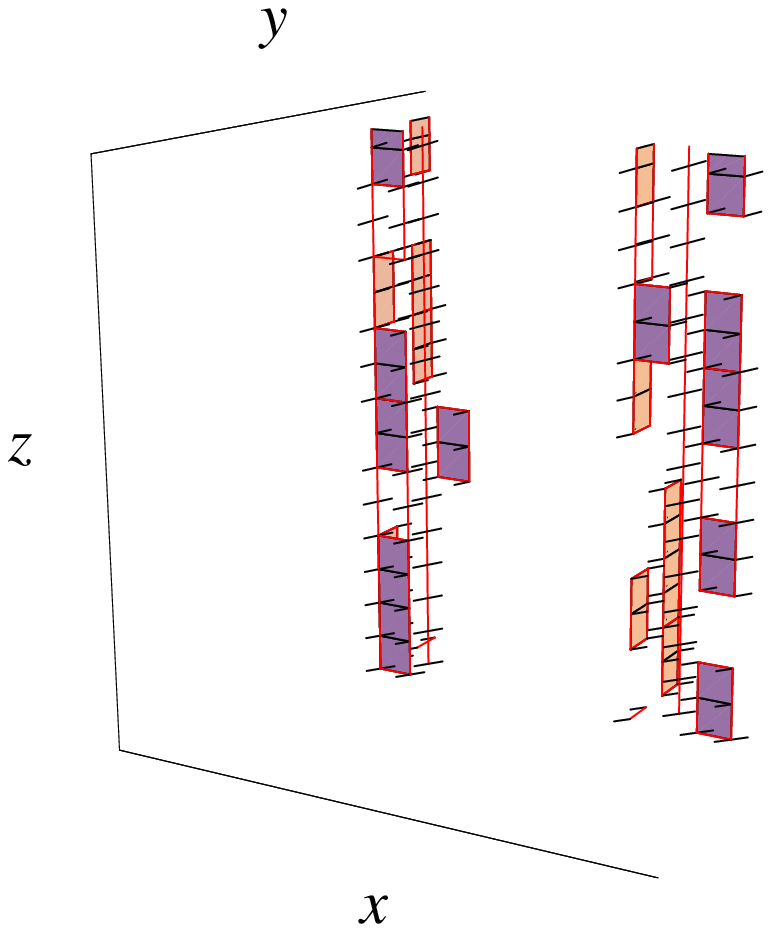}\\
	\includegraphics[width=.22\linewidth]{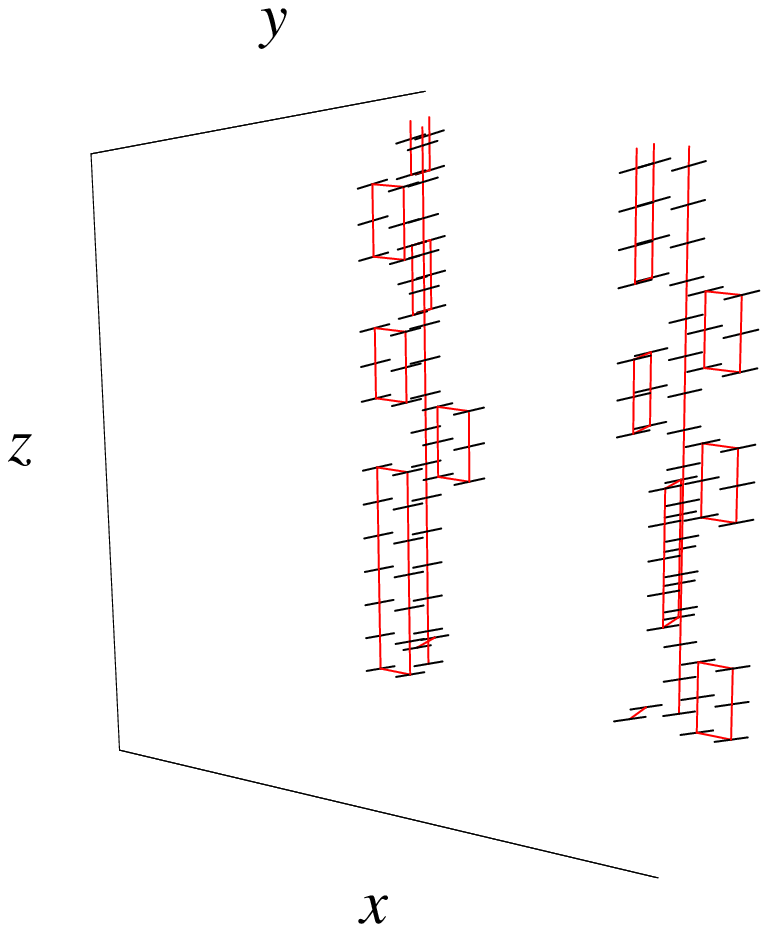}
	\includegraphics[width=.22\linewidth]{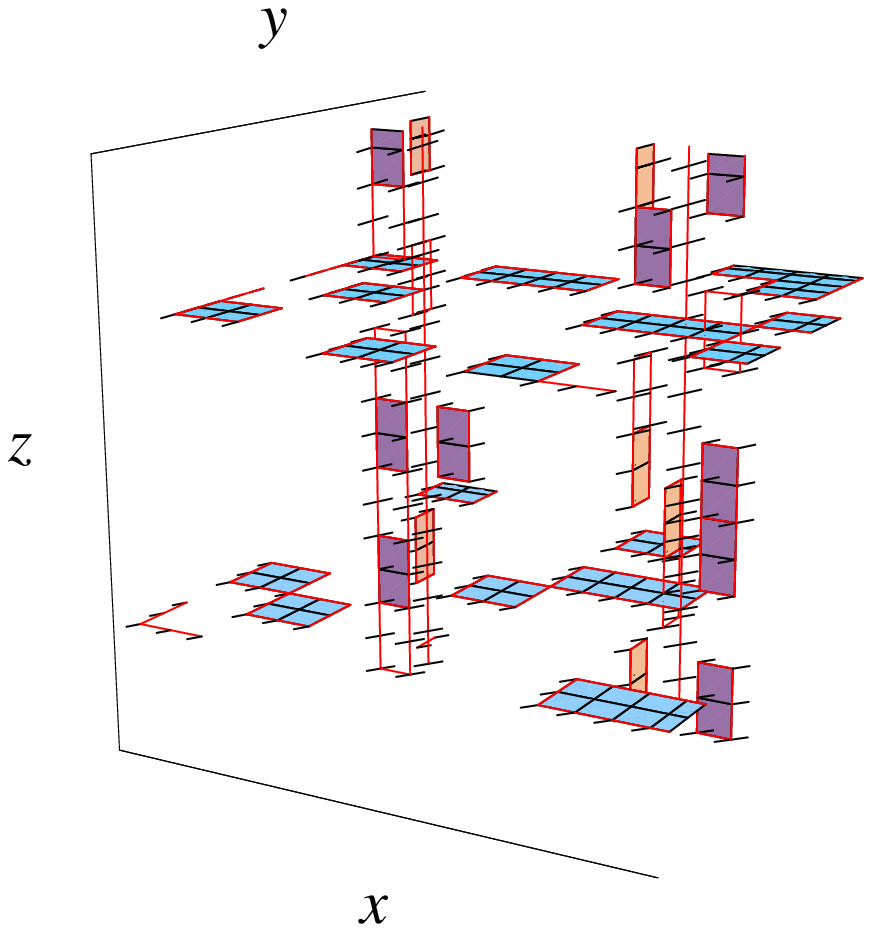}
	\includegraphics[width=.22\linewidth]{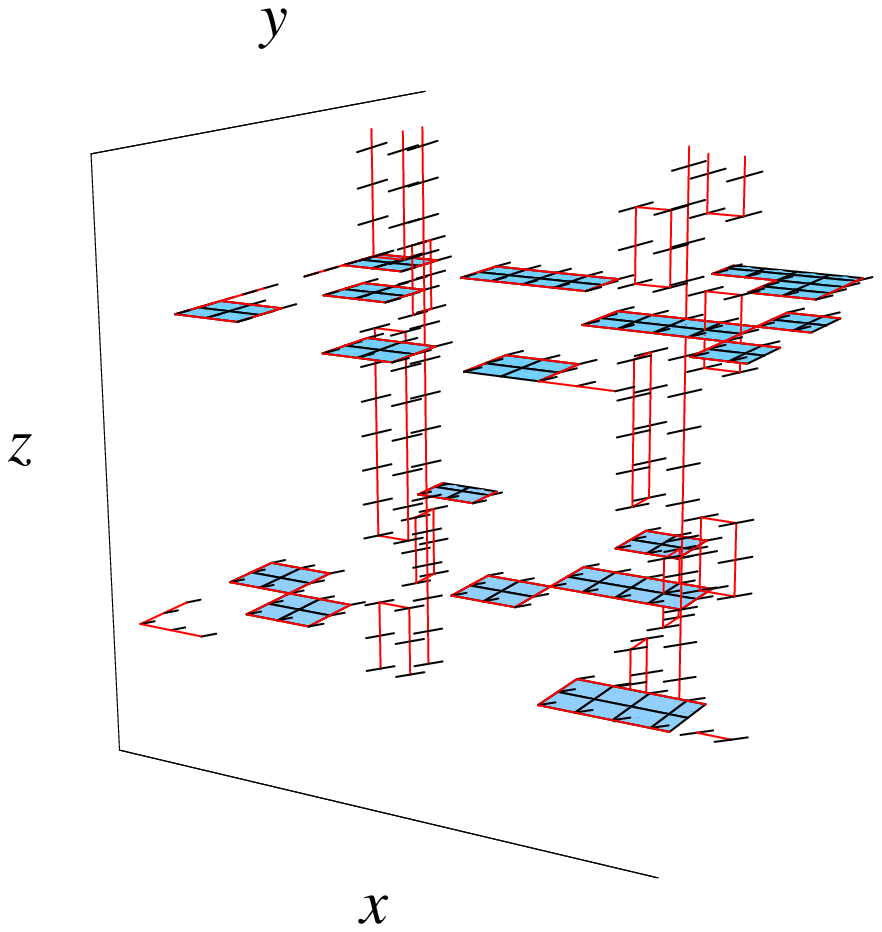}
	\includegraphics[width=.22\linewidth]{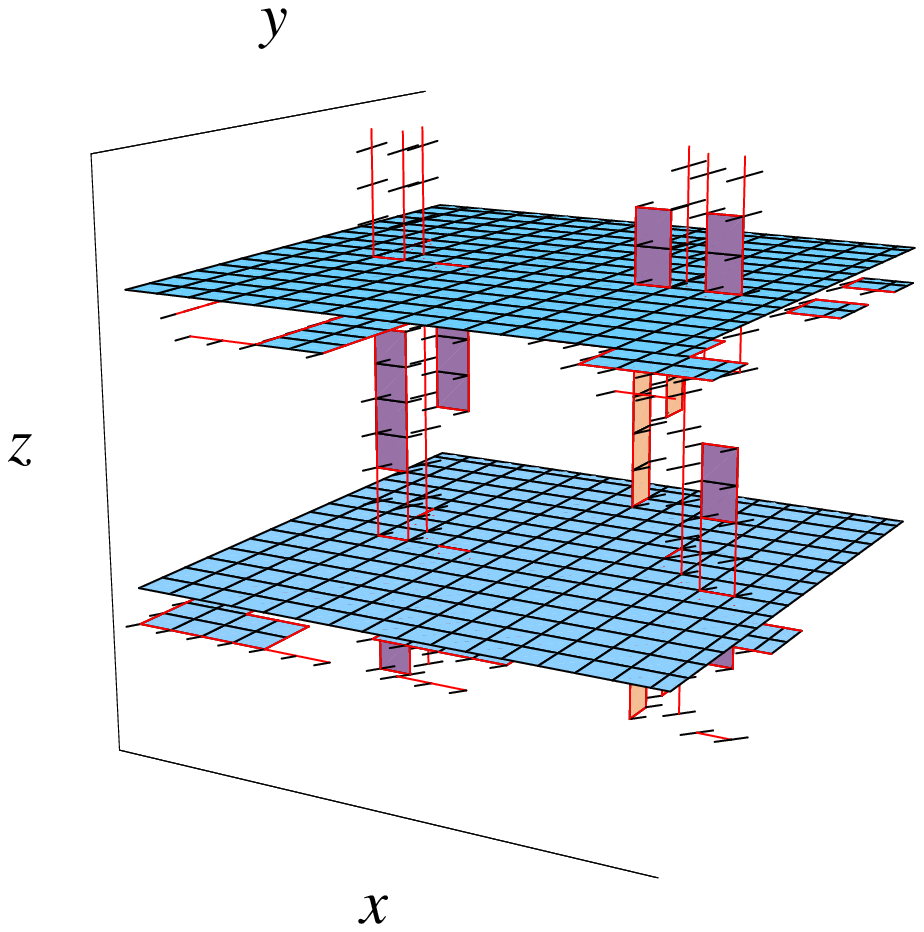}\\
	\includegraphics[width=.22\linewidth]{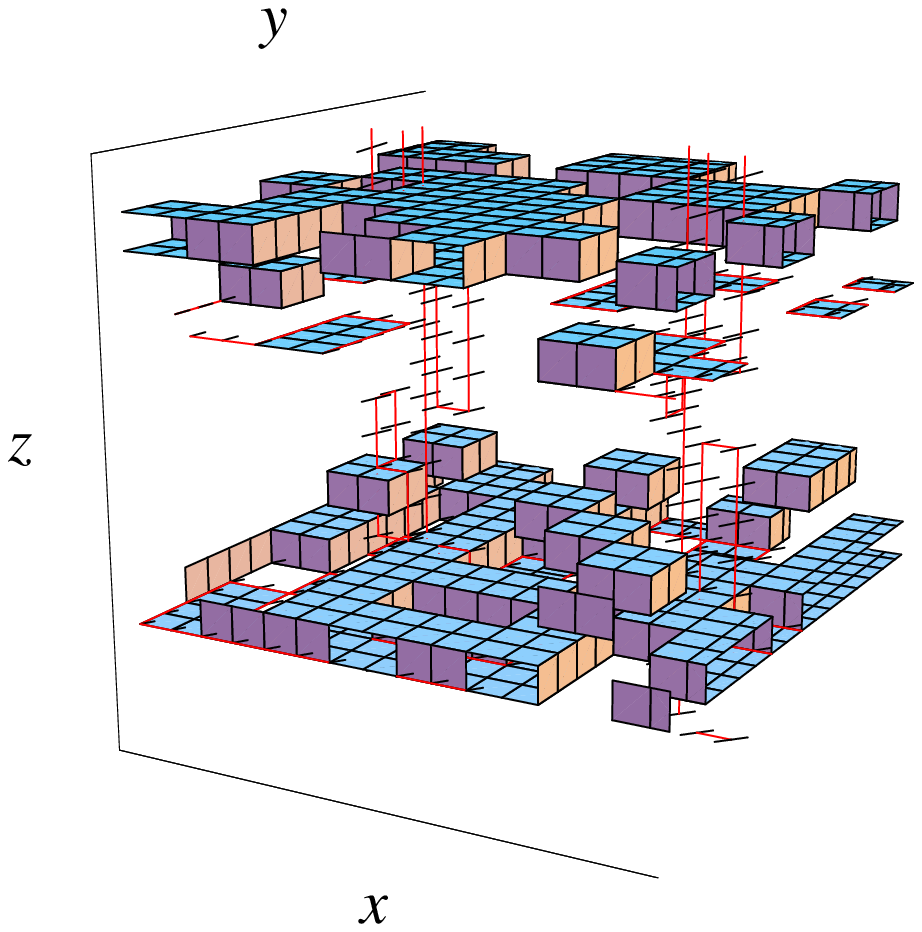}
	\includegraphics[width=.22\linewidth]{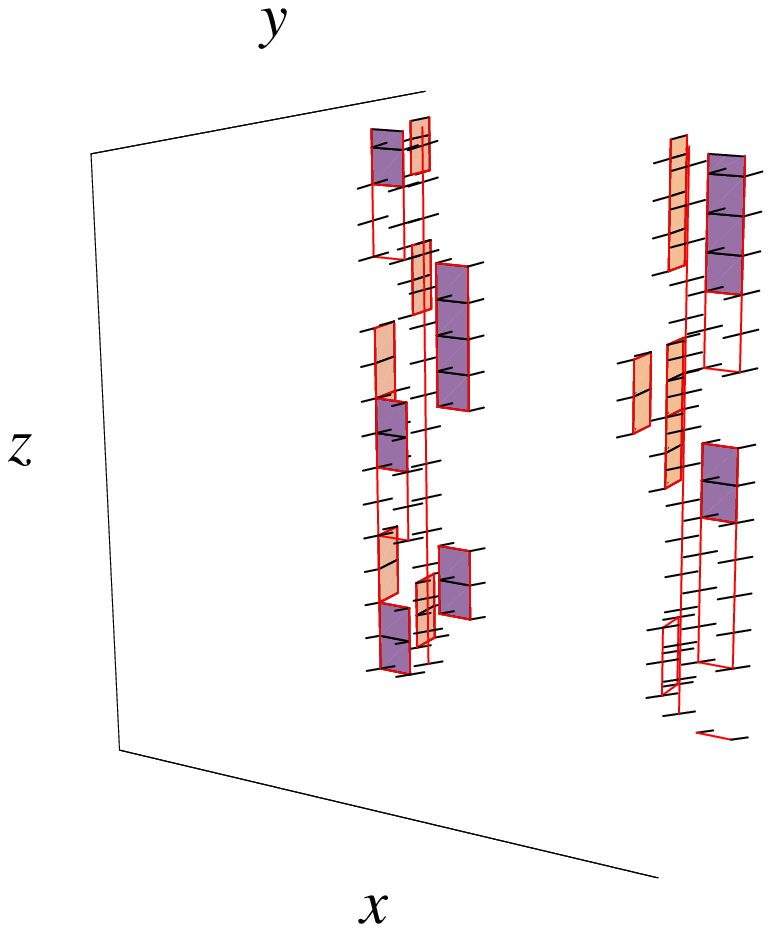}
	\includegraphics[width=.22\linewidth]{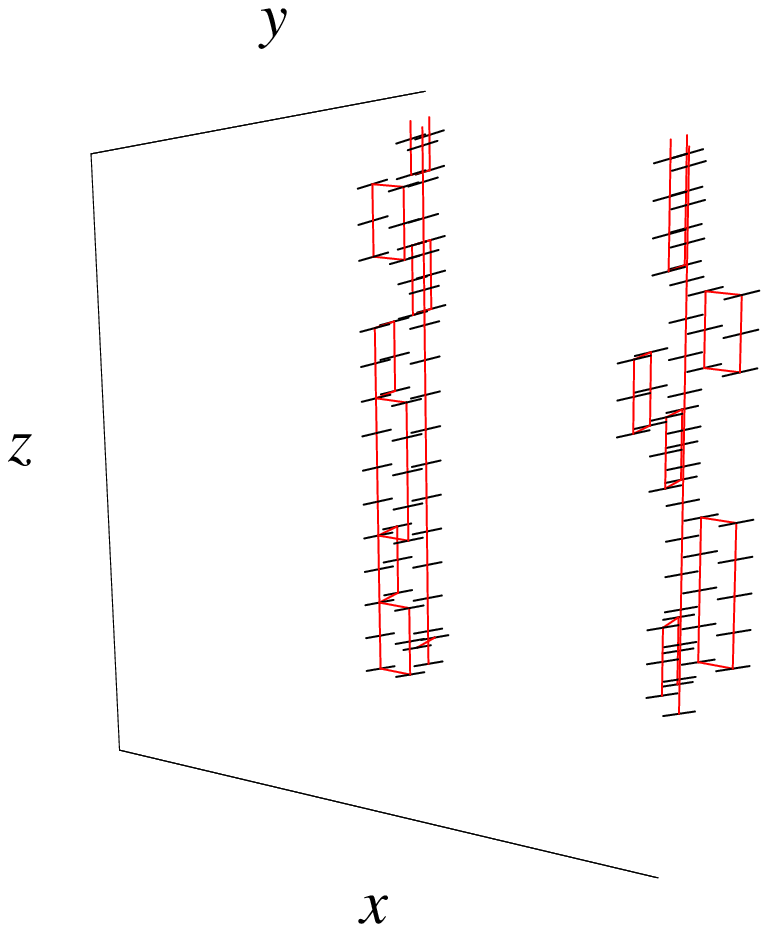}
	\includegraphics[width=.22\linewidth]{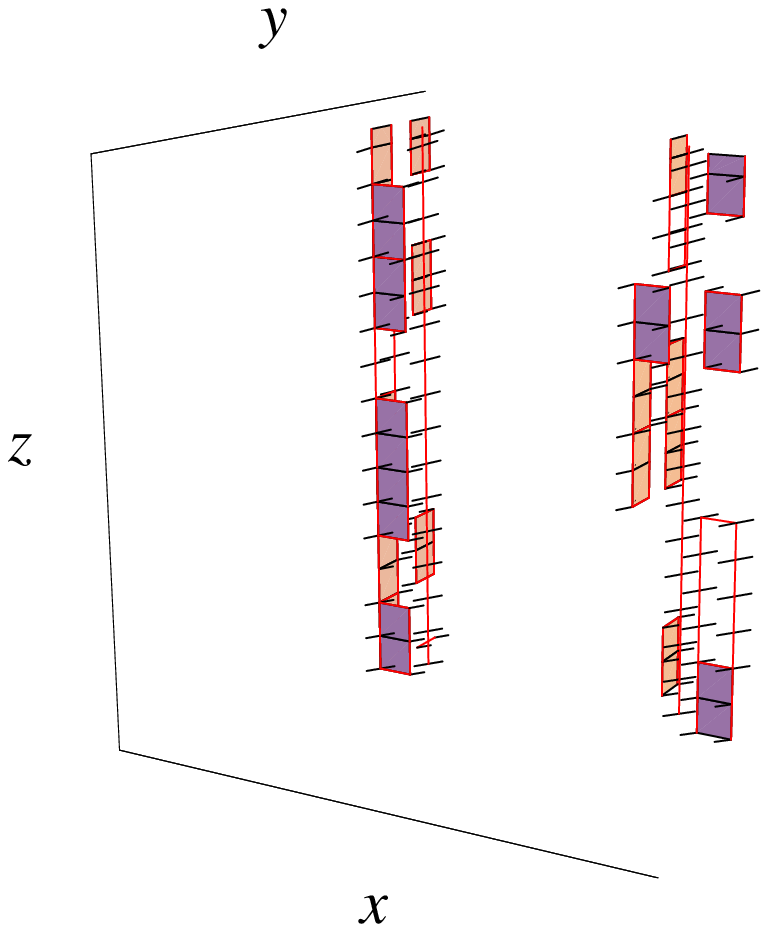}\\
	\includegraphics[width=.22\linewidth]{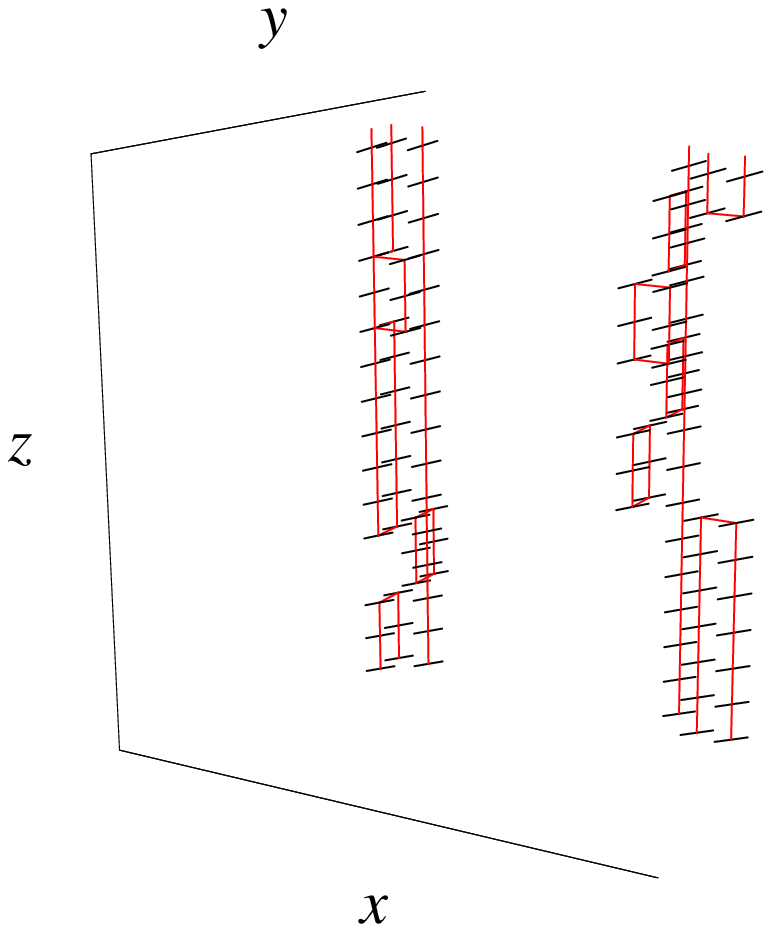}
	\includegraphics[width=.22\linewidth]{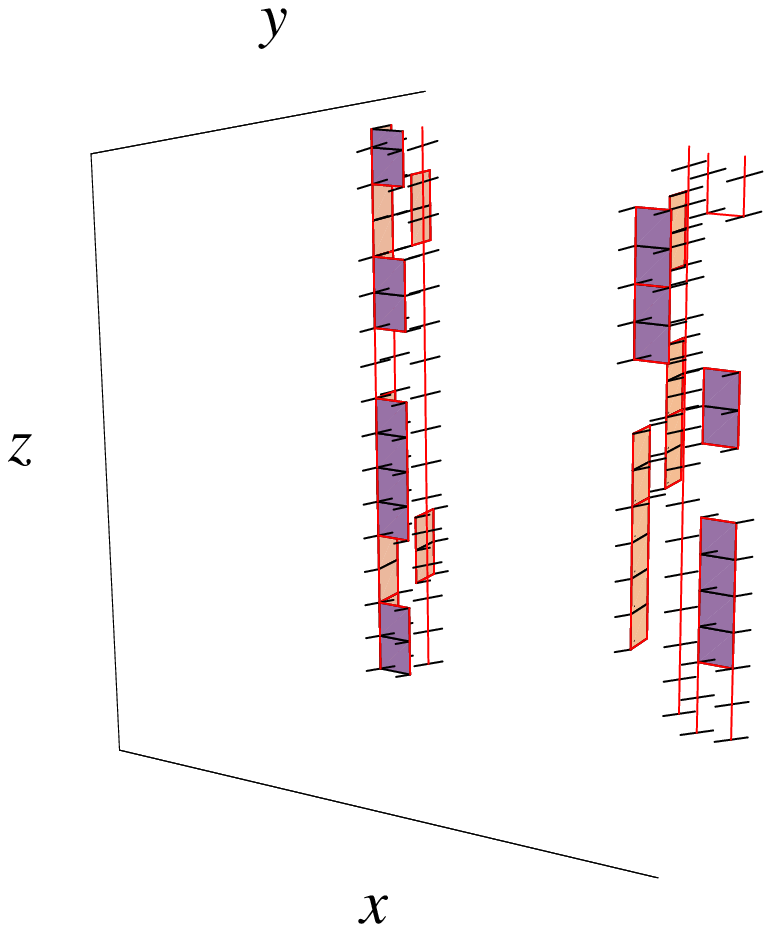}
	\includegraphics[width=.22\linewidth]{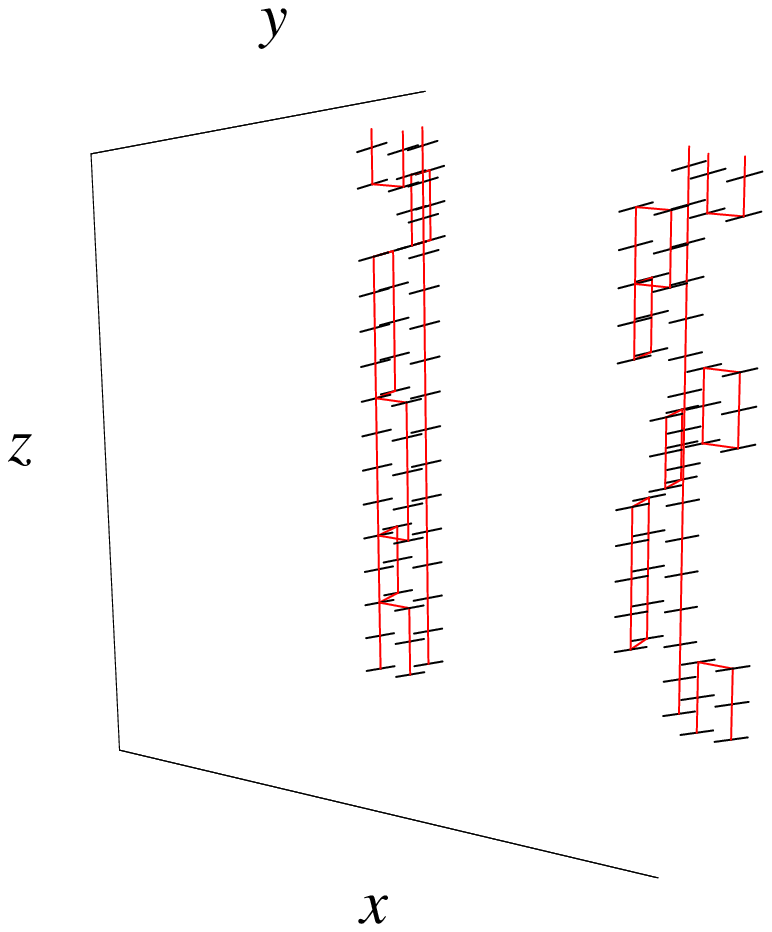}
	\includegraphics[width=.22\linewidth]{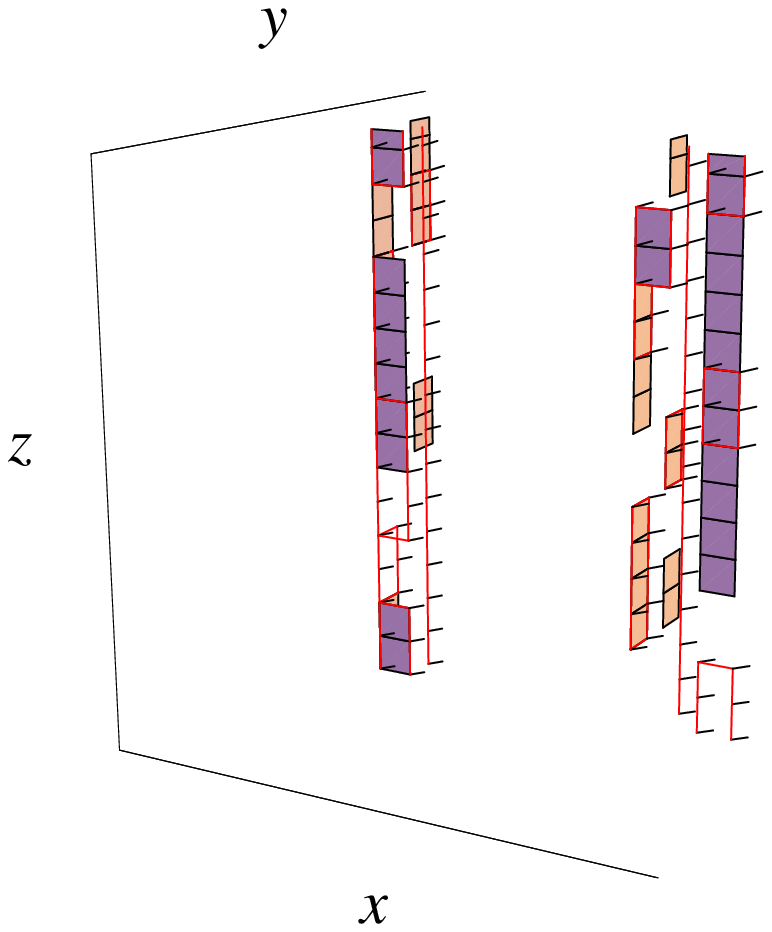}
	\caption{Dual vortex plaquettes of two orthogonal pairs of planar vortices after vortex smearing on $16^4$ lattices from Fig.~\ref{fig:plan}c). From left to right and top
	to bottom we plot the single time slices (t=1-16).}
	\label{fig:plan2}
\end{figure}

\subsection{Spherical Vortex}
\begin{figure}[h]
	\centering
	a)\includegraphics[width=.22\linewidth]{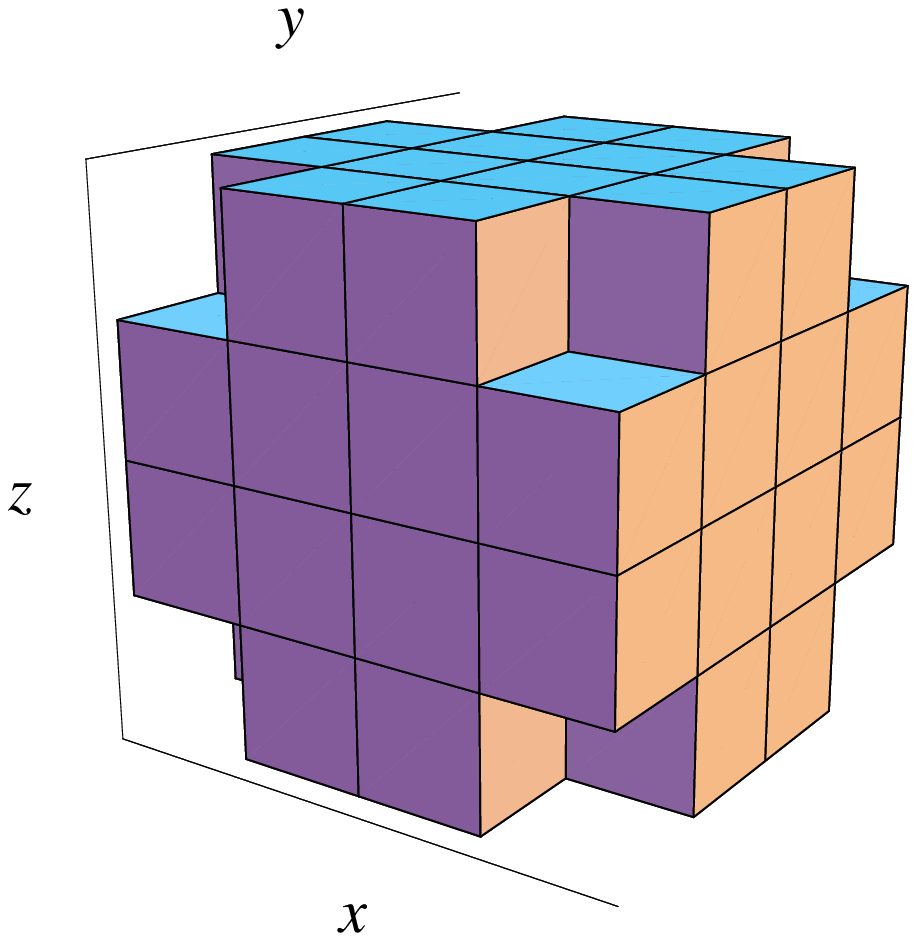}
	b)\includegraphics[width=.22\linewidth]{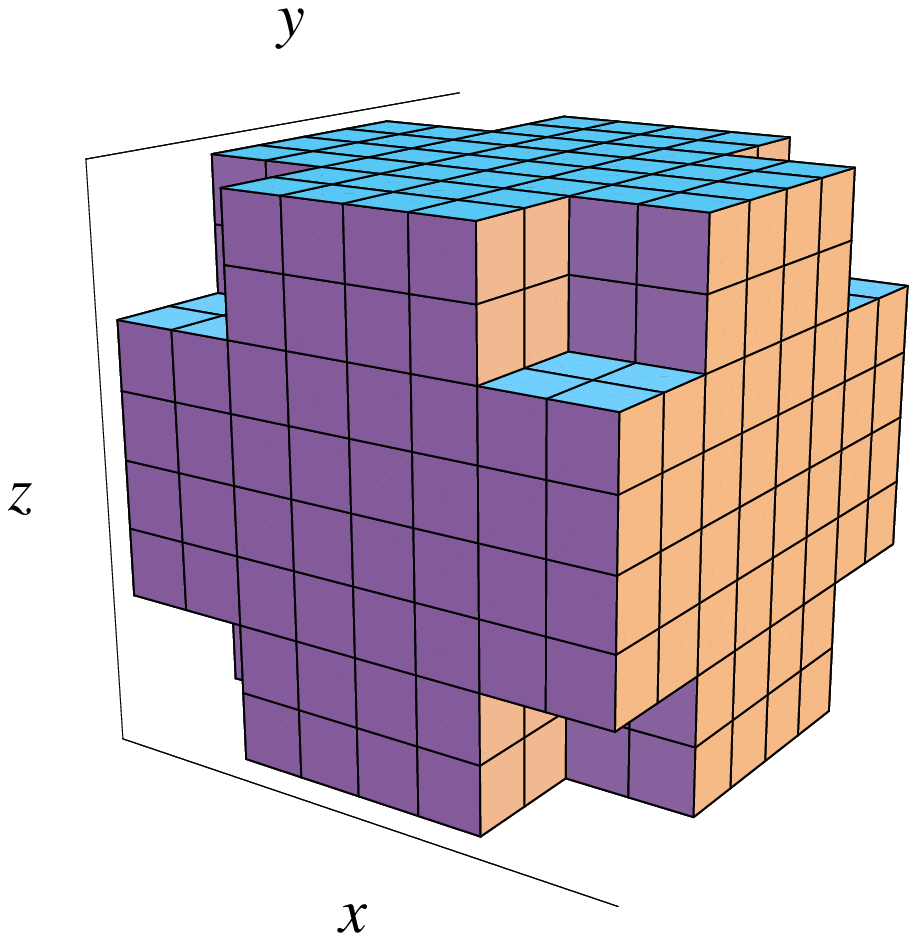}
	c)\includegraphics[width=.22\linewidth]{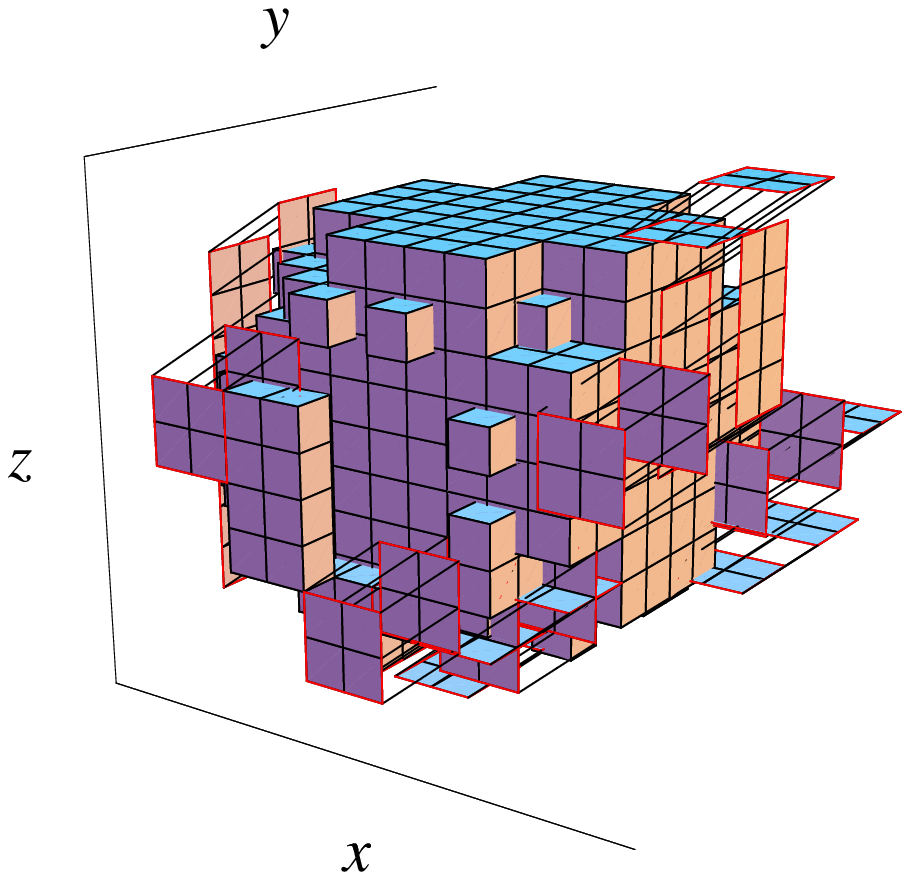}
	d)\includegraphics[width=.22\linewidth]{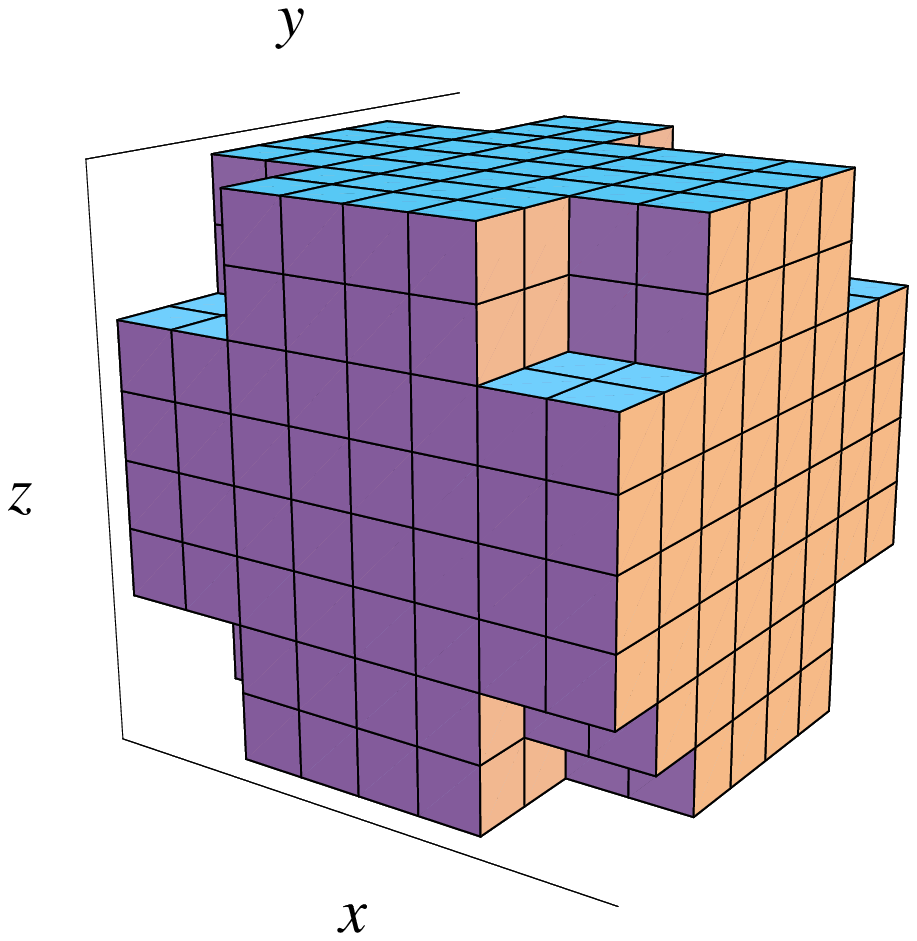}
	\caption{The spherical vortex a) on a $8^4$ lattice and
		after b) refinement, c) vortex smearing and d) smoothing on $16^4$
		lattices. Blocking the configuration in c) (or b or d) gives a). We plot the dual
vortex plaquettes, representing the closed vortex surface.}
	\label{fig:sphr}
\end{figure}

The spherical vortex was introduced in~\cite{Jordan:2007ff} and analyzed in more detail
in~\cite{Hollwieser:2010mj} and~\cite{Schweigler:2012ae}. The thin (Z(2)) vortex surface is given by a three dimensional sphere, which we put in a single time slice. In the thick vortex representation we can define an orientable and a non-orientable spherical
vortex, which are characterized by topological charge $Q=0$ and $Q=\pm1$. The
latter shows a hedgehog like structure of gauge links at the vortex surface, the
3-sphere, defining a map $S^3\to SU(2)$ which is characterized by a winding number
which yields the non-zero topological charge $Q$. Smearing the thin vortex,
without any information on orientation, is very unlikely to reproduce the
hedgehog-like structure and therefore always gives the $Q=0$ case. The vortex structure
itself however, even though distorted during the smearing procedure, is nicely
recovered during smoothing and perfectly after blocking, see
Fig.~\ref{fig:sphr}.

\section{Conclusions}\label{sec:con}
We presented a method to smear $Z(2)$ vortex configurations such as to recover
thick vortices with $SU(2)$ Yang-Mills information. The main goal was to remove the eigenvalue gap observed for overlap
fermions in center projected $Z(2)$ vortex configurations. In order to maintain
the original vortex structure we have to put the $Z(2)$ configurations on finer
lattices, where we implemented two different smearing methods. On the refined
lattice we can distribute the center vortex flux of the (dual) vortex plaquettes
straightforwardly onto the refined plaquettes making up the original vortex
plaquette. On the other hand, we can smear in terms of link variables, applying
a smooth link profile, {\it i.e.}, a "slow" rotation of the links within several
lattice spacings instead of the sudden jump from $+\mathbbm{1}$ to
$-\mathbbm{1}$ or the other way around, which characterizes the vortex surface.
Both approaches thicken the vortices in the sense that the center flux is 
not restricted to a singular surface but spread out over a few lattice spacings.
The refinement procedure applied to the $Z(2)$ configurations, although preserving the
exact vortex structure, causes new obstacles for lattice fermions, especially
for the staggered Dirac operator, which are supposedly related to discretization
effects. For the staggered operator these seem to be caused by its special implementation
with even/odd lattices and are hard to overcome with smearing methods. For the
overlap operator however, we can optimize our smearing routines to close the
eigenvalue gap and reproduce a finite number of actual zero modes. Besides the
refined smearing methods, we also discuss a method to block the smeared lattice
back to its original size. With the various methods we can also reproduce
topological properties of the original gauge fields; on blocked vortex smeared lattices
the different definitions of topological charge, {\it i.e.}, fermionic, gluonic
and vortex topological charge, result in comparable susceptibilities. However,
one-to-one correlations of topological aspects for individual configurations are
not observed. The reason was discussed in detail above and can be
summarized in a simple manner. Gluonic topological charge definitions are
usually applied after cooling or smearing, both transforming Monte Carlo
configurations into smooth gauge fields without center vortex excitations. Thin
center vortex gauge fields, {\it i.e.}, $Z(2)$ configurations, on the other hand
lack any information of the orientation of thick center vortices, which is
crucial for topological charge determination. During vortex smearing or vortex
topological charge determination we apply random orientations to the vortex
sheets and cannot expect to reproduce the original topological charge. However,
earlier results and the analysis here show that vortex gauge fields reproduce
the net topological charge and susceptibility via vortex topological charge
definition and via fermionic or gluonic definitions after the introduced vortex
smearing methods. The vortex smeared configurations, besides preserving the
original vortex structure, also reproduce the asymptotic string tension of the
original gauge field ensemble, which is the basis of the confinement mechanism by center
vortices. It should be stressed that our method is not intended to reproduce full
Yang-Mills dynamics on arbitrarily short length scales, but rather to
encode the infrared dynamics consistently in fields which only vary
appreciably over lengths commensurate with an infrared effective
picture. This consistency is, strictly speaking, not manifest as
long as one remains in a thin P-vortex framework. In the process,
properties are seen to be recovered which are not accessible using
pure $Z(2)$ configurations. In accordance with this, it should be
remarked that, although we primarily have not analyzed the scaling
behavior of our smeared ensembles due to the considerable numerical
effort associated with the refined lattices, we do not envisage
reproducing a particular scaling law with the smeared degrees of
freedom. Rather, the center vortex picture as an infrared
effective model has a fixed scale given by the vortex thickness, which acts as
an ultraviolet cutoff and has a direct relation to $\Lambda_{QCD}$. Starting from
(thin) $Z(2)$ vortices our smearing method tries to regain the finite vortex
thickness, which then sets the initial scale and allows us to extract observables within
these infrared effective degrees of freedom. Going forward in that sense, the
plan is to use the developed tools to analyze topological and fermionic
properties of the $SU(2)$ effective center vortex
model~\cite{Engelhardt:1999wr}. Further, the methods shall be advanced to the
$SU(3)$ gauge group and applied to the corresponding vortex
model~\cite{Engelhardt:2003wm}.

\acknowledgments{We thank Manfried Faber and {\v S}tefan Olejn\'{\i}k for helpful discussions. We also thank Urs M. Heller for his help with various fermion program issues. The numerical simulations were performed at the Phoenix and Vienna Scientific Cluster (VSC) at VUT and the Riddler Cluster at NMSU. This research was supported by the Erwin Schr\"odinger Fellowship program of the Austrian Science Fund FWF (``Fonds zur F\"orderung der wissenschaftlichen Forschung'') under Contract No. J3425-N27 (R.H.) and the
U.S. DOE through the grant DE-FG02-96ER40965 (M.E.).}

\appendix
\begin{center}
	{\bf Appendix A: Overlap Fermion Mode Distributions}
\end{center}

\begin{figure}[h]
	\centering
	a)\includegraphics[width=.48\linewidth]{scovl}
	b)\includegraphics[width=.48\linewidth]{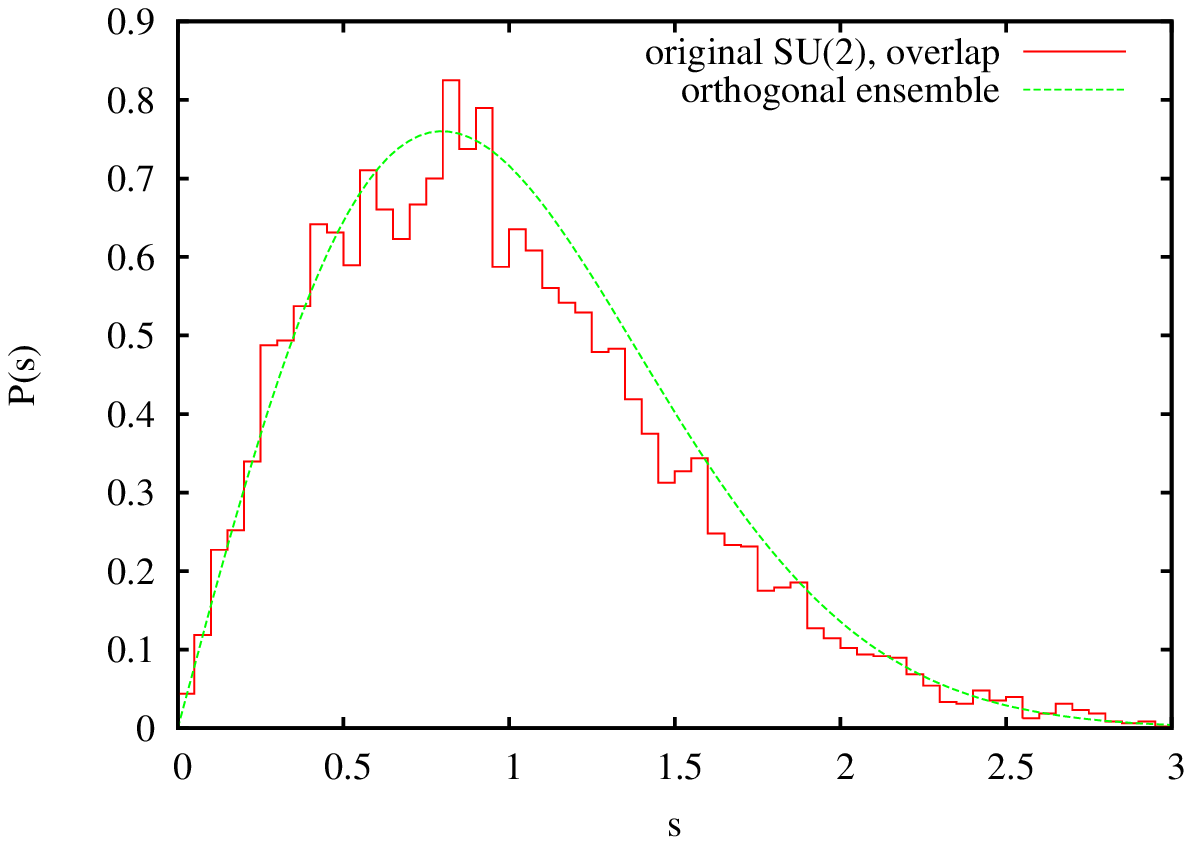}
	c)\includegraphics[width=.48\linewidth]{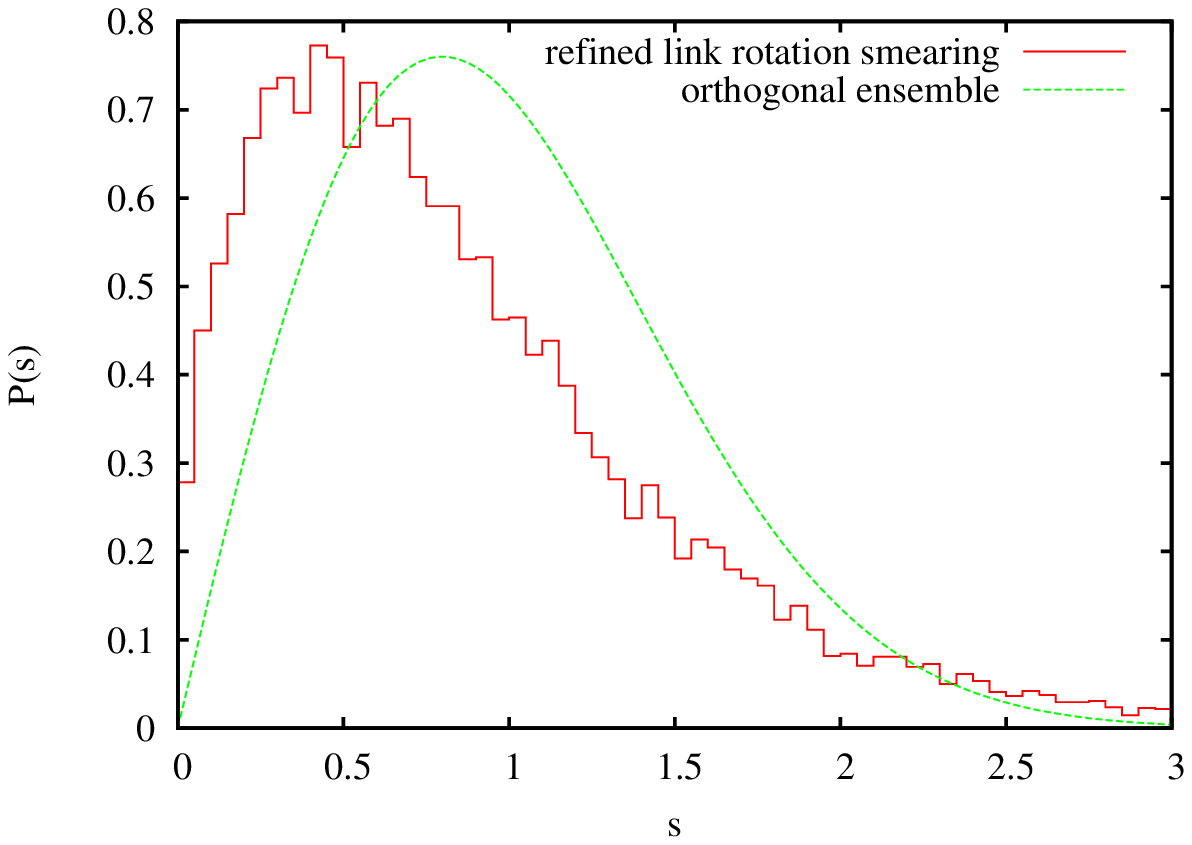}
	d)\includegraphics[width=.48\linewidth]{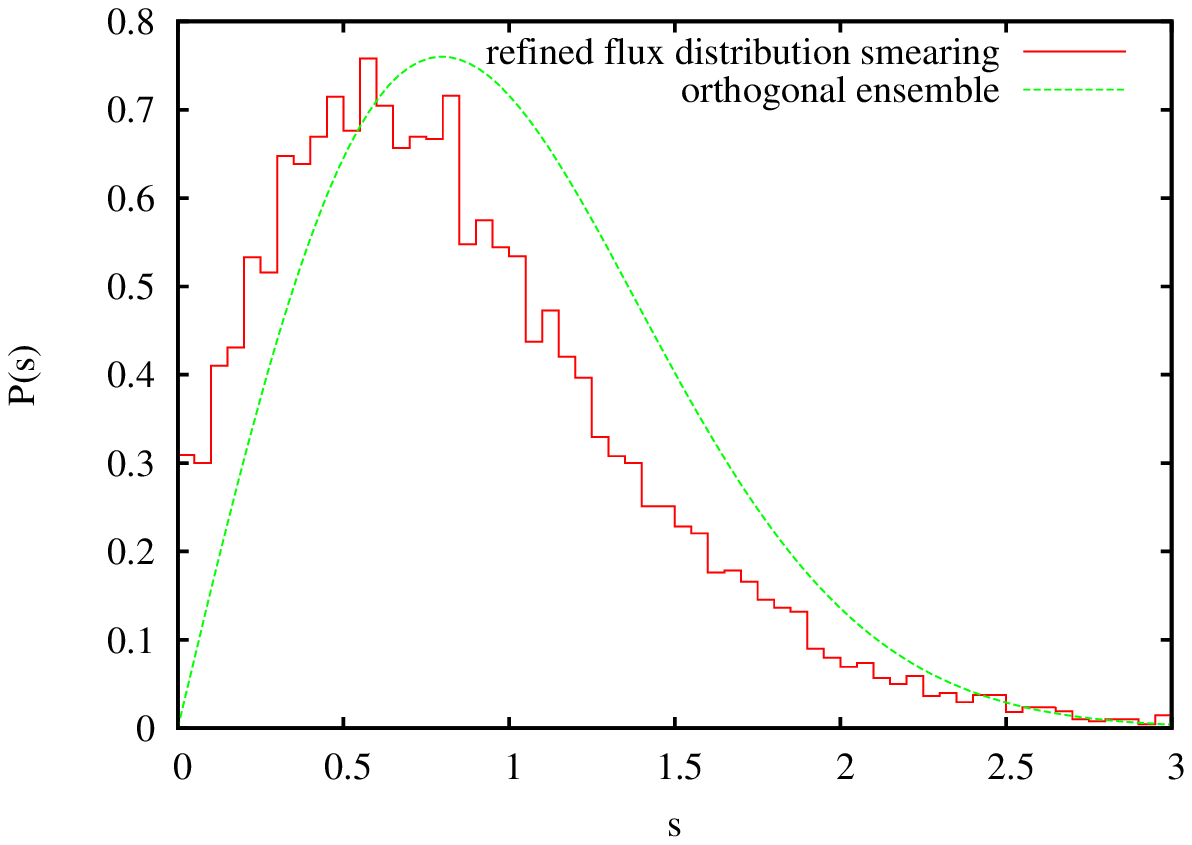}
	e)\includegraphics[width=.48\linewidth]{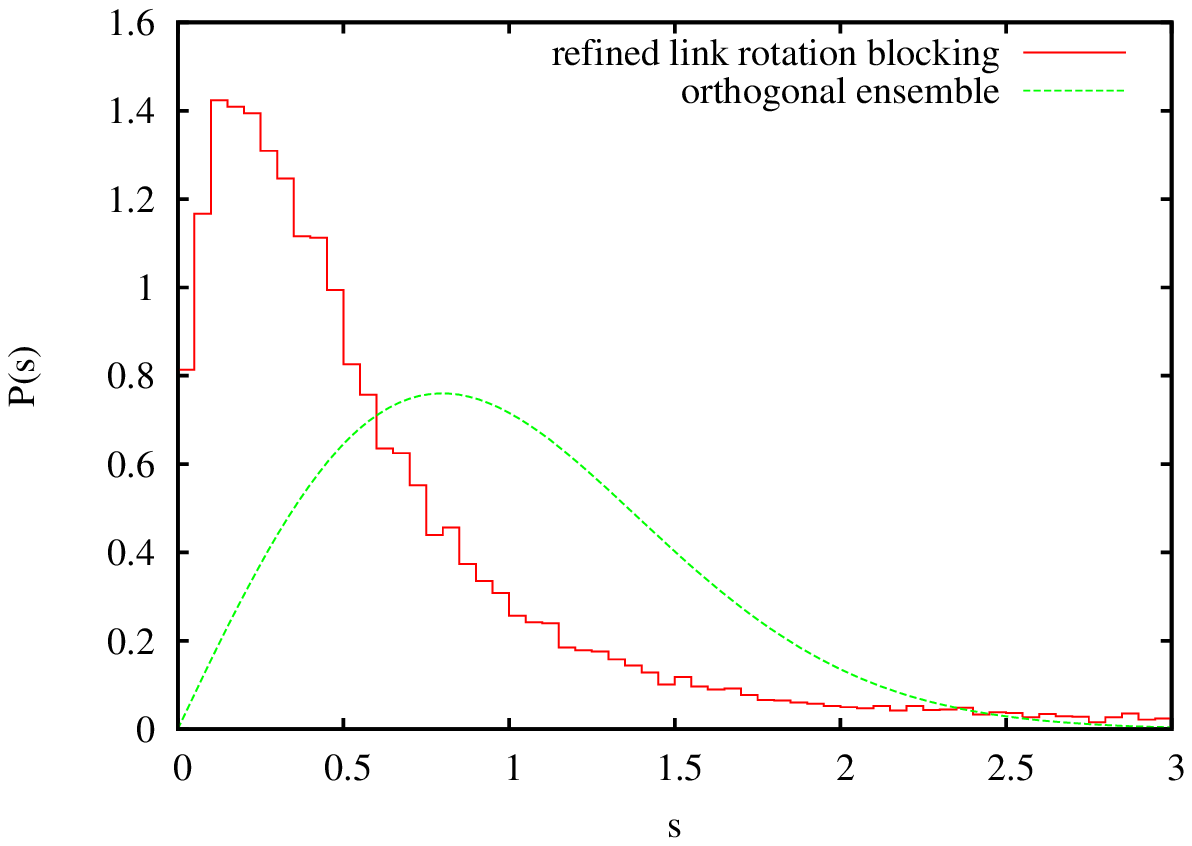}
	f)\includegraphics[width=.48\linewidth]{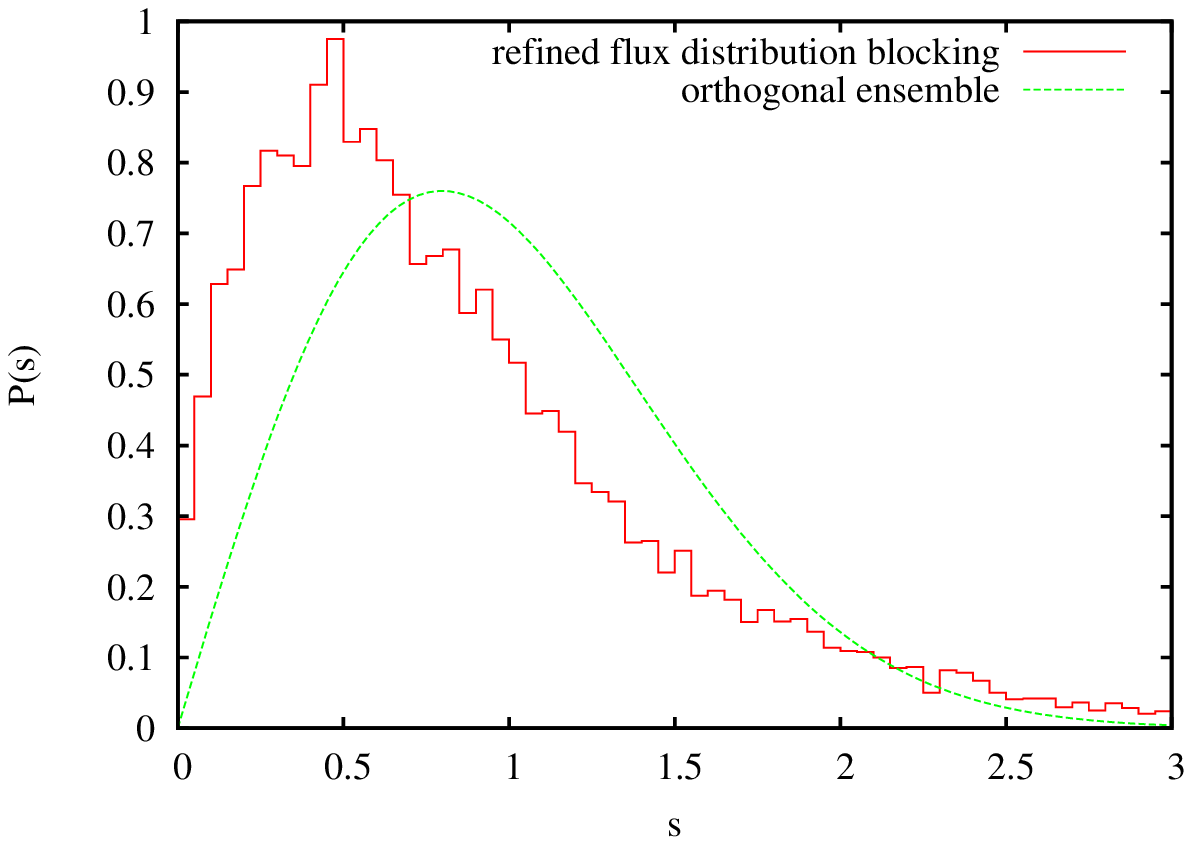}
	\caption{"Unfolded" level spacing of overlap eigenvalues in the fundamental
		representation of SU(2) (orthogonal ensemble) for
		a) all, b) original and c-d) smeared configs: c) link rotation smearing, d) flux
distribution smearing and their blocked versions in e) and f), respectively.}
	\label{fig:zova}
\end{figure}

\begin{figure}[p]
	\centering
	a)\includegraphics[width=.48\linewidth]{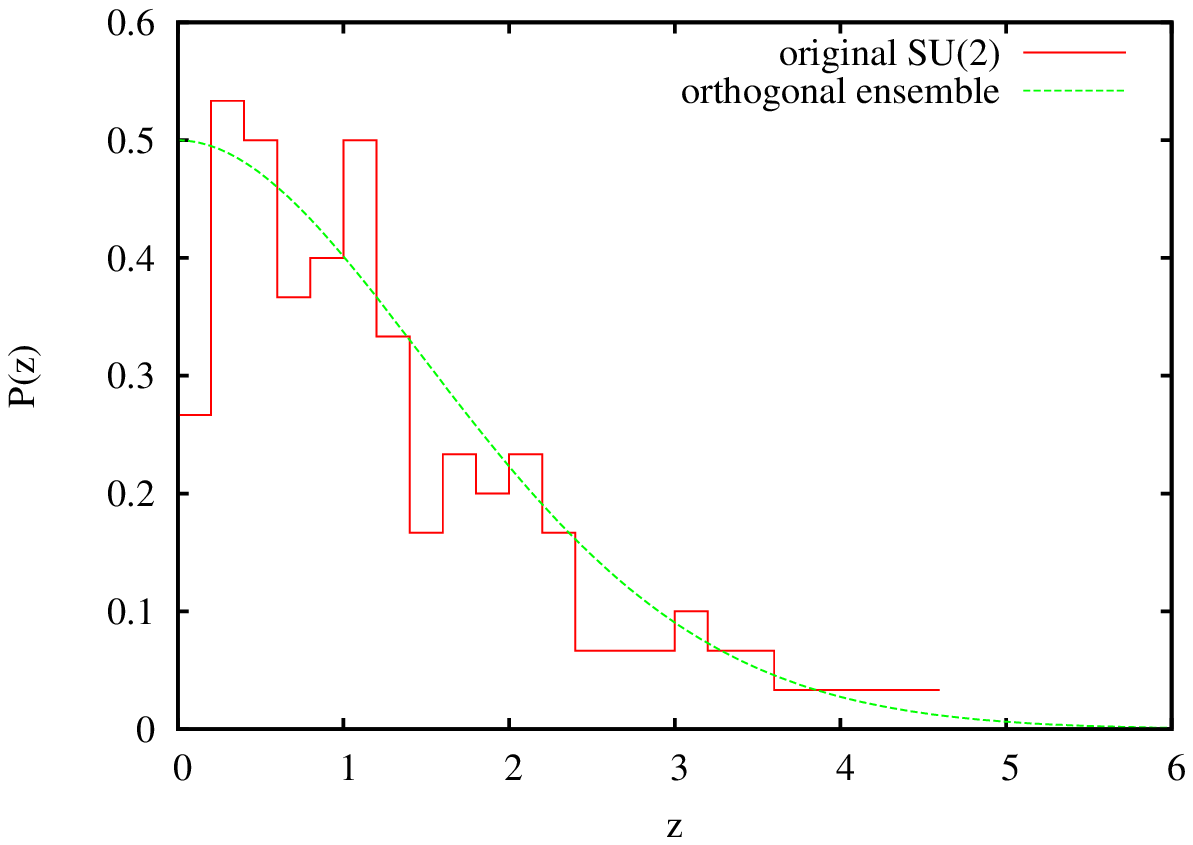}
	b)\includegraphics[width=.48\linewidth]{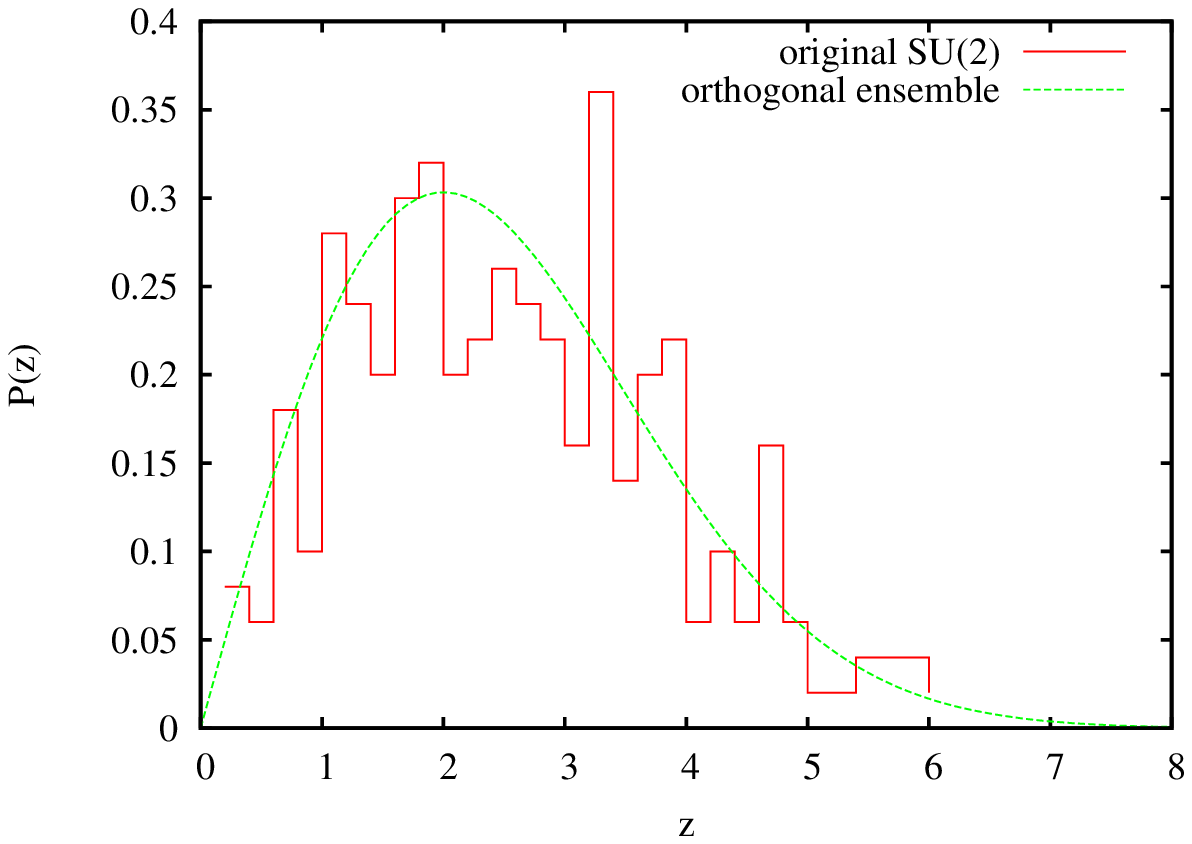}
	c)\includegraphics[width=.48\linewidth]{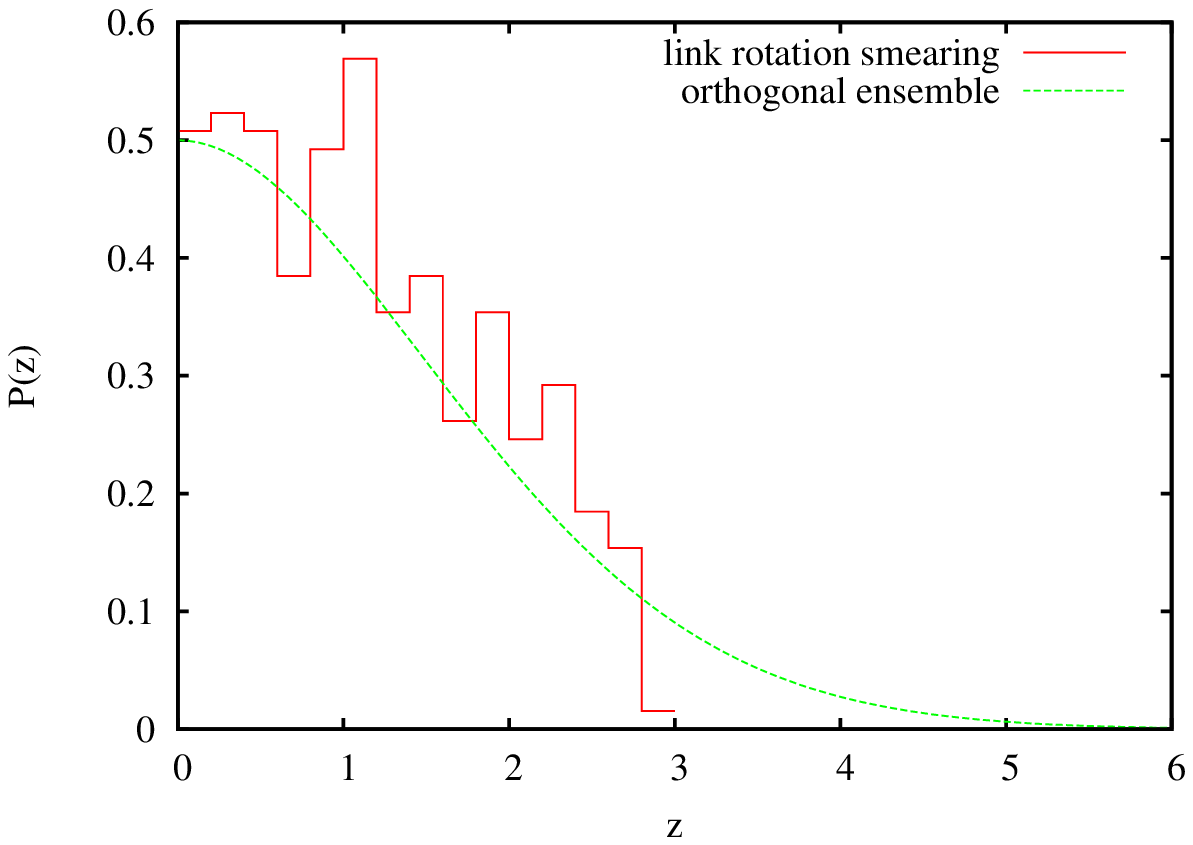}
	d)\includegraphics[width=.48\linewidth]{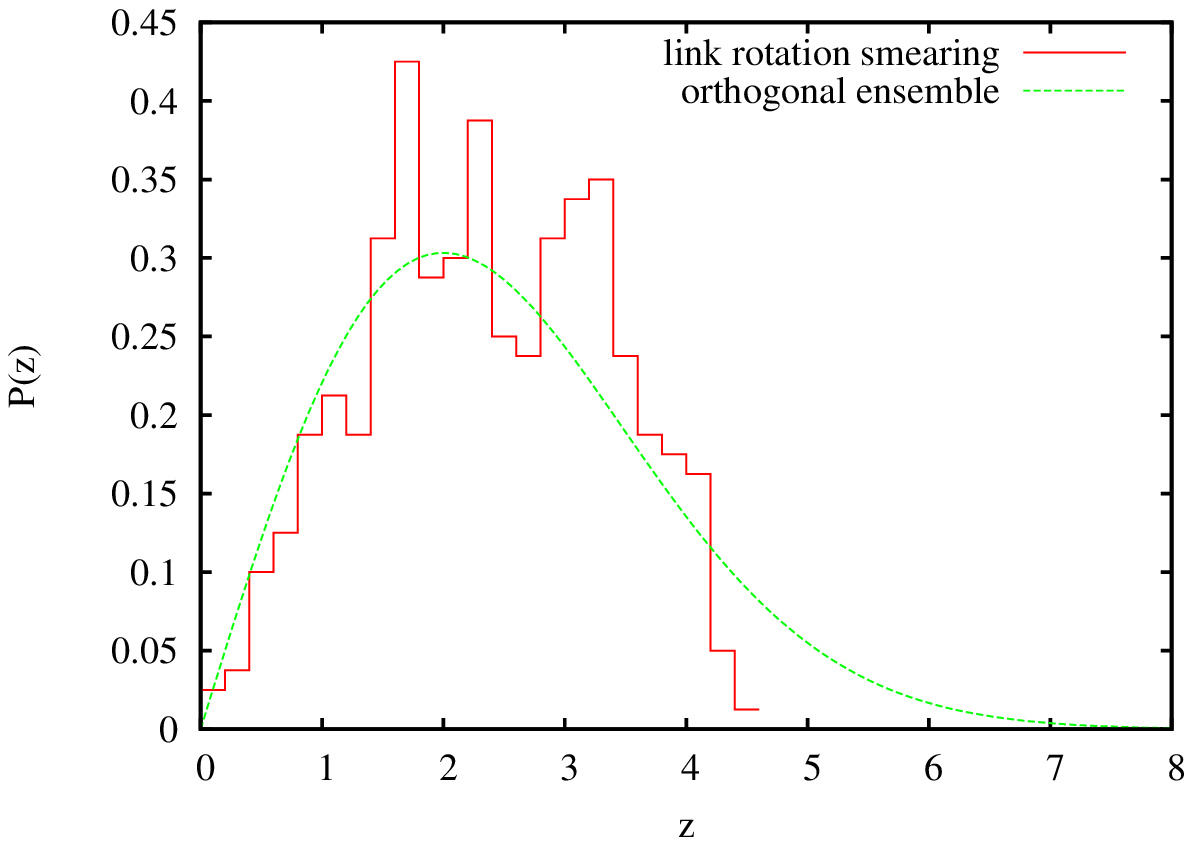}
	e)\includegraphics[width=.48\linewidth]{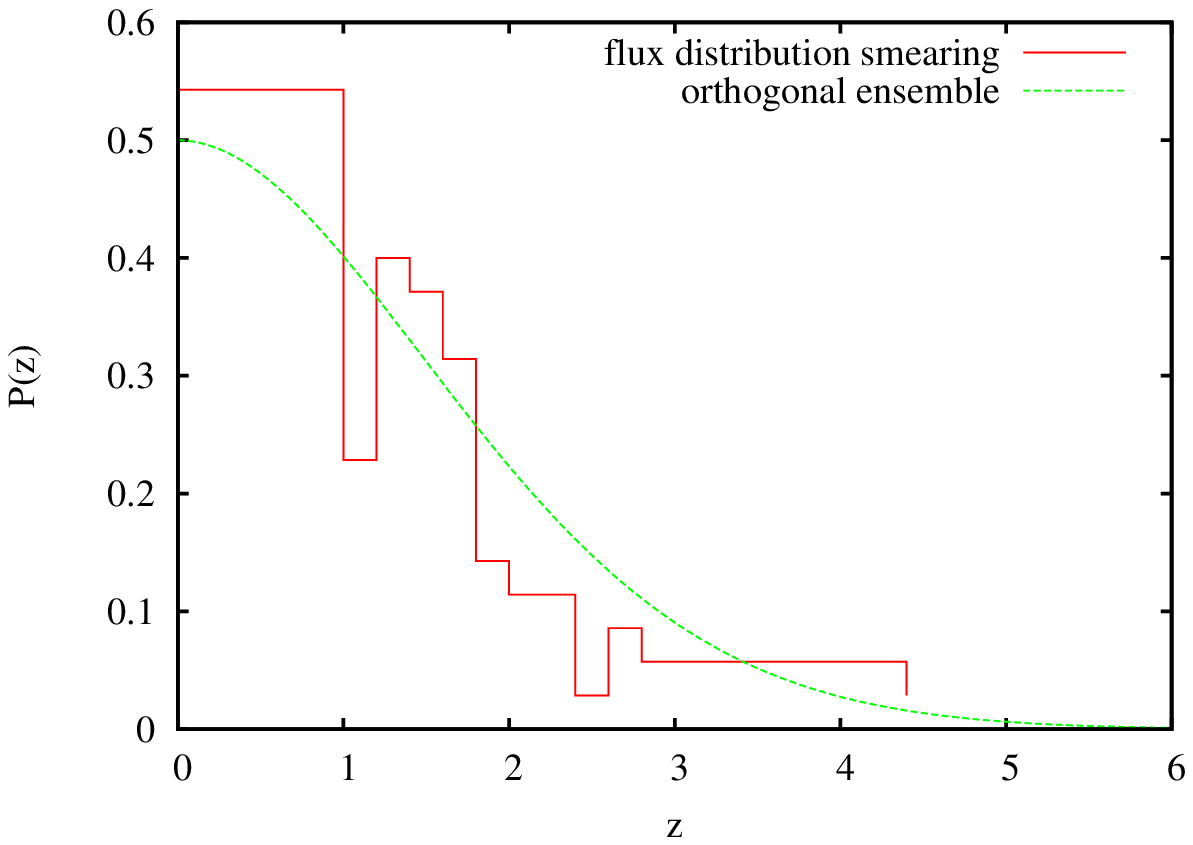}
	f)\includegraphics[width=.48\linewidth]{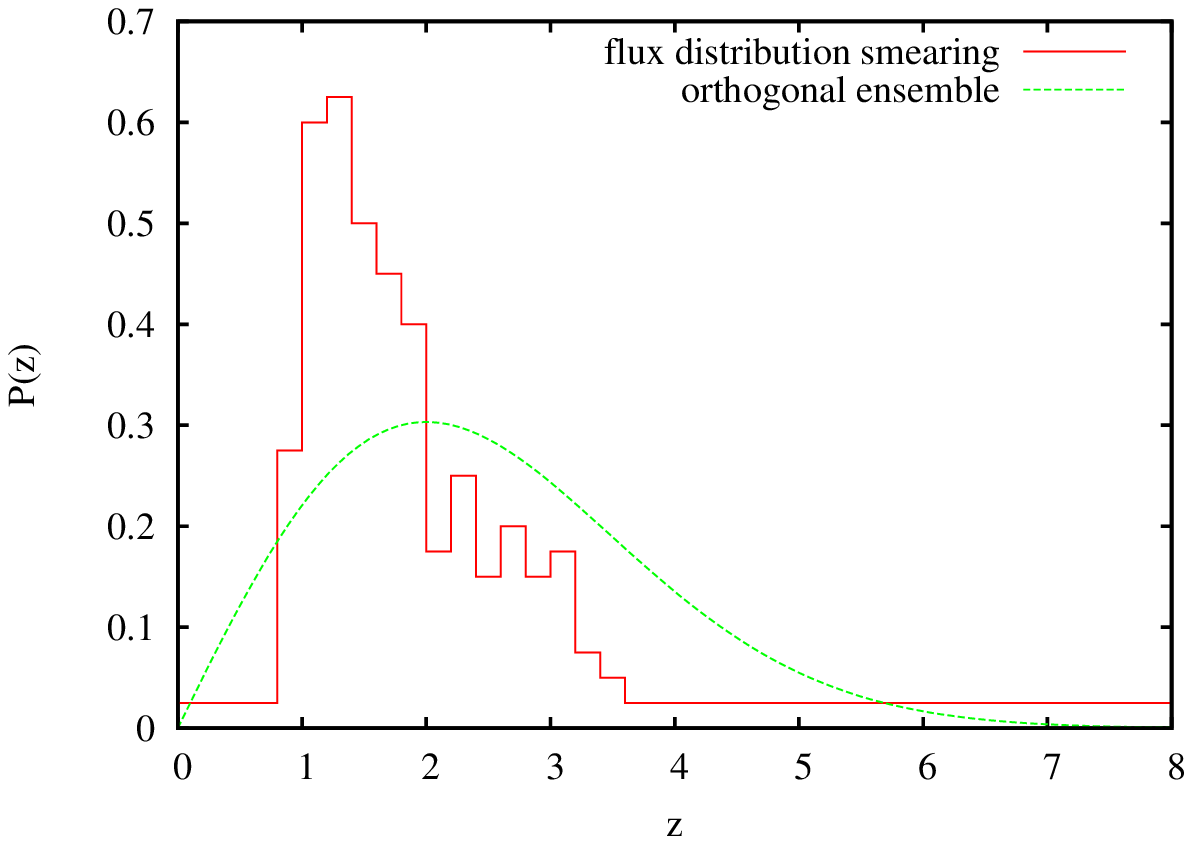}
	\caption{Distribution of lowest overlap eigenvalues in the fundamental
		representation of SU(2) (orthogonal ensemble) for a-b) original, c-d)
	link rotation and e-f) flux distribution smearing in topological sectors
$\nu=0$ (left) and $\nu=1$ (right).}
\label{fig:zovc}
\end{figure}

\begin{figure}[p]
	\centering
	a)\includegraphics[width=.48\linewidth]{z0ovl}
	b)\includegraphics[width=.48\linewidth]{z1ovl}
	c)\includegraphics[width=.48\linewidth]{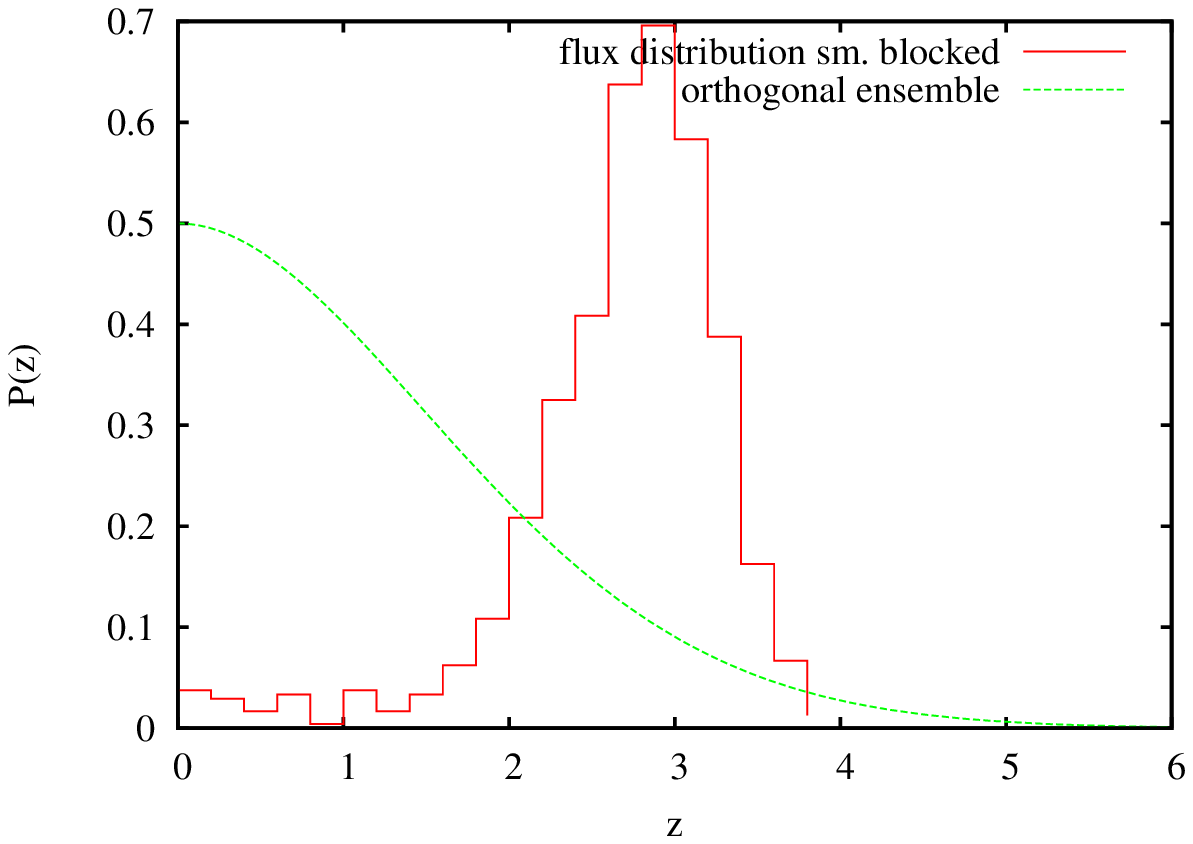}
	d)\includegraphics[width=.48\linewidth]{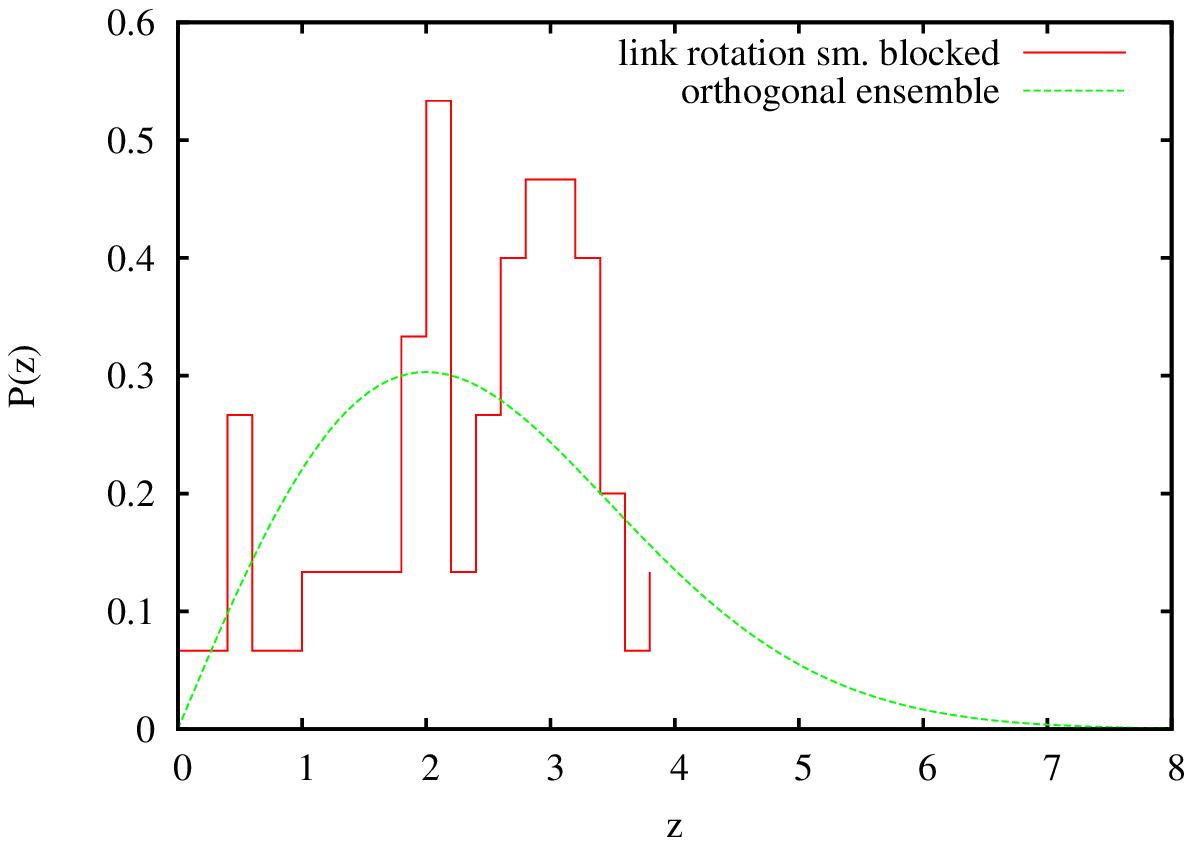}
	e)\includegraphics[width=.48\linewidth]{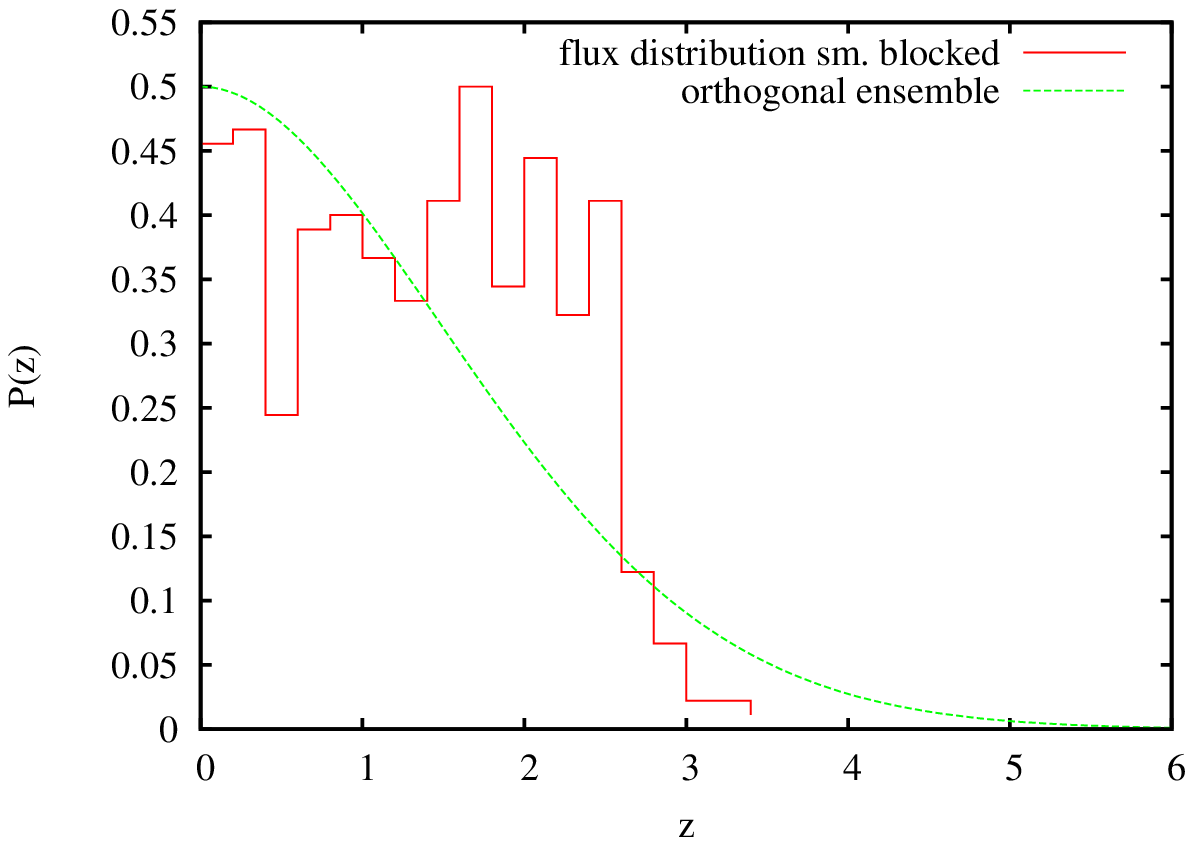}
	f)\includegraphics[width=.48\linewidth]{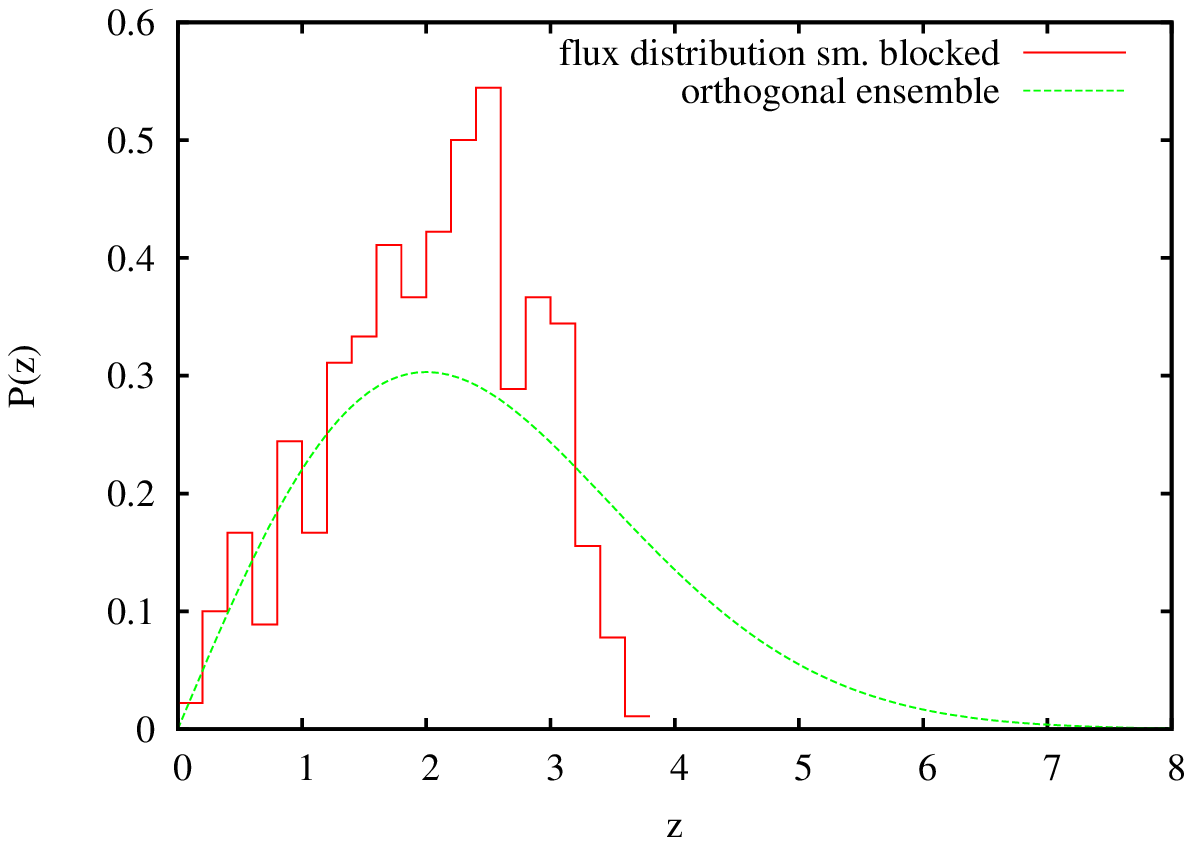}
	\caption{Distribution of lowest overlap eigenvalues in the fundamental
		representation of SU(2) (orthogonal ensemble) for a-b) combined
		configurations, c-d) link rotation and e-f) flux distribution smeared blocking in topological sectors $\nu=0$ (left) and $\nu=1$ (right).}
	\label{fig:zovcc}
\end{figure}

\clearpage
\newpage

\begin{center}
	{\bf Appendix B: Staggered Fermion Mode Distributions}
\end{center}

\begin{figure}[h]
	\centering
	a)\includegraphics[width=.48\linewidth]{sstag}
	b)\includegraphics[width=.48\linewidth]{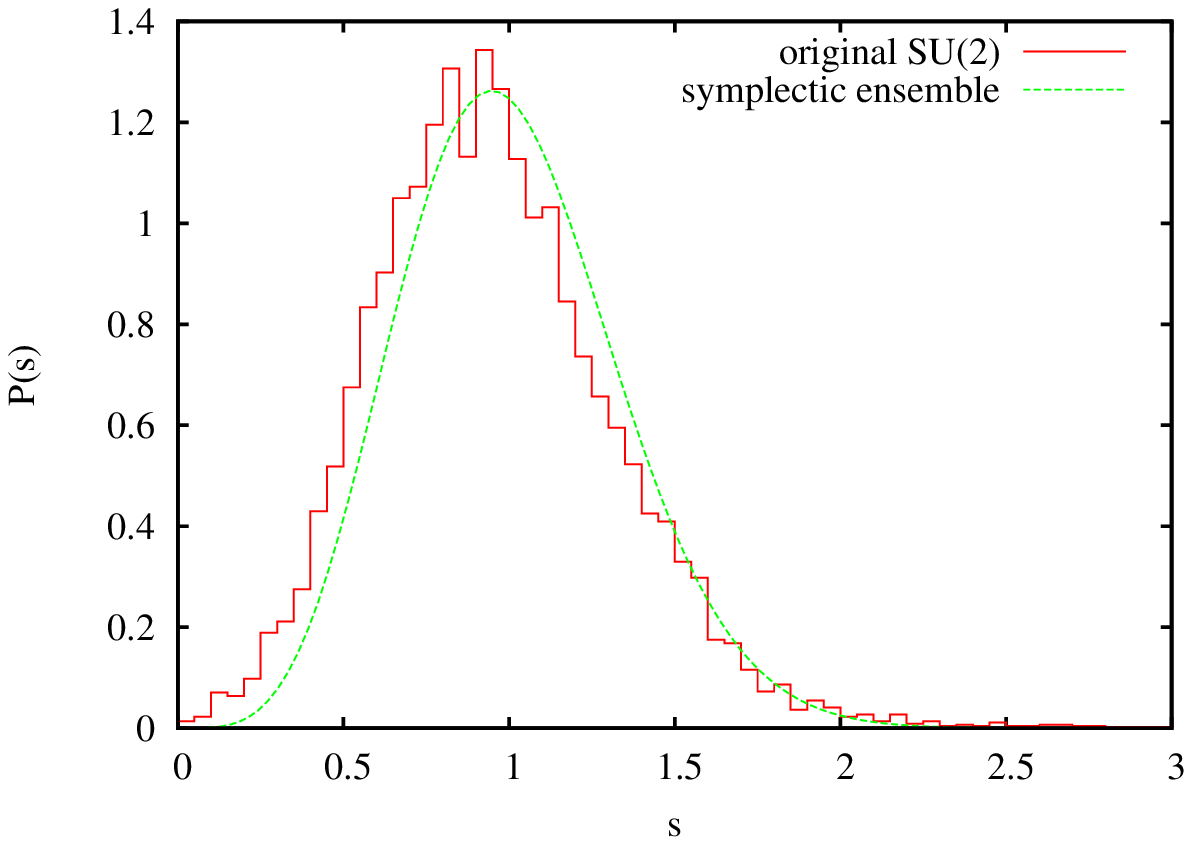}
	c)\includegraphics[width=.48\linewidth]{svsmr}
	d)\includegraphics[width=.48\linewidth]{sfsmr}
	e)\includegraphics[width=.48\linewidth]{svsmrblq}
	f)\includegraphics[width=.48\linewidth]{sfsmrblq}
	\caption{"Unfolded" level spacing of staggered eigenvalues in the fundamental
		representation of SU(2) (symplectic ensemble) for a) all, b) original
	and c-f) smeared configurations: c) link rotation smearing, d) flux
distribution smearing and their blocked versions in e) and f), respectively.}
	\label{fig:zovb}
\end{figure}

\begin{figure}[p]
	\centering
	a)\includegraphics[width=.48\linewidth]{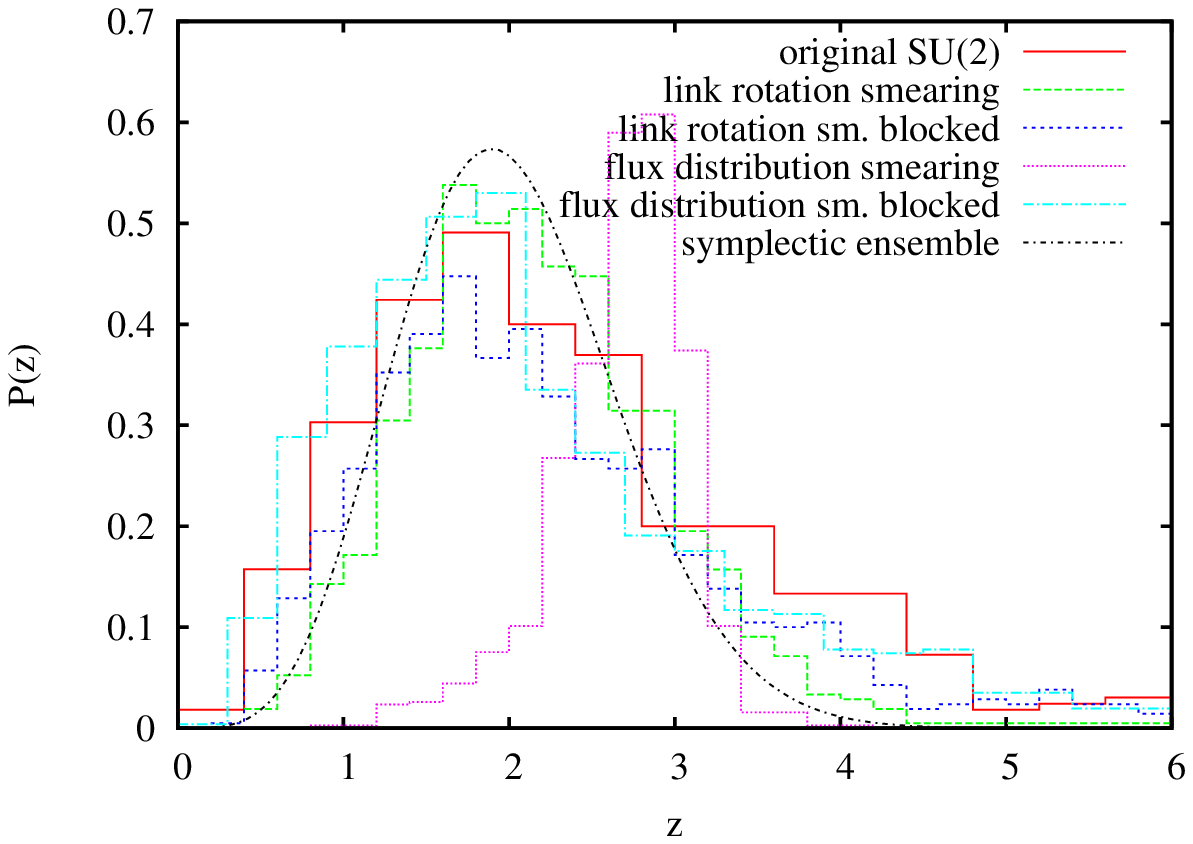}
	b)\includegraphics[width=.48\linewidth]{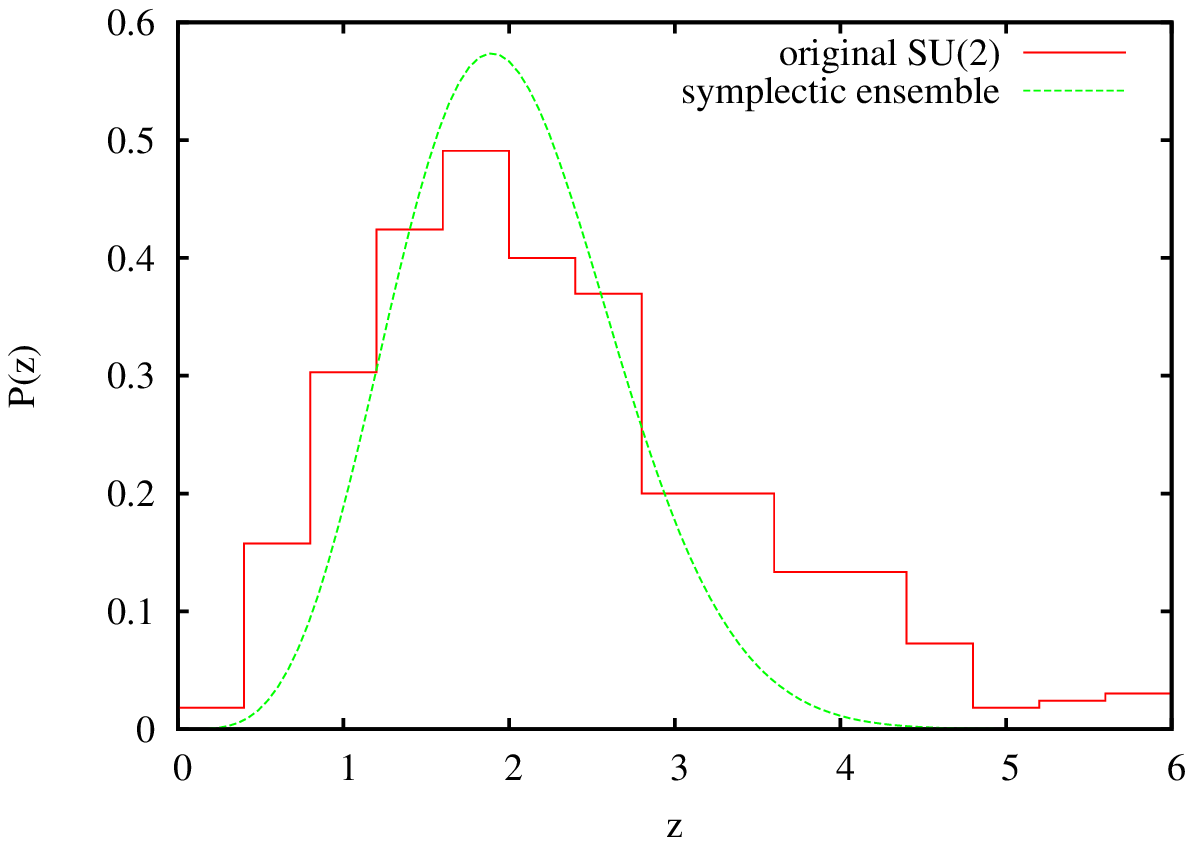}
	c)\includegraphics[width=.48\linewidth]{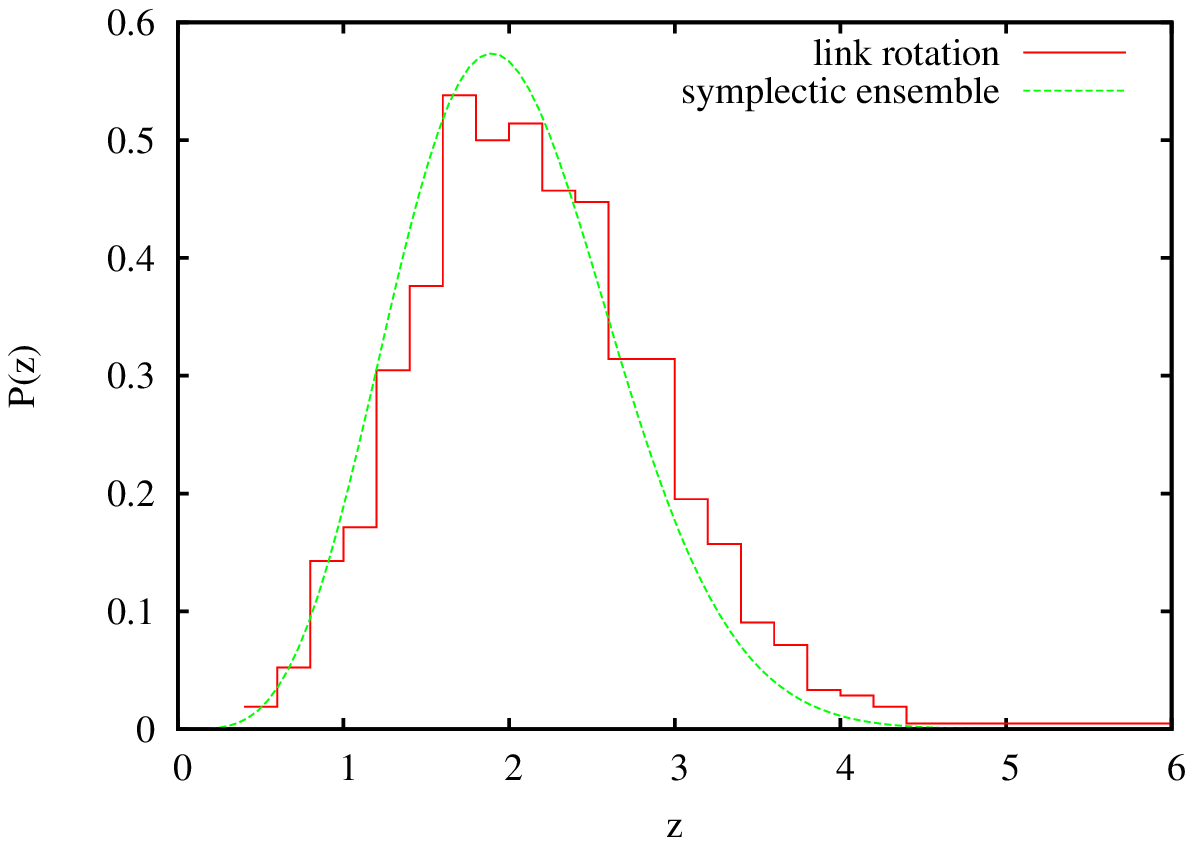}
	d)\includegraphics[width=.48\linewidth]{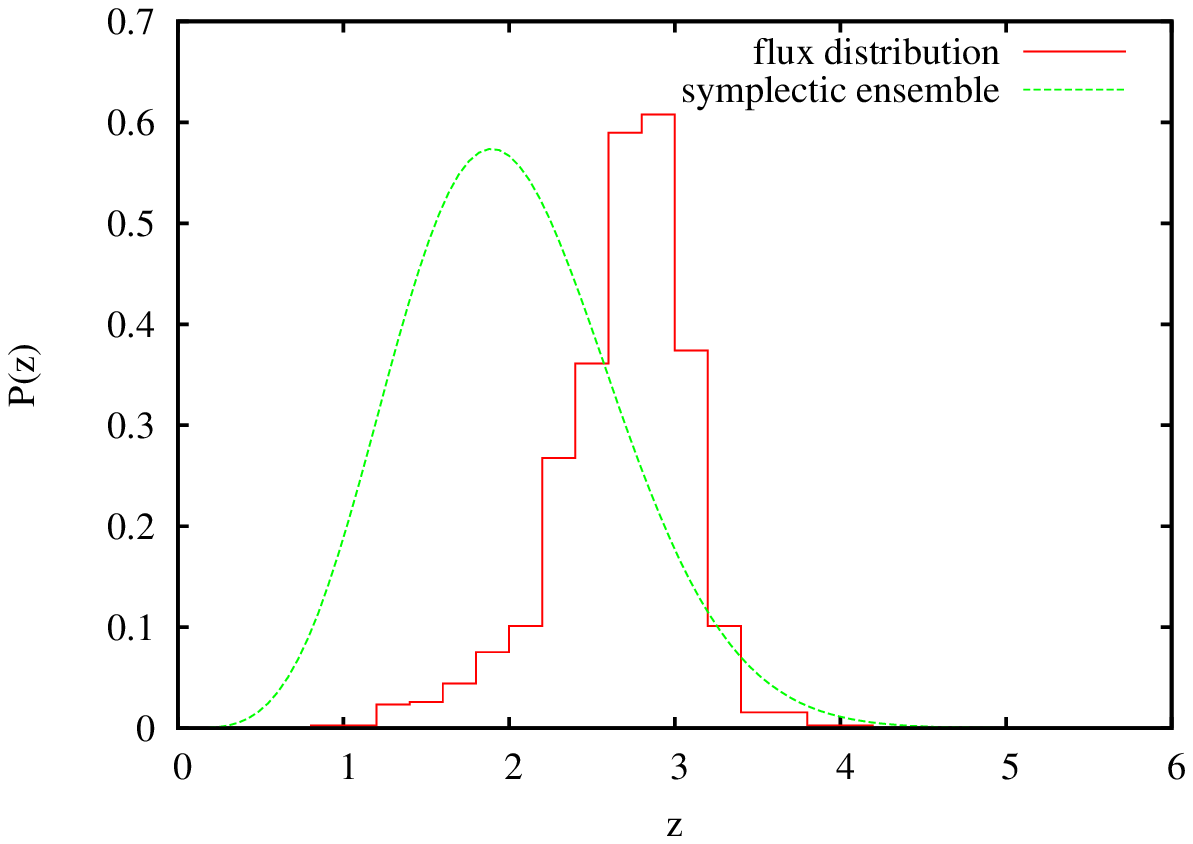}
	e)\includegraphics[width=.48\linewidth]{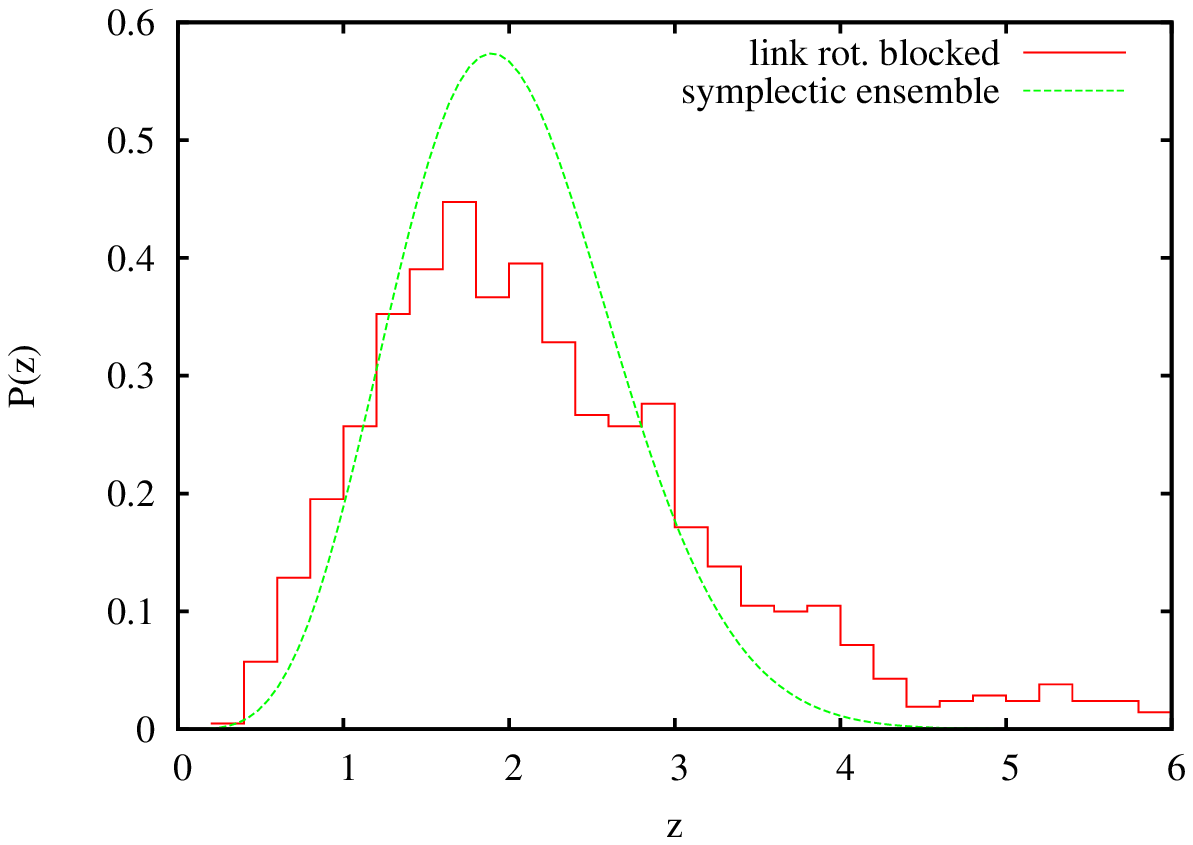}
	f)\includegraphics[width=.48\linewidth]{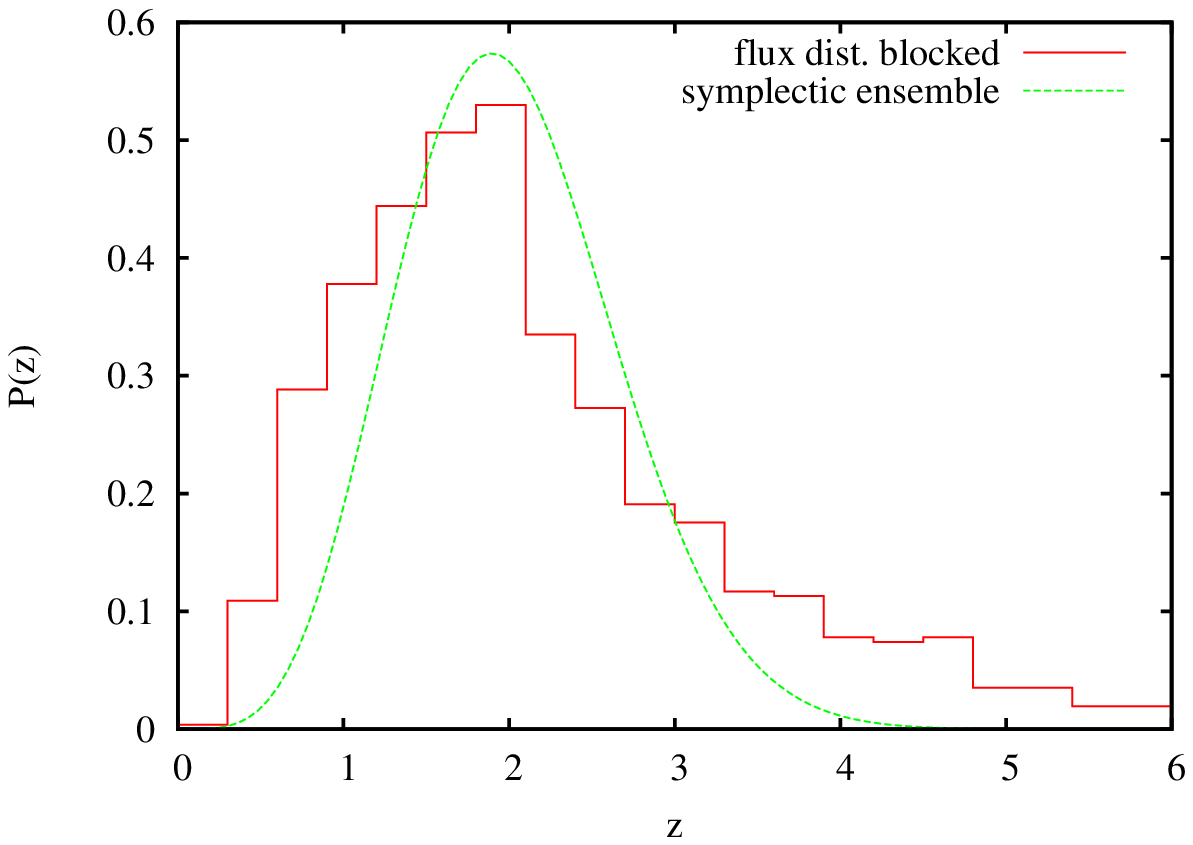}
	\caption{Distribution of lowest staggered eigenvalues in the fundamental
		representation of SU(2) (symplectic ensemble) for a) all, b) original
		and c-f) smeared configurations:  c) link rotation and d) flux
	distribution smearing and their blocked versions in e) and f), respectively.}
	\label{fig:zovbb}
\end{figure}

\clearpage
\newpage

\bibliographystyle{utphys}
\bibliography{../literatur}

\end{document}